\documentclass[acmtocl,acmnow]{acmtrans2m}

\usepackage{times}
\usepackage{amsmath}
\usepackage{amssymb}

\newtheorem{thm}{Theorem}[section]
\newtheorem{cor}{Corollary}[section]
\newtheorem{lem}{Lemma}[section]

\newtheorem{defn}{Definition}[section]
\newtheorem{prop}{Proposition}[section]
\newtheorem{exam}{Example}[section]

\def\>{\ensuremath{\rangle}}
\def\<{\ensuremath{\langle}}

\newcommand{\rto}[1]{\stackrel{#1}\rightarrow}

\title{An Algebra of Quantum Processes}

\author{MINGSHENG YING\\Tsinghua University and University of Technology,
Sydney \and
YUAN FENG and RUNYAO DUAN\\ Tsinghua University \and ZHENGFENG JI\\
Institute of Software, Chinese Academy of Sciences}
\date{}

\begin{abstract}
We introduce an algebra qCCS of pure quantum processes in which
communications by moving quantum states physically are allowed and
computations are modeled by super-operators, but no classical data
is explicitly involved. An operational semantics of qCCS is
presented in terms of (non-probabilistic) labeled transition
systems. Strong bisimulation between processes modeled in qCCS is
defined, and its fundamental algebraic properties are established,
including uniqueness of the solutions of recursive equations. To
model sequential computation in qCCS, a reduction relation between
processes is defined. By combining reduction relation and strong
bisimulation we introduce the notion of strong
reduction-bisimulation, which is a device for observing interaction
of computation and communication in quantum systems. Finally, a
notion of strong approximate bisimulation (equivalently, strong
bisimulation distance) and its reduction counterpart are introduced.
It is proved that both approximate bisimilarity and approximate
reduction-bisimilarity are preserved by various constructors of
quantum processes. This provides us with a formal tool for observing
robustness of quantum processes against inaccuracy in the
implementation of its elementary gates.
\end{abstract}




\begin{document}

\setcounter{page}{1}

\begin{bottomstuff}
This work was partly supported the National Natural Science
Foundation of China (Grant No: 60736011, 60621062) and the National
Key Project for Fundamental Research of China (Grant No:
2007CB807901).
\newline
Author's address: Mingsheng Ying (Corresponding author), State Key
Laboratory of Intelligent Technology and Systems, Tsinghua National
Laboratory for Information Science and Technology, Department of
Computer Science and Technology, Tsinghua University, Beijing
100084, China, and Center of Quantum Computation and Intelligent
Systems, Faculty of Information Technology, University of
Technology, Sydney, City Campus, 15 Broadway, Ultimo, NSW 2007,
Australia, email: yingmsh@tsinghua.edu.cn; Yuan Feng and Runyao
Duan, State Key Laboratory of Intelligent Technology and Systems,
Tsinghua National Laboratory for Information Science and Technology,
Department of Computer Science and Technology, Tsinghua University,
Beijing 100084, China; Zhengfeng Ji, State Key Laboratory of
Computer Science, Institute of Software, Chinese Academy of
Sciences, Beijing 100080, China
\end{bottomstuff}

\maketitle

\section{Introduction}

Quantum information science is usually divided into two subareas:
quantum computation and quantum communication. Quantum computation
offers the possibility of considerable speedup over classical
computation by exploring the power of superposition of quantum
states. Two striking examples of quantum algorithms are Shor's
quantum factoring and Grover's quantum searching. On the other hand,
some communication protocols are proposed by employing quantum
mechanical principles (in particular, the no-cloning property and
entanglement), for example BB84 and B92, which are provably secure.
Quantum communication systems using these protocols are already
commercially available from Id Quantique, MagiQ Technologies and
NEC.

The studies of quantum process algebras allow us to glue the two
subareas of quantum information science. To provide formal
techniques for modeling, analysis and verification of quantum
communication protocols, Gay and Nagarajan~\cite{GN05}, ~\cite{GN06}
defined a language CQP (Communicating Quantum Processes), which is
obtained from the pi-calculus by adding primitives for measurements
and transformations of quantum states and allowing transmission of
qubits. They gave an operational semantics and presented a type
system for CQP, and in particular proved that the semantics
preserves typing and that typing guarantees that each qubit is owned
by a unique process within a system. To model concurrent quantum
computation, Jorrand and Lalire~\cite{JL04}, ~\cite{JL05},
~\cite{L06}, ~\cite{LJ04} defined a language QPAlg (Quantum Process
Algebra). It is obtained by adding primitives expressing unitary
transformations and quantum measurements, as well as communications
of quantum states, to a classical process algebra, which is similar
to CCS. An operational semantics of QPAlg is given, and further a
probabilistic branching bisimulation between quantum processes
modeled in QPAlg is defined.

In this paper, we introduce a new algebra of quantum processes,
qCCS, which is a quantum generalization of CCS. The design decision
of qCCS differs from that of the previous quantum process algebras
in the following two aspects: (1) The driving idea of the design of
CQP is to provide formal model for analyzing quantum communication
protocols. Almost all of the existing quantum protocols involve
transmission of both classical and quantum data. The purpose of
designing QPAlg is to model cooperation between quantum and
classical computations. Thus, these quantum process algebras have to
accommodate quantum communication as well as classical
communication. The aim of the present paper is different, and we
mainly want to provide a suitable framework in which we can
understand the mechanism of quantum concurrent computation and
observe interaction and conjugation of computation and communication
in quantum systems. At the first step, it is reasonable to isolate
quantum data from classical data so that we have a much simpler
model in which a clearer understanding of quantum concurrent
computation may be achieved. So, we decide to focus our attention on
an algebra of purely quantum processes, not involving any classical
information. Of course,  in the future, after we have a thorough
understanding of purely quantum processes, qCCS can be extended by
adding classical ingredients. (2) The mathematical tools used to
describe transformations of quantum states in the previous quantum
process algebras are unitary operators. According the basic
postulates of quantum mechanics, unitary operators are suited to
depict the dynamics of closed quantum systems, but a more suitable
mathematical formalism for evolution of open quantum systems is
given in terms of super-operators. Since quantum process algebras
are mainly applied in modeling quantum concurrent systems in which
interactions between their subsystems happen frequently, and it
seems more reasonable to treat the involved systems as open systems,
we choose to use super-operators in describing transformations of
quantum states. Indeed, the usage of super-operators in qCCS was
influenced by Selinger's denotational semantics for his quantum
functional programming language QPL~\cite{S04}.

There are still some technical differences between qCCS and the
previous quantum process algebras. First, the treatment of quantum
variables and their substitutions is a key ingredient in defining
the operational and bisimulation semantics of qCCS. This was not
addressed in the previous works. It was already realized
in~\cite{FDJY07}, ~\cite{GN05}, ~\cite{GN06}, ~\cite{JL04},
~\cite{JL05}, ~\cite{L06}, ~\cite{LJ04} that one should consider
passing of the quantum systems used to express certain quantum
information instead of passing of the quantum information itself,
due to the no-cloning property of quantum information~\cite{WZ82}.
Hence, quantum variables must be explicitly introduced to denote the
quantum systems under consideration. In treating quantum variables
in qCCS, we follow the way of manipulating names in the
pi-calculus~\cite{MPW92}. But a serious difference is that distinct
quantum variables cannot be substituted by the same quantum
variable, complying with, again, the no-cloning theorem of quantum
information. Second, as in classical process algebras, operational
semantics of quantum processes is presented in terms of transitions
between configurations. However, a quantum variable and its current
state have to be separated in order to avoid abuse of quantum
information which may violate the no-cloning theorem. Thus, a
quantum configuration defined in~\cite{FDJY07}, ~\cite{GN05},
~\cite{GN06}, ~\cite{JL04}, ~\cite{JL05}, ~\cite{L06}, ~\cite{LJ04}
consists of a quantum process together with state information of the
involved quantum variables. In this paper, a configuration is
required to record state information of all quantum variables (not
only those occurring in the process under consideration). Although a
configuration defined in this way includes some unnecessary
information, it allows us to simplify considerably our presentation.
(Note that such a simple idea is widely used in mathematical logic;
for example, it simplifies the presentation of propositional logic
in the following way: in evaluating a given propositional formula we
only need to know the truth values assigned to the propositional
variables occurring in this formula, but a truth valuation is
generally defined to be an assignment of truth values to all
propositional variables.) Third, in the previous
works~\cite{FDJY07}, ~\cite{GN05}, ~\cite{GN06}, ~\cite{JL04},
~\cite{JL05}, ~\cite{L06}, ~\cite{LJ04}, the operational semantics
of a quantum process algebra is always defined to be a probabilistic
transition system, but this paper presents a non-probabilistic
operational semantics of qCCS. This is realized by treating quantum
measurements as super-operators (see Example~\ref{ex-op}(2) and
(3)), and it considerably simplifies the bisimulation semantics of
qCCS. Nevertheless, probabilistic information still can be retrieved
from such a non-probabilistic semantics via Eq.~(\ref{m-prob})
below. Fourth, only the notion of exact bisimulation is generalized
to quantum processes in~\cite{FDJY07}, ~\cite{L06}. Recall that a
set of classical gates is universal if it can be used to compute
exactly an arbitrary boolean function. However, exact universality
does not make sense in quantum computation because all quantum gates
form a continuum which cannot be generated by a finite set of
quantum gates. Instead, a set of quantum gates is said to be
universal provided any quantum gate can be approximated to arbitrary
accuracy by a circuit constructed from the gates in this set. To
describe approximation between quantum processes and, in particular,
implementation of a quantum process by some (usually finitely many)
special quantum gates, an approximate version of bisimulation (or
equivalently, bisimulation distance) is still missing. Recently, the
first author~\cite{Y01}, ~\cite{Y02}, ~\cite{YM00} and van
Breugel~\cite{VB05} among others introduced the notion of
approximate bisimulation for classical processes in which a distance
between actions is presumed. In the present paper, both exact and
approximate bisimulations are defined in qCCS, the latter using a
distance between super-operators induced naturally from the trace
distance of quantum states. We believe that approximate
bisimulations are appropriate formal tools for analyzing robustness
of quantum processes against inaccuracy in the implementation of its
elementary gates.

This paper is organized as follows: Section 2 reviews some basic
notions, needed in the subsequent sections, from quantum theory. In
Section 3 we define the syntax and an operational semantics of qCCS
and give some simple examples to illustrate the expressive power of
qCCS. The notion of strong bisimulation between quantum processes is
introduced, monoid and expansion laws as well as congruence and
recursive properties of strong bisimilarity are established, and
uniqueness of solutions of equations with respect to strong
bisimilarity is presented in Section 4. In Section 5, we first
define a reduction relation between strings of quantum operations
and then extend it to a reduction between quantum processes. The
notion of strong reduction-bisimilarity is defined by combining
reduction relation and strong bisimilarity, and it is shown to be
congruent under the process constructors in qCCS. In Section 6, the
notions of approximate strong bisimilarity and
reduction-bisimilarity are proposed and their corresponding metrics
are defined. It is proved that all process constructors are
non-expansive with respect to both strong bisimulation metric and
reduction-bisimulation metric. Section 7 is the concluding section
where we draw a brief conclusion and mention some topics for further
studies. For readability, we put the detailed proofs of some
propositions in the Appendix.

\section{Preliminaries}

For convenience of the reader we briefly recall some basic notions
from quantum theory and fix the notations needed in the sequel. We
refer to \cite{NC00} for more details.

\subsection{Hilbert spaces}

An isolated physical system is associated with a Hilbert space which
is called the state space of the system. In this paper, we mainly
consider finite-dimensional and countably infinite-dimensional
Hilbert spaces. A finite-dimensional Hilbert space is a complex
vector space $\mathcal{H}$ together with an inner product which is a
mapping $\langle\cdot|\cdot\rangle:\mathcal{H}\times
\mathcal{H}\rightarrow \mathbf{C}$ satisfying the following
properties: \begin{enumerate}\item
$\langle\varphi|\varphi\rangle\geq 0$ with equality if and only if
$|\varphi\rangle =0$; \item
$\langle\varphi|\psi\rangle=\langle\psi|\varphi\rangle^{\ast}$;
\item $\langle\varphi|\lambda_1\psi_1+\lambda_2\psi_2\rangle=
\lambda_1\langle\varphi|\psi_1\rangle+\lambda_2\langle\varphi|\psi_2\rangle$,\end{enumerate}
where $\mathbf{C}$ is the set of complex numbers, and
$\lambda^{\ast}$ stands for the conjugate of $\lambda$ for each
complex number $\lambda\in \mathbf{C}$. All countably
infinite-dimensional Hilbert spaces considered in this paper will be
simply treated as tensor products of countably infinitely many
finite-dimensional Hilbert spaces (see Subsection~\ref{tensor}
below).

\begin{exam} Let $n\geq 1$. For any
$|\varphi\rangle=(x_1,...,x_n)^{T}, |\psi\rangle =
(y_1,...,y_n)^{T}\in \mathbf{C}^{n}$ and $\lambda\in \mathbf{C}$, we
define: $$|\varphi\rangle +|\psi\rangle
=(x_1+y_1,...,x_n+y_n)^{T},$$ $$\lambda |\varphi\rangle =(\lambda
x_1, ..., \lambda x_n)^{T},$$ where $^{T}$ stands for transpose.
Then $\mathbf{C}^{n}$ is a vector space. We often write $\langle
\varphi|$ for the adjoint $|\varphi\rangle^{\dag}$ of
$|\varphi\rangle$. Furthermore, we define $\langle
\cdot|\cdot\rangle$ in $\mathbf{C}^{n}$ as follows: $$\langle
\varphi|\psi\rangle =\sum_{i=1}^{n}x_i^{\ast}y_i.$$ Then
$(\mathbf{C}^{n},\langle\cdot|\cdot\rangle)$ is an $n-$dimensional
Hilbert space. Indeed, each $n-$dimensional Hilbert space is
isometric to $\mathbf{C}^{n}$. In particular, a qubit is a physical
system whose state space is
$\mathcal{H}_2=\mathbf{C}^{2}$. If we write $|0\rangle=\left(\begin{array}{cc}1\\
0\end{array}\right)$ and $|1\rangle=\left(\begin{array}{cc}0\\
1\end{array}\right),$ corresponding to one-bit classical values and
called the computational basis, then a qubit has state $\alpha
|0\rangle +\beta |1\rangle$ with $\alpha,\beta\in \mathbf{C}$ and
$|\alpha|^{2}+|\beta|^{2}=1$. The Hadamard basis consists of the
following two states: $$|+\rangle =\frac{1}{\sqrt{2}}(|0\rangle +
|1\rangle),\hspace{2em} |-\rangle =\frac{1}{\sqrt{2}}(|0\rangle -
|1\rangle).$$
\end{exam}

For any vector $|\psi\rangle$ in $\mathcal{H}$, its length
$||\psi||$ is defined to be $\sqrt{\langle\psi|\psi\rangle}$. A pure
state of a quantum system is a unit vector in its state space; that
is, a vector $|\psi\rangle$ with $||\psi||=1$. An orthonormal basis
of a Hilbert space $\mathcal{H}$ is a basis $\{|i\rangle\}$ with
$$\langle i|j\rangle=\begin{cases}1, & \mbox{if }i=j,\\ 0, &
\mbox{otherwise}.\end{cases}$$ Then the trace of a linear operator
$A$ on $\mathcal{H}$ is defined to be
$$tr(A)=\sum_{i}\langle i|A|i\rangle.$$ A mixed state of quantum
system is represented by a density operator. A density operator in a
Hilbert space $\mathcal{H}$ is a linear operator $\rho$ on it
fulfilling the following conditions: \begin{enumerate}\item $\rho$
is positive in the sense that $\langle \psi|\rho|\psi\rangle \geq 0$
for all $|\psi\rangle$; \item $tr(\rho)=1$.\end{enumerate} An
equivalent concept of density operator is an ensemble of pure
states. An ensemble is a set of the form $\{(p_i,|\psi_i\rangle)\}$
such that $p_i \geq 0$ and $|\psi_i\rangle$ is a pure state for each
$i$, and $\sum_{i}p_i=1$. Then
$$\rho=\sum_{i}p_i|\psi_i\rangle\langle\psi_i|$$ is a density
operator, and conversely each density operator can be generated by
an ensemble of pure states in this way. A positive operator $\rho$
is called a partial density operator if $tr(\rho)\leq 1$. We write
$\mathcal{D}(\mathcal{H})$ for the set of partial density operators
on $\mathcal{H}$.

\subsection{Unitary operators}
The evolution of a closed quantum system is described by a unitary
operator on its state space. A linear operator $U$ on a Hilbert
space $\mathcal{H}$ is said to be unitary if
$U^{\dag}U=I_\mathcal{H},$ where $I_\mathcal{H}$ is the identity
operator on $\mathcal{H}$, and $U^{\dag}$ is the adjoint of $U$. If
the states of the system at times $t_1$ and $t_2$ are $\rho_1$ and
$\rho_2$, respectively, then $$\rho_2=U\rho_1U^{\dag}$$ for some
unitary operator $U$ which depends only on $t_1$ and $t_2$. In
particular, if $\rho_1$ and $\rho_2$ are pure states
$|\psi_1\rangle$ and $|\psi_2\rangle$, respectively; that is,
$\rho_1=|\psi_1\rangle\langle \psi_1|$ and
$\rho_2=|\psi_2\rangle\langle \psi_2|$, then we have $|\psi_2\rangle
=U|\psi_1\rangle$.

\begin{exam}The most frequently used unitary operators on qubits are the Hadamard transformation:
$$H=\frac{1}{\sqrt{2}}\left(\begin{array}{cc}1 & 1\\
1 & -1\end{array}\right),$$ and the Pauli matrices:
$$I=\left(\begin{array}{cc}1 & 0\\ 0 & 1\end{array}\right),\hspace{2em} \sigma_x=\left(\begin{array}{cc}0 & 1\\ 1 & 0\end{array}\right),$$
$$\sigma_y=\left(\begin{array}{cc}0 & -i\\ i & 0\end{array}\right),\hspace{2em} \sigma_z=\left(\begin{array}{cc}1 & 0\\ 0 & -1\end{array}\right).$$
\end{exam}

\subsection{Quantum measurement}

A quantum measurement is described by a collection $\{M_m\}$ of
measurement operators, where the indexes $m$ refer to the
measurement outcomes. It is required that the measurement operators
satisfy the completeness equation
$$\sum_{m}M_m^{\dag}M_m=I_\mathcal{H}.$$ If the system is in state
$\rho$, then the probability that measurement result $m$ occurs is
given by $$p(m)=tr(M_m^{\dag}M_m\rho),$$ and the state of the system
after the measurement is $$\frac{M_m\rho M_m^{\dag}}{p(m)}.$$ For
the case that $\rho$ is a pure state $|\psi\rangle$, we have
$p(m)=||M_m|\psi\rangle||^{2}$, and the post-measurement state is
$$\frac{M_m|\psi\rangle}{\sqrt{p(m)}}.$$

\begin{exam}The measurement on qubits in the computational basis
consists of $P_0=|0\rangle \langle 0|$ and $P_1=|1\rangle \langle
1|$. If we perform it on a qubit which is in state $\alpha |0\rangle
+\beta |1\rangle$, then either the result $0$ will be obtained, with
probability $|\alpha|^{2}$, or the result $1$, with probability
$|\beta|^{2}$.
\end{exam}

\subsection{Tensor products}\label{tensor}

The state space of a composite system is the tensor product of the
state spaces of its components. Let $\mathcal{H}_1$ and
$\mathcal{H}_2$ be two Hilbert spaces. Then their tensor product
$\mathcal{H}_1\otimes \mathcal{H}_2$ consists of linear combinations
of vectors $|\psi_1\psi_2\rangle =|\psi_1\rangle\otimes
|\psi_2\rangle$ with $|\psi_1\rangle\in \mathcal{H}_1$ and
$|\psi_2\rangle\in \mathcal{H}_2$.

For any linear operator $A_1$ on $\mathcal{H}_1$ and $A_2$ on
$\mathcal{H}_2$, $A_1\otimes A_2$ is an operator on
$\mathcal{H}_1\otimes \mathcal{H}_2$ and it is defined by
$$(A_1\otimes A_2)|\psi_1\psi_2\rangle = A_1|\psi_1\rangle\otimes
A_2|\psi_2\rangle$$ for each $|\psi_1\rangle \in \mathcal{H}_1$ and
$|\psi_2\rangle \in \mathcal{H}_2$.

Let
$|\varphi\rangle=\sum_{i}\alpha_i|\varphi_{1i}\varphi_{2i}\rangle$
and $|\psi\rangle =\sum_{j}\beta_j|\psi_{1j}\psi_{2j}\rangle\in
\mathcal{H}_1\otimes \mathcal{H}_2$. Then their inner product is
defined as follows: $$\langle
\varphi|\psi\rangle=\sum_{i,j}\alpha_i^{\ast}\beta_j\langle\varphi_{1i}|\psi_{1j}\rangle\langle
\varphi_{2i}|\psi_{2j}\rangle.$$

\begin{exam}A composite quantum system can exhibit the phenomenon of
entanglement. A state of a composite system is an entangled state if
it cannot be written as a product of states of its component
systems. The following are maximally entangled states of two-qubits,
called Bell states:
$$|\beta_{00}\rangle =\frac{1}{\sqrt{2}}(|00\rangle +
|11\rangle),\hspace{2em} |\beta_{01}\rangle
=\frac{1}{\sqrt{2}}(|01\rangle + |10\rangle),$$
$$|\beta_{10}\rangle =\frac{1}{\sqrt{2}}(|00\rangle -
|11\rangle),\hspace{2em} |\beta_{11}\rangle
=\frac{1}{\sqrt{2}}(|01\rangle - |10\rangle).$$
\end{exam}

The notion of tensor product may be easily generalized to the case
of any finite number of Hilbert spaces. The tensor product of
countably infinitely many finite-dimensional Hilbert spaces is a
countably infinite-dimensional Hilbert space isometric to $l^{2}$ of
sequences $\{x_n\}_{n=0}^{\infty}$ of complex numbers such that
$\sum_{n=0}^{\infty}|x_n|^{2}$ converges. The vector addition,
scalar multiplication and inner product are defined as follows:
$$|\varphi\rangle+|\psi\rangle=\{x_n+y_n\}_{n=0}^{\infty},$$
$$\lambda |\varphi\rangle =\{\lambda x_n\}_{n=0}^{\infty},$$
$$\langle \varphi |\psi\rangle=\sum_{n=0}^{\infty}x_n^{\ast}y_n$$
for any $|\varphi\rangle=\{x_n\}_{n=0}^{\infty},
|\psi\rangle=\{y_n\}_{n=0}^{\infty}\in l^{2}$ and
$\lambda\in\mathbf{C}$. It is easy to see that $l^{2}$ enjoys the
following completeness: if $\{|\varphi_n\rangle\}_{n=0}^{\infty}$ is
a Cauchy sequence in $l^{2}$, i.e. for any $\epsilon>0$, there
exists positive integer $N$ such that
$||\varphi_m-\varphi_n||<\epsilon$ for all $m,n\geq N$, then exists
$|\varphi\rangle\in l^{2}$ with
$\lim_{n\rightarrow\infty}|\varphi_n\rangle=|\varphi\rangle$, i.e.
for any $\epsilon>0$, there exists positive integer $N$ such that
$||\varphi_n-\varphi||<\epsilon$ for all $n\geq N$.

The notion of tensor product of two linear operators can be
generalized to the case of more than two operators and the case of
countably infinitely many operators in a natural way. Since density
operators are special linear operators, their tensor product is then
well-defined. A basic postulate of quantum mechanics asserts that if
component system $i$ is in state $\rho_i$ for each $i$, then the
state of the composite system is $\bigotimes_i\rho_i$.

\subsection{Super-operators}

The dynamics of open quantum systems cannot be described by unitary
operators, and one of its mathematical formalisms is the notion of
super-operator. A super-operator on a Hilbert space $\mathcal{H}$ is
a linear operator $\mathcal{E}$ from the space of linear operators
on $\mathcal{H}$ into itself which satisfies the following two
conditions: \begin{enumerate} \item $tr[\mathcal{E}(\rho)]\leq
tr(\rho)$ for each $\rho\in \mathcal{D}(\mathcal{H})$; \item
Complete positivity: for any extra Hilbert space $\mathcal{H}_R$,
$(\mathcal{I}_R\otimes \mathcal{E})(A)$ is positive provided $A$ is
a positive operator on $\mathcal{H}_R\otimes \mathcal{H}$, where
$\mathcal{I}_R$ is the identity operation on
$\mathcal{H}_R$.\end{enumerate} If (1) is strengthened to
$tr[\mathcal{E}(\rho)]=tr(\rho)$ for all $\rho\in
\mathcal{D}(\mathcal{H})$, then $\mathcal{E}$ is said to be
trace-preserving.

\begin{exam}\label{ex-op}\begin{enumerate}\item Let $U$ be a unitary operator on Hilbert space $\mathcal{H}$,
and $\mathcal{E}(\rho)=U\rho U^{\dag}$ for any $\rho\in
\mathcal{D}(\mathcal{H})$. Then $\mathcal{E}$ is a trace-preserving
super-operator.
\item Let $\{M_m\}$ be a quantum measurement on $\mathcal{H}$. For each $m$, we
define $\mathcal{E}_m(\rho)=M_m\rho M_m^{\dag}$ for any $\rho\in
\mathcal{D}(\mathcal{H})$. Then $\mathcal{E}_m$ is a super-operator,
which is not necessarily trace-preserving. If the state of the
system immediately before the measurement is $\rho$, then the
probability of obtaining measurement result $m$ is
\begin{equation}\label{m-prob}p(m)=tr(\mathcal{E}_m(\rho)),\end{equation}
and the state of the system immediately after the measurement is
$$\mathcal{E}_m(\rho)/tr(\mathcal{E}_m(\rho)).$$
\item As in (2), let $\{M_m\}$ be a quantum measurement on $\mathcal{H}$. If $\mathcal{E}$ is given by this
measurement, with the result of the measurement unknown, i.e.,
$$\mathcal{E}(\rho)=\sum_{m}M_m\rho M_m^{\dag}$$ for each $\rho\in
\mathcal{D}(\mathcal{H})$, then $\mathcal{E}$ is a trace-preserving
super-operator.\end{enumerate}\end{exam}

The following theorem gives two elegant representations of
super-operators.

\begin{lem}\label{kraus} (\cite{NC00}, Section 8.2.3; Theorem 8.1)
The following three statements are equivalent:
\begin{enumerate}
\item $\mathcal{E}$ is a super-operator on Hilbert space $\mathcal{H}$;
\item (System-environment model) There are an environment system $E$
with state space $\mathcal{H}_E$, and a unitary transformation $U$
and a projector $P$ on $\mathcal{H}\otimes \mathcal{H}_E$ such that
$$\mathcal{E}(\rho)=tr_E[PU(\rho\otimes |e_0\rangle\langle
e_0|)U^{\dag}P]$$ for any $\rho\in \mathcal{D}(\mathcal{H})$, where
$\{|e_k\rangle\}$ is an orthonormal basis of $\mathcal{H}_E$, and
$tr_E(\cdot)$ is defined by $$tr_E(\sigma)=\sum_{k}\langle
e_k|\sigma|e_k\rangle$$ for any $\sigma\in
\mathcal{D}(\mathcal{H}\otimes \mathcal{H}_E)$;
\item (Kraus operator-sum representation) There exists a set of
operators $\{E_i\}$ on $\mathcal{H}$ such that
$\sum_{i}E_i^{\dag}E_i\sqsubseteq I$ and
$$\mathcal{E}(\rho)=\sum_{i}E_i\rho E_i^{\dag}$$ for all density
operators $\rho\in \mathcal{D}(\mathcal{H})$, where $\sqsubseteq$
stands for the L$\ddot{o}$wner order; that is, $A\sqsubseteq B$ if
and only if $B-A$ is a positive operator. We often say that
$\mathcal{E}$ is represented by the set $\{E_i\}$ of operators, or
$\{E_i\}$ are operation elements giving rise to $\mathcal{E}$ when
$\mathcal{E}$ is given by the above
equation.\end{enumerate}\end{lem}

\subsection{Diamond distance between super-operators}\label{dis}

We shall need a distance between super-operators in defining
approximate bisimulation between quantum processes. We choose to use
a natural extension of trace distance between mixed quantum states.
For any positive operator $A$, if $A=\sum_{i}\lambda_i
|i\rangle\langle i|$, $\lambda_i\geq 0$ for all $i$, is a spectral
decomposition of $A$, then we define
$$\sqrt{A}=\sum_{i}\sqrt{\lambda_i}|i\rangle\langle i|.$$ Furthermore,
for any operator $A$, we define $|A|=\sqrt{A^{\dag}A}$. One of the
most popular metrics measuring how close two quantum states are,
used by the quantum information community, is trace distance. For
any $\rho,\sigma\in \mathcal{D}(\mathcal{H})$, their trace distance
is defined to be $$D(\rho,\sigma)=\frac{1}{2}tr|\rho-\sigma|.$$
$D(\rho,\sigma)$ quantifies the distinguishability between mixed
states $\rho$ and $\sigma$. The following property of trace distance
is needed in the sequel.

\begin{lem}\label{op-dist} (\cite{NC00}, Theorem 9.2) If $\mathcal{E}$ is a trace-preserving super-operator on $\mathcal{H}$,
then $$D(\mathcal{E}(\rho),\mathcal{E}(\sigma))\leq D(\rho,\sigma)$$
for any $\rho,\sigma\in \mathcal{D}(\mathcal{H}).$
\end{lem}

The notion of trace distance can be extended to the case of
super-operators in a natural way~\cite{K97}. For any super-operators
$\mathcal{E}_1$, $\mathcal{E}_2$ on $\mathcal{H}$, their diamond
trace distance is defined to be $$D_\diamond
(\mathcal{E}_1,\mathcal{E}_2)  =\sup\{D((\mathcal{E}_1 \otimes
\mathcal{I}_{\mathcal{H}^{\prime}})(\rho), (\mathcal{E}_2\otimes
\mathcal{I}_{\mathcal{H}^{\prime}})(\rho)):
\rho\in\mathcal{D}(\mathcal{H}\otimes \mathcal{H}^{\prime})\}$$
where $\mathcal{H}^{\prime}$ ranges over all finite-dimensional
Hilbert spaces. $D_\diamond(\mathcal{E}_1,\mathcal{E}_2)$
characterizes the maximal probability that the outputs of
$\mathcal{E}_1$ and $\mathcal{E}_2$ can be distinguished for the
same input where auxiliary systems are allowed.

\section{Syntax and Operational Semantics}

\subsection{Syntax}

Let $Chan$ be the set of names for quantum channels, and let $Var$
be the set of quantum variables. It is assumed that $Var$ is a
countably infinite set. We shall use meta-variables $c,d,...$ to
range over $Chan$ and $x,y,z,...$ to range over $Var$. Let $\tau$ be
the name of silent action.

For each quantum variable $x\in Var$, imagine that we have a quantum
system named by $x$. Let $\mathcal{H}_x$ be a finite-dimensional
complex Hilbert space, which is the state space of the $x-$system.
For any $x, y\in Var$, if $\mathcal{H}_x=\mathcal{H}_y$, then it is
said that $x$ and $y$ have the same type. Imagine further that there
is a big quantum system composed of all $x-$systems, $x\in Var$, in
which all of our quantum processes live. We call this composed
system the environment of our calculus. Put
$$\mathcal{H}_X=\bigotimes_{x\in X}\mathcal{H}_x$$ for any $X\subseteq
Var$. Then $\mathcal{H}=\mathcal{H}_{Var}$ is the state space of the
environment. Note that $\mathcal{H}$ is a countably
infinite-dimensional Hilbert space.

We assume a set of process constant schemes, ranged over by
meta-variables $A, B, ...$. For each process constant $A$, a
nonnegative arity $ar(A)$ is assigned to it. Let
$\widetilde{x}=x_1,...,x_{ar(A)}$ be a tuple of distinct quantum
variables. Then $A(\widetilde{x})$ is called a process constant.

We write $\mathcal{P}$ for the set of quantum processes, and we
write $fv(P)$ for the set of free quantum variables in $P$ for each
quantum process $P\in\mathcal{P}$. Now we are ready to present the
syntax of qCCS.

\begin{defn}\label{syn}Quantum processes are defined inductively by the
following formation rules:
\begin{enumerate}\item each process constant $A(\widetilde{x})$ is in $\mathcal{P}$
and $fv(A(\widetilde{x}))=\{\widetilde{x}\}$; \item
$\mathbf{nil}\in\mathcal{P}$ and $fv(\mathbf{nil})=\emptyset;$ \item
if $P\in \mathcal{P}$, then $\tau.P\in \mathcal{P}$ and
$fv(\tau.P)=fv(P)$; \item if $P\in\mathcal{P}$, $X$ is a finite
subset of $Var$, and $\mathcal{E}$ is a super-operator on
$\mathcal{H}_X$, then $\mathcal{E}[X].P\in \mathcal{P}$ and
$fv(\mathcal{E}[X].P)=fv(P)\cup X$; \item if $P\in\mathcal{P}$, then
$c?x.P\in \mathcal{P}$, and $fv(c?x.P)=fv(P)-\{x\}$; \item if
$P\in\mathcal{P}$ and $x\notin fv(P)$, then $c!x.P\in \mathcal{P}$,
and $fv(c!x.P)=fv(P)\cup \{x\}$; \item if $P,Q\in \mathcal{P}$, then
$P+Q\in \mathcal{P}$ and $fv(P+Q)=fv(P)\cup fv(Q)$; \item if
$P,Q\in\mathcal{P}$ and $fv(P)\cap fv(Q)=\emptyset$, then
$P\|Q\in\mathcal{P}$ and $fv(P\|Q)=fv(P)\cup fv(Q)$; \item if $P\in
\mathcal{P}$ and $L\subseteq Chan$, then $P\backslash
L\in\mathcal{P}$ and $fv(P\backslash L)=fv(P)$.
\end{enumerate}
\end{defn}

Using the standard BNF grammar the syntax of qCCS can be summarized
as follows:
$$P::=A(\widetilde{x})\ |\ \mathbf{nil}\ |\ \tau.P\ |\ \mathcal{E}[X].P\ |\ c?x.P\ |\ c!x.P\ |\ P+P\ |\ P||P\
|\ P\backslash L.$$ It is similar to the syntax of classical CCS,
and the only differences between them are:
\begin{itemize}\item \ Clause 4 in the above definition allows us to perform quantum
operations on some involved systems; \item \ Condition $x\notin
fv(P)$ in clause 6 and condition $fv(P)\cap fv(Q)=\emptyset$ in
clause 8 are required due to the well-known fact that unknown
quantum information cannot be perfectly
cloned~\cite{WZ82}.\end{itemize} It is worth noting that these
conditions force us to assign a set of free quantum variables to
each process constant in advance. Quantum operations described in
clause 4 may be thought of as constructs for sequential quantum
computation. There are also constructs for sequential computation in
the value-passing CCS, but they are not explicitly given. There such
constructs are implicitly assumed in value expressions (see
\cite{M89}, page 55) so that one can focus his attention on
examining communication behaviors between processes. However, we
explicitly present the constructs for sequential quantum computation
in the syntax of qCCS, and it is one of our main purposes to observe
interaction between sequential quantum computation and communication
of quantum information.

There are two kinds of binding in our language for quantum
processes: the restriction $\backslash L$ binds all channel names in
$L$, and the input prefix $c?x$ binds quantum variable $x$. The
symbol $\equiv_\alpha$ will be used to denote alpha-convertibility
on processes defined by replacing bound quantum variables in the
standard way.

For each process constant scheme $A$, a defining equation of the
form $$A(\widetilde{x})\stackrel{def}{=}P$$ is assumed, where $P$ is
a process with $fv(P)\subseteq \{\widetilde{x}\}$. Recursive
definition in qCCS is different from that in classical CCS in some
intricate way. For example, in qCCS,
$$A(x)\stackrel{def}{=}c!x.A(x)$$ is not allowed to be the defining
equation of process constant scheme $A$. In fact, if $x\in fv(A(x))$
then $c!x.A(x)$ is not a process, and if $x\notin fv(A(x))$ then
$fv(c!x.A(x))\not\subseteq fv(A(x))$. However,
$$A(y)\stackrel{def}{=}c?x.c!x.A(y)$$ is a legitimate defining
equation of $A$.

It is well-known that in the pi-calculus one has to treat
substitution of names very carefully. However, we need to treat
substitution of quantum variables in an even more careful way due to
the fact that arbitrary cloning of quantum information is
prohibited~\cite{WZ82},~\cite{D82}. In particular, we have:

\begin{defn}A substitution of quantum variables is a one-to-one mapping $f$
from $Var$ into itself satisfying \begin{enumerate}\item $x$ and
$f(x)$ have the same type for all $x\in Var$; and
\item
$f|_{Var-X}=Id_{Var-X}$ for some finite subset $X$ of $Var$, where
$Id_Y$ stands for the identity function on $Y$.
\end{enumerate}
\end{defn}

It is common that two different classical variables can be
substituted by the same variable. But it is not the case in qCCS
because a substitution is required to be a bijection. Such a
requirement comes reasonably from our intention that different
variables are references to different quantum systems. Since quantum
variable $f(x)$ will be used to substitute quantum variable $x$, it
is reasonable to require that the $x-$system and the $f(x)-$system
have the same state space. This is exactly condition (1) in the
above definition.

Let $P\in \mathcal{P}$ and $f$ be a substitution. Then $Pf$ denotes
the process obtained from $P$ by simultaneously substituting $f(x)$
for each free occurrence of $x$ in $P$ for all $x$. To give a
precise definition of $Pf$, we need to introduce the notion of
application of a substitution on a super-operator. If $f$ is a
one-to-one mapping from $Var$ into itself, then $f$ induces
naturally an isomorphism from $\mathcal{H}$ onto
$\mathcal{H}_{f(Var)}$, which is a subspace of $\mathcal{H}$. For
simplicity, it is also denoted by $f$. Precisely, the isomorphism
$f: \mathcal{H}\rightarrow \mathcal{H}_{f(Var)}$ is defined as
follows:
$$f(\bigotimes_{x\in Var}|\varphi_x\rangle_x) =\bigotimes_{x\in
Var}|\varphi_{x}\rangle_{f(x)}$$ for any $|\varphi_x\rangle\in
\mathcal{H}_x$, $x\in Var$. Applying $f$ to a state which is not a
tensor product of states in $\mathcal{H}_x$ ($x\in Var$) may be
carried out simply by linearity. Furthermore, it induces a bijection
$f: \mathcal{D}(\mathcal{H})\rightarrow
\mathcal{D}(\mathcal{H}_{f(Var)})$. For any
$\rho=\sum_{i}p_i|\varphi_i\rangle\langle\varphi_i|\in
\mathcal{D}(\mathcal{H})$, where $|\varphi_i\rangle\in\mathcal{H}$
for all $i$, we have: $$f(\rho)=\sum_ip_i|f(\varphi_i)\rangle\langle
f(\varphi_i)|.$$ In particular, if $f(x)=y$, $f(y)=x$ and $f(z)=z$
for all $z\neq x,y$, then $f(\rho)$ is often written as
$\rho\{y/x\}$.

For any super-operator $\mathcal{E}$ on $\mathcal{H}_X$, we define
super-operator $\mathcal{E}f$ on $\mathcal{H}_{f(X)}$ by
$$\mathcal{E}f=f|_X\circ \mathcal{E}\circ (f|_X)^{-1},$$ where $f|_X$
is the restriction of $f$ on $X$, which is obviously a bijection
from $X$ onto $f(X)$.

$$
 \begin{array}{ccc}
 \mathcal{D}(\mathcal{H}_X) & \overset{\mathcal{E}}{\longrightarrow } &
 \mathcal{D}
(\mathcal{H}_X)  \\
    &  &  \\
 f\downarrow  &  & \downarrow f \\
    &  &  \\
   \mathcal{D}(\mathcal{H}_{f(X)}) & \underset{\mathcal{E}f}{\longrightarrow } &
   \mathcal{D}(\mathcal{H}_{f(X)})\
 \end{array}
$$

With the above preliminaries, now we are able to define substitution
of quantum variables in a quantum process.

\begin{defn} For any $P\in \mathcal{P}$ and substitution $f$, $Pf$ is defined recursively as follows:\begin{enumerate}
\item if $P$ is a process constant $A(x_1,...,x_n)$ then
$$Pf=A(f(x_1),...,f(x_n));$$
\item if
$P=\mathbf{nil}$ then $Pf=\mathbf{nil}$; \item if
$P=\tau.P^{\prime}$ then $Pf=\tau.P^{\prime}f$; \item if
$P=\mathcal{E}[X].P^{\prime}$ then
$Pf=(\mathcal{E}f)[f(X)].P^{\prime}f$; \item if $P=c?x.P^{\prime}$
then $Pf=c?y.P^{\prime}\{y/x\}f_y$, where $y\notin
fv(c?x.P^{\prime})\cup fv(P^{\prime}f)$, and $f_y$ is the
substitution with $f_y(y)=y$, $f_y(f^{-1}(y))=f(y)$ and
$f_y(z)=f(z)$ for all $z\neq y, f^{-1}(y)$;
\item if $P=c!x.P^{\prime}$ then $Pf=c!f(x).P^{\prime}f$;
\item if $P=P_1+P_2$ then $Pf=P_1
f+P_2 f$; \item if $P=P_1\|P_2$ then $Pf=P_1 f\|P_2 f$; \item if
$P=P^{\prime} \backslash L$ then $Pf=P^{\prime}f\backslash L$.
\end{enumerate}
\end{defn}

Note that in clause 4 a corresponding modification on super-operator
$\mathcal{E}$ is made when substituting quantum variables in $X$. In
addition, the requirement that $f$ is one-to-one becomes vital when
we consider substitution of output prefix in clauses 6 and of
parallel composition in clause 8; for example, if $f(x)=f(y)=x$, and
$$P_1=c!x.d!y.\mathbf{nil},\ \
P_2=c!x.\mathbf{nil}\|d!y.\mathbf{nil},$$ then the following two
expressions
$$P_1f=c!x.d!x.\mathbf{nil},\ \
P_2f=c!x.\mathbf{nil}\|d!x.\mathbf{nil}$$ are not processes.

If $(Pf)f^{-1}\equiv_\alpha P$; that is, there is no variable
conflict where $f(x)\in fv(P)-\{x\}$ for some $x\in fv(P)$, then
$Pf$ is said to be well-defined. In what follows we always assume
that $Pf$ is well-defined whenever it occurs.

Let $\widetilde{x}=x_1,...,x_n$ and $\widetilde{y}=y_1,...,y_n$. If
$f(x_i)=y_i$ $(1\leq i\leq n)$, we write
$P\{\widetilde{y}/\widetilde{x}\}$ or $P\{y_1/x_1,...,y_n/x_n\}$ for
$Pf$.

\subsection{Operational Semantics}

The operational semantics of qCCS will be given by transitions
between configurations, labeled by actions. A configuration is
defined to be a pair $\langle P,\rho\rangle$ where $P\in\mathcal{P}$
is a process, and $\rho\in \mathcal{D}(\mathcal{H})$ specifies the
current state of the environment. Intuitively, $\rho$ is an
instantiation (or valuation) of quantum variables. Instantiations of
classical variables can be made independently from each other, but
quantum systems represented by different variables may be correlated
because $\rho$ is allowed to be an entangled state. The set of
configurations is written $Con$.

We set $$Act =\{\tau\} \cup Act_{op} \cup Act_{com}$$ for the set of
actions, where $$Act_{op}=\{\mathcal{E}[X]:X\ {\rm is\ a\ finite\
subset\ of}\ Var\ {\rm and}\ \mathcal{E}\ {\rm is\ a}\ {\rm
super-operator\ on}\ \mathcal{H}_X\}$$ is the set of quantum
operations, and $$Act_{com} = \{c?x,c!x: c\in Chan\ {\rm and}\ x\in
Var\}$$ is the set of communication actions, including inputs and
outputs. The set $Act$ will be ranged over by meta-variables
$\alpha, \beta,...$. We need the following notations for
actions:\begin{itemize} \item\ For each $\alpha\in Act$, we use
$cn(\alpha)$ to stand for the channel name in action $\alpha$; that
is, $cn(c?x)=cn(c!x)=c$, and $cn(\tau)$ and $cn(\mathcal{E}[X])$ are
not defined. \item\ We write $fv(\alpha)$ for the set of free
variables in $\alpha$; that is, $fv(c!x)=\{x\}$,
$fv(\mathcal{E}[X])=X,$ $fv(\tau)=fv(c?x)=\emptyset$. \item\ We
define $bv(\alpha)$ to be the bound variable in $\alpha$; that is,
$bv(c?x)=x$, and $bv(\tau)$, $bv(\mathcal{E}[X])$ and $bv(c!x)$ are
not defined.\end{itemize}

To present the operational semantics of qCCS, we need one more
auxiliary notation. For any $X\subseteq Var$ and super-operator
$\mathcal{E}$ on $\mathcal{H}_X$, the cylindric extension of
$\mathcal{E}$ on $\mathcal{H}$ is defined to be
\begin{equation}\label{EX}\mathcal{E}_X\stackrel{def}{=}\mathcal{E}\otimes
\mathcal{I}_{\mathcal{H}_{Var-X}}\end{equation} where
$\mathcal{I}_{\mathcal{H}_{Var-X}}$ is the identity operator on
$\mathcal{H}_{Var-X}$. In what follows we always assume that $X$ is
a finite subset of $Var$ and $\mathcal{E}$ is a super-operator on
$\mathcal{H}_X$ whenever $\mathcal{E}_X$ is encountered.

Then the operational semantics of qCCS is given as a transition
system $(Con,Act,\rightarrow )$, where the transition relation
$\rightarrow$ is defined by the following rules:

\[
\begin{array}{rl}
\mbox{\textbf{Tau}}: & \frac{}{\displaystyle \langle
\tau.P,\rho\rangle \rto{\tau} {\langle P,
\rho\rangle}}\\
\\
\mbox{\textbf{Oper}}: & \frac{}{\displaystyle \langle
\mathcal{E}[X].P,\rho\rangle \rto{\mathcal{E}[X]} {\langle P,
\mathcal{E}_X(\rho)\rangle}}\\
\\
\mbox{\textbf{Input}}: & \frac{}{\displaystyle \langle
c?x.P,\rho\rangle \rto{c?y} {\langle P\{y/x\},\rho\rangle}}
\hspace{1em} y\notin fv(c?x.P)\\
\\
\mbox{\textbf{Output}}: & \frac{} {\displaystyle \langle
c!x.P,\rho\rangle \rto{c!x} {\langle P,\rho\rangle}}
\\
\\
 \mbox{\textbf{Choice}}: & \frac{\displaystyle \langle P,\rho\rangle \rto{\alpha}{\langle P^{\prime},\rho^{\prime}\rangle}} {\displaystyle
\langle P+Q,\rho\rangle \rto{\alpha} {\langle
P^{\prime},\rho^{\prime}\rangle}}
\\
\\
 \mbox{\textbf{Intl1}}: & \frac{\displaystyle \langle P,\rho\rangle \rto{c?x}{\langle
 P^{\prime},\rho^{\prime}\rangle}} {\displaystyle
\langle P\|Q,\rho\rangle \rto{c?x} {\langle
P^{\prime}\|Q,\rho^{\prime}\rangle}} \hspace{1em} x\notin fv(Q)
\\
\\
 \mbox{\textbf{Intl2}}: & \frac{\displaystyle \langle P,\rho\rangle \rto{\alpha}{\langle
 P^{\prime},\rho^{\prime}\rangle}} {\displaystyle
\langle P\|Q,\rho\rangle \rto{\alpha} {\langle
P^{\prime}\|Q,\rho^{\prime}\rangle}} \hspace{1em} \alpha\ {\rm is\
not\ an\ input}
\\
\\
 \mbox{\textbf{Comm}}: & \frac{\displaystyle \langle P,\rho\rangle
\rto{c?x}{\langle
 P^{\prime},\rho\rangle} \hspace{2em} \langle Q,\rho\rangle \rto{c!x}{\langle Q^{\prime},\rho\rangle}} {\displaystyle
\langle P\|Q,\rho\rangle \rto{\tau} {\langle
P^{\prime}\|Q^{\prime},\rho\rangle}} \\
\\
 \mbox{\textbf{Res}}: & \frac{\displaystyle \langle P,\rho\rangle
\rto{\alpha}{\langle
 P^{\prime},\rho^{\prime}\rangle}} {\displaystyle
\langle P\backslash L,\rho\rangle \rto{\alpha} {\langle
P^{\prime}\backslash L,\rho^{\prime}\rangle}} \hspace{1em}
cn(\alpha)\notin L\\
\\
\mbox{\textbf{Def}}: & \frac{\displaystyle \langle
P\{\widetilde{y}/\widetilde{x}\},\rho\rangle \rto{\alpha}{\langle
 P^{\prime},\rho^{\prime}\rangle}} {\displaystyle
\langle A(\widetilde{y}),\rho\rangle \rto{\alpha} {\langle
P^{\prime} ,\rho^{\prime}\rangle}} \hspace{1em}
A(\widetilde{x})\stackrel{def}{=}P
\end{array}\]

\smallskip\

\smallskip\

The symmetric forms of the \textbf{Choice}, \textbf{Intl1},
\textbf{Intl2} and \textbf{Comm} rules are omitted in the above
table.

The operator $\mathcal{E}_X(\cdot)$ in the \textbf{Oper} rule was
defined by Eq.~(\ref{EX}). In the output transition $\langle
c!x.P,\rho\rangle\rto{c!x}{\langle P,\rho\rangle}$, the $x-$system
is sent out through channel $c$. Note that the current state of the
$x-$system is specified in $\rho$. But $\rho$ is not necessary to be
a separable state, and it is possible that the $x-$system is
entangled with the $y-$system for some $y\in Var-\{x\}$. Moreover,
the entanglement between the $x-$system and the $y-$systems
$(y\notin Var-\{x\})$ is preserved after the action $c!x$. The input
transition $\langle c?x.P,\rho\rangle \rto{c?y}{\langle
P\{y/x\},\rho\rangle}$ means that the $y-$system is received from
channel $c$ and then it is put into the (free) occurrences of $x$ in
$P$ (There may be more than one free occurrences of a single
variable $x$ in $P$ because it is not required that $fv(P)\cap
fv(Q)=\emptyset$ in sum $P+Q$). It should be noted that in $c?x.P$
the variable $x$ is bound and it does not represent concretely the
$x-$system. Instead it is merely a reference to the place where the
received system will go. Thus, $c?x.P$ can perform action $c?y$ with
$y\neq x$. The side condition $y\notin fv(c?x.P)$ for the input
transition is obviously to avoid variable name conflict, and it also
makes that $P\{y/x\}$ is well-defined. During performing both the
input and output actions, the state of the environment is not
changed. Passing quantum systems happens in a communication
described by the \textbf{Comm} rule, but it is realized in a
\textquotedblleft call-by-name\textquotedblright scheme and does not
change the state of the environment.

From Definition~\ref{syn}(8) we note that it is required that
$fv(P^{\prime})\cap fv(Q^{\prime})=\emptyset$ to guarantee that the
\textbf{Comm} rule is reasonable. However, we do not need to impose
this condition into the \textbf{Comm} rule because it is a
consequence of the other rules. The verification of this condition
is postponed to the end of Lemma~\ref{var}. The same happens to the
\textbf{Intl1} and \textbf{Intl2} rules.

\subsection{Examples}

To illustrate the transition rules introduced in the last
subsection, we give some simple examples.

In the first example, we shall use the language of qCCS to describe
how quantum systems are passed between processes and how a unitary
transformation or a quantum measurement is performed on some quantum
systems. The most interesting thing is to observe how entangled
systems behave during computation and communication.

\begin{exam}Let $$P_1=c?y.P_1^{\prime},\ \ P_2=c!x.P_2^{\prime}$$ and
$P=(P_1\|P_2)\backslash c$, where $x\notin fv(P_1)$. Then for any
$\rho$, the only possible transition of $P$ is $$\langle
P,\rho\rangle\rto{\tau}{\langle
(P_1^{\prime}\{x/y\}\|P_2^{\prime})\backslash c,\rho\rangle}.$$ Note
that in this transition the $x-$system is passed from $P_2$ to $P_1$
but the state $\rho$ of the environment is not changed. This is
reasonable because $\rho$ does not contain any position information
of the quantum systems under consideration; more precisely, in a
configuration $\langle Q,\rho\rangle$, for all quantum variables
$x$, $\rho$ only describes the state of the $x-$system, but it does
not indicate any subprocess of $Q$ by which the $x-$system is
possessed.

If $$Q_1=c?y.H[y].Q_1^{\prime},\ \ Q=(Q_1\|P_2)\backslash c,$$ and
$\rho=|0\rangle_x\langle 0|\otimes \rho^{\prime}$ where
$\rho^{\prime}\in \mathcal{D}(\mathcal{H}_{Var-\{x\}})$ and $x\notin
fv(Q_1)$, then
\begin{equation*}\begin{split}\langle Q,\rho\rangle & \rto{\tau}{\langle
(H[x].Q_1^{\prime}\{x/y\}\|P_2^{\prime})\backslash c,\rho\rangle}\\
& \rto{H[x]}{\langle (Q_1^{\prime}\{x/y\}\|P_2^{\prime})\backslash
c,|+\rangle_x\langle
+|\otimes\rho^{\prime}\rangle}.\end{split}\end{equation*} At the
beginning of the transition the state of the $x-$system is
$|0\rangle$. Then the $x-$system is passed from $P_2$ to $Q_1$ and
the Hadamard transformation is performed on it at $Q_1$. The state
of the $x-$system becomes $|+\rangle$ after the transition.

Suppose that $$R_1=c?y.CNOT[y,z].R_1^{\prime}$$ and
$R=(R_1\|P_2)\backslash c$ and $\sigma = |+\rangle_x\langle
+|\otimes |0\rangle_z\langle 0|\otimes \sigma^{\prime}$ where
$\sigma^{\prime}\in \mathcal{D}(\mathcal{H}_{Var-\{x,z\}})$ and
$x\notin fv(R_1)$. Then
\begin{equation*}\begin{split}\langle R,\sigma\rangle & \rto{\tau}{\langle
(CNOT[x,z].R_1^{\prime}\{x/y\}\|P_2^{\prime})\backslash
c,\sigma\rangle}\\ & \rto{CNOT[x,z]}{\langle
(R_1^{\prime}\{x/y\}\|P_2^{\prime})\backslash
c,|\beta_{00}\rangle_{xz}\langle
\beta_{00}|\otimes\sigma^{\prime}\rangle}.\end{split}\end{equation*}
The $x-$system is passed from $P_2$ to $R_{1}$, and then the $CNOT$
operator is applied to it and the $z-$system together. It is worth
noting that the state of the $xz-$system is separable before the
transition, but an entanglement between the $x-$system and the
$z-$system is created at the end of the transition.

Let $$S_1=c?y.CNOT[y,z].\mathcal{M}_{0,1}[z].S_1^{\prime}$$ and
$S=(S_1\|P_2)\backslash c,$ where $\mathcal{M}_{0,1}$ is the
operation generated by the measurement of single qubit in the
computational basis $|0\rangle$, $|1\rangle$, with the measurement
result unknown; that is, $\mathcal{M}_{0,1}(\rho)=P_0\rho
P_0+P_1\rho P_1$ for each $\rho\in \mathcal{D}(\mathcal{H}_2)$,
where $P_0=|0\rangle\langle 0|$ and $P_1=|1\rangle\langle 1|$. Then
\begin{equation*}\begin{split} & \langle S,\sigma\rangle \rto{\tau}{\langle (CNOT[x,z].\mathcal{M}_{0,1}[z].S_1^{\prime}\{x/y\}\|
P_2^{\prime})\backslash c, \sigma\rangle}\\
& \rto{CNOT[x,z]}{\langle
(\mathcal{M}_{0,1}[z].S_1^{\prime}\{x/y\}\| P_2^{\prime})\backslash
c, |\beta_{00}\rangle_{xz}\langle \beta_{00}|\otimes
\sigma^{\prime}\rangle}\\
& \rto{\mathcal{M}_{0,1}[z]}{\langle
(S_1^{\prime}\{x/y\}\|P_2^{\prime})\backslash c,
\frac{1}{2}(|00\rangle_{xz}\langle 00|+|11\rangle_{xz}\langle
11|)\otimes \sigma^{\prime}\rangle}.\end{split}\end{equation*} In
the last transition the measurement in computational basis
$|0\rangle$, $|1\rangle$ is performed on the $z-$system. We can see
that the $x-$system and the $z-$system are always in the same state
in the last configuration. This is because they are entangled before
the measurement.
\end{exam}

The communication channels in qCCS (named by elements of $Chan$) are
implicitly assumed to be noiseless. However, the major part of
quantum information theory is devoted to solve the problem of
transmitting reliably information through noisy quantum channels
(see~\cite{NC00}, Chapter 12). The next example shows how we can
formally describe noisy quantum channels in qCCS by combining
noiseless communications and quantum operations on the passed
systems.

\begin{exam}Quantum noisy channel. We imagine a simple scenario
where Alice sends quantum information to Bob through a quantum noisy
channel. Usually, a quantum noisy channel is represented by a
super-operator $\mathcal{E}$ (see Chapters 8 and 12 of~\cite{NC00}).
Thus, Alice and Bob may be described as processes:
$$P=c_1!x.P^{\prime},\ \ Q=c_2?z.Q^{\prime}$$ respectively, and the
channel is described as a nullary process constant scheme $C$ whose
defining equation is
$$C\stackrel{def}{=}c_1?y.\mathcal{E}[y].c_2!y.C.$$ Put
$S=(P\|C\|Q)\backslash \{c_1,c_2\}.$ If information that Alice wants
to send is expressed by a quantum state $\rho$ of the $x-$system,
then for any $\rho^{\prime}\in\mathcal{D}(\mathcal{H}_{Var-\{x\}})$,
we have:
\begin{equation*}\begin{split}\langle S,\rho\otimes \rho^{\prime}\rangle & \rto{\tau}{\langle (P^{\prime}\|\mathcal{E}[x].c_2!x.C\|Q)\backslash
\{c_1,c_2\},\rho\otimes \rho^{\prime}\rangle}\\ &
\rto{\mathcal{E}[x]}{\langle (P^{\prime}\|c_2!x.C\|Q)\backslash
\{c_1,c_2\},\mathcal{E}(\rho)\otimes \rho^{\prime}\rangle}\\
& \rto{\tau}{\langle (P^{\prime}\|C\|Q^{\prime}\{x/z\})\backslash
\{c_1,c_2\},\mathcal{E}(\rho)\otimes
\rho^{\prime}\rangle}.\end{split}
\end{equation*}
Note that $fv(C)$ does not contain $y$; otherwise $C$ is not a
process. Thus, $C\{x/y\}=C$.

Moreover, suppose that a system-environment model of $\mathcal{E}$
is given as in Lemma~\ref{kraus}(2). Let $\mathcal{E}_U$,
$\mathcal{E}_P$ and $\mathcal{E}_{tr_E}$ be super-operators on
$\mathcal{H}_x\otimes\mathcal{H}_E$ and be defined as follows:
$\mathcal{E}_U(\sigma)=U\sigma U^{\dag},$
$\mathcal{E}_P(\sigma)=P\sigma P$, and
$$\mathcal{E}_{tr_E}(\sigma)=\sum_k \langle
e_k|\sigma|e_k\rangle\otimes |e_0\rangle\langle e_0|,$$
respectively, for all $\sigma\in \mathcal{D}(\mathcal{H}_x\otimes
\mathcal{H}_E)$. We define process constant scheme $C^{\prime}$ by
$$C^{\prime}(E)\stackrel{def}{=}c_1?y.\mathcal{E}_U[\{y,E\}].\mathcal{E}_P[\{y,E\}].
\mathcal{E}_{tr_E}[\{y,E\}].c_2!y.C^{\prime}(E),$$ and put
$S^{\prime}=(P\|C^{\prime}(E)\|Q)\backslash \{c_1,c_2\}.$ Then for
all $\rho\in\mathcal{D}(\mathcal{H}_x)$ and $\rho^{\prime\prime}\in
\mathcal{D}(\mathcal{H}_{Var-\{x,E\}})$, the transitions of
$S^{\prime}$ are displayed in Eq.~(\ref{figure}).

\begin{figure*}[!t]
\begin{eqnarray}\label{figure}
\begin{split} &\langle S^{\prime},\rho\otimes |e_0\rangle\langle
e_0| \otimes \rho^{\prime\prime}\rangle  \rto{\tau}{\langle
(P^{\prime}\|\mathcal{E}_U[\{x,E\}].\mathcal{E}_P[\{x,E\}].
\mathcal{E}_{tr_E}[\{x,E\}].c_2!x.C^{\prime}(E)}\\ & \ \ \ \ \ \ \ \
\ \ \ \ \ \ \ \ \ \ \ \ \ \ \ \ \ \ \ \ \ \ \ \ \ \ \ \ \ \ \ \ \ \
\ \ \ \ \ \ \ \ \ \ \ \ \ \ \ \ \ \ \ \ \ \ \ \ \ \ \ \ \ \ \ \ \ \
\ \|Q)\backslash \{c_1,c_2\},\rho\otimes|e_0\rangle\langle
e_0|\otimes \rho^{\prime\prime}\rangle\\ &
\rto{\mathcal{E}_U[\{x,E\}]}{\langle
(P^{\prime}\|\mathcal{E}_P[\{x,E\}].
\mathcal{E}_{tr_E}[\{x,E\}].c_2!x.C^{\prime}(E)\|Q)\backslash
\{c_1,c_2\},}\\ & \ \ \ \ \ \ \ \ \ \ \ \ \ \ \ \ \ \ \ \ \ \ \ \ \
\ \ \ \ \ \ \ \ \ \ \ \ \ \ \ \ \ \ \ \ \ \ \ \ \ \ \ \ \ \ \ \ \ \
\ \ \ \ \ \ \ \ \ \ \ \ \ \ \ \ \ \ U(\rho\otimes|e_0\rangle\langle
e_0|)U^{\dag}\otimes
\rho^{\prime\prime}\rangle\\
& \rto{\mathcal{E}_P[\{x,E\}]}{\langle (P^{\prime}\|
\mathcal{E}_{tr_E}[\{x,E\}].c_2!x.C^{\prime}(E)\|Q)\backslash
\{c_1,c_2\},PU(\rho\otimes|e_0\rangle\langle e_0|)U^{\dag}P\otimes
\rho^{\prime\prime}\rangle}\\
& \rto{\mathcal{E}_{tr_E}[\{x,E\}]}{\langle (P^{\prime}\|
c_2!x.C^{\prime}(E)\|Q)\backslash
\{c_1,c_2\},\mathcal{E}(\rho)\otimes|e_0\rangle\langle e_0| \otimes
\rho^{\prime\prime}\rangle}\\
& \rto{\tau}{\langle
(P^{\prime}\|C^{\prime}(E)\|Q^{\prime}\{x/z\})\backslash
\{c_1,c_2\},\mathcal{E}(\rho)\otimes|e_0\rangle\langle e_0| \otimes
\rho^{\prime\prime}\rangle}.\end{split}
\end{eqnarray}\hrulefill\vspace*{2pt}
\end{figure*}
\end{exam}

Quantum copying has attracted considerably much interest in the
community of quantum information. The aim is to find some physical
devices, called quantum-copying machines, which can produce two
copies of a unknown input quantum state at their output. However,
Wootters and Zurek~\cite{WZ82} and Dieks~\cite{D82} proved the
no-cloning theorem which asserts that such an ideal quantum-copying
machine does not exist. Now the no-cloning is widely recognized as
one of the essential differences between classical and quantum
information. Although there is not a perfect cloning machine, the
no-cloning theorem does not forbid imperfect copying of arbitrary
quantum states. Indeed, an approximate quantum copier was first
designed by Buzek and Hillery~\cite{BH96}. The following example
gives a formal description of it in the language of qCCS.

\begin{exam} Approximate quantum copier. Suppose that an agent $Q$
wants to copy a (unknown) quantum state, $Q$ sends the state to a
copier $P$ through channel $c$, and $P$ receives it and puts it at
place $y$ (the original mode). First, $P$ has to ask for a new place
from another agent $R$ as the copy mode. Then $P$ performs a copying
operation on the original and copy modes together, which is
represented by a unitary transformation $U$, independent of the
input state. Finally, $P$ will send two (approximate) copies of the
original state back to $Q$ through channel $c$. So, $P$, $Q$, $R$
and the whole system $S$ may be described as follows:
$$P=c?y.d?z.U[y,z].c!y.c!z.P,\ \ Q=c!x.c?u.c?v.Q^{\prime},\
\ R=d!x_0.\mathbf{nil},$$ and $S=(P\|Q\|R)\backslash \{c,d\}.$ Note
that $P$ is a nullary process constant scheme and $y,z\notin fv(P)$.

Let $\rho=|\varphi\rangle_x\langle \varphi|\otimes
|0\rangle_{x_0}\langle 0|\otimes \sigma$, where $\sigma\in
\mathcal{D}(\mathcal{H}_{Var-\{x,x_0\}})$, $|\varphi\rangle$ is the
state to be copied, and the initial state of the copy mode is
assumed to be $|0\rangle$. Then the copying process is described by
the following transitions:
\begin{equation*}\begin{split}\langle S,\rho\rangle & \rto{\tau}{\langle (d?z.U[x,z].c!x.c!z.P\|c?u.c?v.Q^{\prime}
\|R)\backslash \{c,d\},\rho\rangle}\\ & \rto{\tau}{\langle
(U[x,x_0].c!x.c!x_0.P\|c?u.c?v.Q^{\prime}\|\mathbf{nil})\backslash
\{c,d\},\rho\rangle}\\ & \rto{U[x,x_0]}{\langle
(c!x.c!x_0.P\|c?u.c?v.Q^{\prime}\|\mathbf{nil})\backslash \{c,d\},
\rho^{\prime}\rangle}\\ & \rto{\tau}{\langle
(c!x_0.P\|c?v.Q^{\prime}\{x/u\}\|\mathbf{nil})\backslash
\{c,d\},\rho^{\prime}\rangle}\\ & \rto{\tau}{\langle
P\|Q^{\prime}\{x/u\}\{x_0/v\}\|\mathbf{nil})\backslash
\{c,d\},\rho^{\prime}\rangle},
\end{split}\end{equation*}
where it is supposed that
$U|\varphi\rangle_x|0\rangle_{x_0}=|\varphi^{\prime}\rangle_x|\varphi^{\prime}\rangle_{x_0}$,
and $\rho^{\prime}=|\varphi^{\prime}\rangle_x\langle
\varphi^{\prime}|\otimes |\varphi^{\prime}\rangle_{x_0}\langle
\varphi^{\prime}|\otimes \sigma$. The Wootters-Zurek-Dieks
no-cloning theorem~\cite{WZ82},~\cite{D82} excludes the possibility
that for all $|\varphi\rangle\in\mathcal{H}_x$,
$|\varphi^{\prime}\rangle=|\varphi\rangle$. But it is shown by Buzek
and Hillery~\cite{BH96} that there exists a (universal) copier $P$
which approximately copies the input state $|\varphi\rangle$ such
that the quality of the output state $|\varphi^{\prime}\rangle$,
measured by the Hilbert-Schmidt norm of the difference between
$|\varphi\rangle$ and $|\varphi^{\prime}\rangle$, does not depend on
$|\varphi\rangle$.
\end{exam}

\subsection{Properties of Transitions}

We now present some basic properties of the transition relation
defined in Section 3.2. Their proofs can be carried out by induction
on the depth of inference. Some of them need very careful analysis,
we put them into the Appendix to improve readability of the paper.

First, we observe how does the environment of a configuration change
in a transition.

\begin{lem}\label{action}\begin{enumerate}\item If
$\langle P,\rho\rangle\rto{\mathcal{E}[X]}{\langle
P^{\prime},\rho^{\prime}\rangle}$, then
\begin{enumerate}\item $\rho^{\prime}=\mathcal{E}_X(\rho)$; and \item $\langle
P,\sigma\rangle\rto{\mathcal{E}[X]}{\langle
P^{\prime},\mathcal{E}_X(\sigma)\rangle}$ holds for all $\sigma\in
\mathcal{D}(\mathcal{H})$.\end{enumerate}
\item If
$\langle P,\rho\rangle\rto{\alpha}{\langle
P^{\prime},\rho^{\prime}\rangle}$ and $\alpha$ is not of the form
$\mathcal{E}[X]$, then
\begin{enumerate}\item $\rho=\rho^{\prime}$; and \item $\langle
P,\sigma\rangle\rto{\alpha}{\langle P^{\prime},\sigma\rangle}$ holds
for all $\sigma\in \mathcal{D}(\mathcal{H})$. Thus, we can simply
write $P\rto{\alpha}{P^{\prime}}$.
\end{enumerate}\end{enumerate}
\end{lem}

Next we see how are the variables in an action related to the free
variables of a process performing this action and those of the
process immediately after it.

\begin{lem}\label{var}If $\langle P,\rho\rangle\rto{\alpha}{\langle
P^{\prime},\rho^{\prime}\rangle}$, then
\begin{enumerate}\item $fv(\alpha)\subseteq fv(P)-fv(P^{\prime})$; and \item
$fv(P^{\prime})\subseteq fv(P)\cup \{bv(\alpha)\}$.
\end{enumerate}
\end{lem}

This lemma enables us to verify that the \textbf{Intl1, Intl2} and
\textbf{Comm} rules are well-defined. We only consider \textbf{Comm}
for instance: if $\langle P,\rho\rangle\rto{c?x}{\langle
P^{\prime},\rho\rangle}$ and $\langle Q,\rho\rangle\rto{c!x}{\langle
Q^{\prime},\rho\rangle}$, then using the above lemma we obtain
$fv(P^{\prime})\subseteq fv(P)\cup \{x\}$, $fv(Q^{\prime})\subseteq
fv(Q)$ and $x\notin fv(Q^{\prime})$. If $P\|Q\in\mathcal{P}$, this
obviously leads to $fv(P^{\prime})\cap fv(Q^{\prime})=\emptyset$
because $fv(P)\cap fv(Q)=\emptyset$. The other two rules can be
dealt with in a similar way.

The next lemma shows that the variable in an input can be changed in
a transition provided a corresponding modification of the process
after the transition is made.

\begin{lem}\label{input}If $\langle P,\rho\rangle\rto{c?x}{\langle
P^{\prime},\rho\rangle}$ and $y\notin fv(P)$, then $\langle
P,\rho\rangle \rto{c?y}{\langle P^{\prime\prime},\rho\rangle}$ for
some $P^{\prime\prime}\equiv_\alpha P^{\prime}\{y/x\}$.
\end{lem}

The following two lemmas carefully examine interference of
substitution and transition. Let $f$ be a substitution. Then we
define its extension on actions by $f(\tau) =\tau,$
$f(\mathcal{E}[X])=\mathcal{E}f[f(X)],$ $f(c?x)=c?x,$ and
$f(c!x)=c!f(x).$

\begin{lem}\label{sub1}If $\langle
P,\rho\rangle\rto{\alpha}{\langle P^{\prime},\rho^{\prime}\rangle}$
and $f(bv(\alpha))=bv(\alpha)$, then $\langle
Pf,f(\rho)\rangle\rto{f(\alpha)}{\langle
P^{\prime\prime},f(\rho^{\prime})\rangle}$ for some
$P^{\prime\prime}\equiv_\alpha P^{\prime}f$.\end{lem}

\begin{lem}\label{sub2}If $\langle Pf,f(\rho)\rangle\rto{\alpha}{\langle
Q,\sigma\rangle}$ and $f(bv(\alpha))=bv(\alpha)$, then for some
$\beta$, $P^{\prime}$ and $\rho^{\prime}$, $\langle
P,\rho\rangle\rto{\beta}{\langle P^{\prime},\rho^{\prime}\rangle}$,
$Q\equiv_\alpha P^{\prime}f$, $\sigma=f(\rho^{\prime})$ and
$\alpha=f(\beta).$
\end{lem}

Finally, we exhibit a certain invariance of transitions under
$\alpha-$conversion.

\begin{lem}\label{alpha}Let $P_1\equiv_\alpha P_2$. Then \begin{enumerate}\item
if $\langle P_1,\rho\rangle \rto{\alpha}{\langle
P_1^{\prime},\rho^{\prime}\rangle}$ and $\alpha$ is not an input,
then $\langle P_2,\rho\rangle \rto{\alpha}{\langle
P_2^{\prime},\rho^{\prime}\rangle}$ for some
$P_2^{\prime}\equiv_\alpha P_1^{\prime}$; \item if $\langle
P_1,\rho\rangle \rto{c?x}{\langle P_1^{\prime},\rho\rangle}$, then
for any $y\notin fv(P_2)$, $\langle P_2,\rho\rangle
\rto{c?y}{\langle P_2^{\prime},\rho\rangle}$ for some
$P_2^{\prime}\equiv_\alpha P_1^{\prime}\{y/x\}$.
\end{enumerate}
\end{lem}

\section{Strong Bisimulations}

\subsection{Basic Definitions}

We first introduce the notion of strong bisimulation on
configurations.

\begin{defn}\label{strongdef}A symmetric relation $\mathcal{R}\subseteq Con\times
Con$ is called a strong bisimulation if for any $\langle
P,\rho\rangle, \langle Q,\sigma\rangle\in Con$, $\langle
P,\rho\rangle\mathcal{R}\langle Q,\sigma\rangle$ implies,
\begin{enumerate}\item whenever $\langle P,\rho\rangle\rto {\alpha}{\langle
P^{\prime}, \rho^{\prime}\rangle}$ and $\alpha$ is not an input,
then for some $Q^{\prime}$ and $\sigma^{\prime},$ $\langle
Q,\sigma\rangle\rto{\alpha}{\langle
Q^{\prime},\sigma^{\prime}\rangle}$ and $\langle
P^{\prime},\rho^{\prime}\rangle\mathcal{R}\langle
Q^{\prime},\sigma^{\prime}\rangle;$
\item whenever $\langle P,\rho\rangle\rto {c?x}{\langle
P^{\prime}, \rho\rangle}$ and $x\notin fv(P)\cup fv(Q)$, then for
some $Q^{\prime}$, $\langle Q,\sigma\rangle\rto{c?x}{\langle
Q^{\prime},\sigma\rangle}$ and for all $y\notin fv(P^{\prime})\cup
fv(Q^{\prime})-\{x\}$, $\langle
P^{\prime}\{y/x\},\rho\rangle\mathcal{R}\langle
Q^{\prime}\{y/x\},\sigma\rangle.$
\end{enumerate}
\end{defn}

It should be noted that in Clause 2 we require $y\notin
fv(P^{\prime})\cup fv(Q^{\prime})-\{x\}$. If we would not put this
requirement, then two previously different quantum states may become
the same state after substitution $\{y/x\}$. This is forbidden by
the no-cloning theorem of quantum information.

Then we are able to define strong bisimilarity between
configurations in a familiar way.

\begin{defn}For any $\langle P,\rho\rangle,
\langle Q,\sigma\rangle\in Con$, we say that $\langle P,\rho\rangle$
and $\langle Q,\sigma\rangle$ are strongly bisimilar, written
$\langle P,\rho\rangle\sim \langle Q,\sigma\rangle$, if $\langle
P,\rho\rangle\mathcal{R}\langle Q,\sigma\rangle$ for some strong
bisimulation $\mathcal{R}$; that is, strong bisimilarity on $Con$ is
the greatest strong bisimulation:
$$\sim\ =\bigcup \{\mathcal{R}: \mathcal{R}\ {\it is\
a\ strong\ bisimulation}\}.$$\end{defn}

Now strong bisimilarity between processes may be defined by
comparing two processes in the same environment.

\begin{defn}For any quantum processes $P,Q\in \mathcal{P}$, we say that $P$
and $Q$ are strongly bisimilar, written $P\sim Q$, if $\langle
P,\rho\rangle \sim \langle Q,\rho\rangle$ for all $\rho\in
\mathcal{D}(\mathcal{H})$.
\end{defn}

The following lemma gives a recursive characterization of strong
bisimilarity between configurations, and it is useful in
establishing strong bisimilarity between some processes.

\begin{lem}\label{sim}For any $\langle P,\rho\rangle, \langle
Q,\sigma\rangle\in Con$, $\langle P,\rho\rangle\sim \langle
Q,\sigma\rangle$ if and only if,
\begin{enumerate}\item whenever $\langle P,\rho\rangle \rto{\alpha}{\langle
P^{\prime},\rho^{\prime}\rangle}$ and $\alpha$ is not an input, then
for some $Q^{\prime}$ and $\sigma^{\prime}$, $\langle
Q,\sigma\rangle\rto{\alpha}{\langle
Q^{\prime},\sigma^{\prime}\rangle}$ and $\langle
P^{\prime},\rho^{\prime}\rangle\sim \langle
Q^{\prime},\sigma^{\prime}\rangle$; \item whenever $\langle
P,\rho\rangle \rto{c?x}{\langle P^{\prime},\rho\rangle}$ and
$x\notin fv(P)\cup fv(Q)$, then for some $ Q^{\prime}$, $\langle
Q,\sigma\rangle\rto{c?x}{\langle Q^{\prime},\sigma\rangle}$ and for
all $y\notin fv(P^{\prime})\cup fv(Q^{\prime})-\{x\}$, $\langle
P^{\prime}\{y/x\},\rho\rangle\sim \langle
Q^{\prime}\{y/x\},\sigma\rangle,$
\end{enumerate} and the symmetric forms of 1 and 2.
\end{lem}

\textit{Proof}. Similar to the proof of Proposition 4.4
in~\cite{M89}. $\Box$

\smallskip\

In the remainder of this section we are going to present some
fundamental properties of strong bisimilarity. First, we show that
strong bisimilarity is preserved by $\alpha-$conversion.

\begin{prop}\label{alpha-sim}If $P_1\equiv_\alpha P_2$, then $P_1\sim P_2$.
\end{prop}

\textit{Proof}. It is easy to show that $$\mathcal{R}=\{(\langle
P_1,\rho\rangle,\langle P_2,\rho\rangle): P_1\equiv_\alpha P_2\}$$
is a strong bisimulation by using Lemma~\ref{alpha}. $\Box$

\subsection{Monoid Laws, Expansion Law and Congruence}

The monoid laws and the static laws in classical CCS can be easily
generalized to qCCS.

\begin{prop}For any $P,Q,R\in \mathcal{P}$, and
$K,L\subseteq Chan$, we have:\begin{enumerate}\item $P+Q\sim Q+P;$
\item $P+(Q+R)\sim (P+Q)+R;$ \item $P+P\sim P;$ \item
$P+\mathbf{nil}\sim P;$ \item $P\|Q\sim Q\|P;$ \item $P\|(Q\|R)\sim
(P\|Q)\|R;$ \item $P\|\mathbf{nil} \sim P;$ \item $P\backslash L\sim
P$ if $cn(P)\cap L=\emptyset$, where $cn(P)$ is the set of free
channel names in $P$; \item $(P\backslash K)\backslash L\sim
P\backslash (K\cup L)$.\end{enumerate}\end{prop}

\textit{Proof.} The items (1)-(4) may be proved by using
Lemma~\ref{sim}, and the items (5)-(9) may be proved by constructing
appropriate strong bisimulation. Here, we only prove (6) as an
example. Put
\begin{equation*}\mathcal{R}=\{(\langle P\|(Q\|R),
\rho\rangle, \langle ( P\|Q)\|R,\rho\rangle): P,Q,R\in \mathcal{P}\
{\rm and}\ \rho\in \mathcal{D}(\mathcal{H})\}\end{equation*} It
suffices to show that $\mathcal{R}$ is a strong bisimulation.
Suppose that
\begin{equation}\label{monoid}\langle
P\|(Q\|R),\rho\rangle\rto{\alpha}{\langle
S,\rho^{\prime}\rangle}.\end{equation} We only consider the
following two cases, and the others are easy or similar.

Case 1. The transition Eq.~(\ref{monoid}) is derived from $\langle
Q,\rho\rangle\rto{c?x}{\langle Q^{\prime},\rho\rangle}$ and $\langle
R,\rho\rangle\rto{c!x}{\langle R^{\prime},\rho\rangle}$ by
\textbf{Comm}. Then $\alpha=\tau$, $\rho^{\prime}=\rho$ and
$S=P\|(Q^{\prime}\|R^{\prime})$. It follows from Lemma~\ref{var}
that $x\in fv(R)$. This leads to $x\notin fv(P\|Q)=fv(P)\cup fv(Q)$
because $(P\|Q)\|R\in \mathcal{P}$. Consequently, we may apply the
\textbf{Intl1} rule to assert that $\langle
P\|Q,\rho\rangle\rto{c?x}{\langle P\|Q^{\prime},\rho\rangle}$, and
furthermore by the \textbf{Comm} rule we obtain $$\langle
(P\|Q)\|R,\rho\rangle\rto{\tau}{\langle
(P\|Q^{\prime})\|R^{\prime},\rho\rangle}.$$ Now it suffices to note
that $\langle S,\rho\rangle\mathcal{R}\langle
(P\|Q^{\prime})\|R^{\prime},\rho\rangle$.

Case 2. $\alpha=c?x$, $$x\notin fv(P\|(Q\|R))\cup
fv((P\|Q)\|R)=fv(P)\cup fv(Q)\cup fv(R)$$ and the transition
Eq.~(\ref{monoid}) is derived from $\langle
P,\rho\rangle\rto{c?x}{\langle P^{\prime},\rho\rangle}$ by
\textbf{Intl1}. Then $\rho^{\prime}=\rho$ and
$S=P^{\prime}\|(Q\|R)$. Since $x\notin fv(Q)$, it follows from the
\textbf{Intl1} rule that $\langle P\|Q,\rho\rangle\rto{c?x}{\langle
P^{\prime}\|Q,\rho\rangle}$. We also have $x\notin fv(R)$. Then
using the \textbf{Intl1} rule once again we obtain $$\langle
(P\|Q)\|R,\rho\rangle\rto{c?x}{\langle
(P^{\prime}\|Q)\|R,\rho\rangle}.$$ Finally, we note that for each
$$y\notin fv(P^{\prime}\|(Q\|R))\cup fv((P^{\prime}\|Q)\|R)- \{x\},$$
$$S\{y/x\}=P^{\prime}\{y/x\}\|(Q\|R),$$
$$((P^{\prime}\|Q)\|R)\{y/x\}=(P^{\prime}\{y/x\}\|Q)\|R,$$ and it
follows that
$$\langle S\{y/x\},\rho\rangle\mathcal{R}\langle
((P^{\prime}\|Q)\|R)\{y/x\},\rho\rangle.\ \Box$$

\smallskip\

\begin{prop} (Expansion law) For any $P,Q\in \mathcal{P}$, we have:
\begin{equation*}\begin{split}(P\|Q)\backslash L  \sim & \sum
\{\alpha.(P^{\prime}\|Q)\backslash L:P\rto{\alpha}{P^{\prime}}\ {\rm
and}\ cn(\alpha)\notin L\}\\ & + \sum
\{\alpha.(P\|Q^{\prime})\backslash L:Q\rto{\alpha}{Q^{\prime}}\ {\rm
and}\
cn(\alpha)\notin L\}\\
& + \sum \{\tau.(P^{\prime}\|Q^{\prime})\backslash L:
P\rto{c?x}{P^{\prime}}\ {\rm and}\ Q\rto{c!x}{Q^{\prime}},\\ & \ \ \
\ \ \ \ \ \ \ \ \ \ \ {\rm or}\ P\rto{c!x}{P^{\prime}}\ {\rm and}\
Q\rto{c?x}{Q^{\prime}}\}.\end{split}\end{equation*}
\end{prop}

\textit{Proof}. Write $S$ for the process in the right-hand side.
Then we can show that $\langle (P\|Q)\backslash L,\rho\rangle \sim
\langle S,\rho\rangle$ for all $\rho\in\mathcal{D}(\mathcal{H})$ in
a way similar to that in classical CCS (see~\cite{M89}, Proposition
4.9). $\Box$

\smallskip\

The following lemma indicates that strong bisimilarity is preserved
by substitution. Its proof requires careful manipulation of
variables, and it is put into the Appendix for readability of the
paper.

\begin{lem}\label{sub-sim}For any $P,Q\in \mathcal{P}$ and for any substitution $f$, $P\sim Q$ if and only if $Pf\sim Qf$.\end{lem}

Now we are ready to show one of the major results in this paper that
strong bisimilarity is a congruence relation with respect to all
combinators in qCCS. It is well-known that congruence is a key
property in all classical process algebras. However, to our best
knowledge, full congruence has not been established for quantum
processes in the previous works. For example, Lalire~\cite{L06}
introduced probabilistic rooted branching bisimilarity between
quantum processes in QPAlg and proved that it is an equivalence
relation and preserved by variable declaration, action prefix,
nondeterministic choice, probabilistic choice, conditional choice
and restriction. But she also gave a counterexample to show that
probabilistic rooted branching (strong) bisimilarity is not
preserved by parallel composition. In~\cite{FDJY07}, a notion of
probabilistic (strong/weak) bisimilarity between processes in a
quantum extension of classical value-passing CCS was proposed by the
authors of this paper. Again, there are some evidences showing that
in general such a probabilistic bisimilarity might not be preserved
by parallel composition. If we write $\sim_p$ for this probabilistic
bisimilarity, then what was achieved in~\cite{FDJY07} is that
$P\sim_p Q$ implies $P\|R\sim_p Q\|R$ when $P$ and $Q$ are free of
quantum input, or $R$ is free of unitary transformation and
measurement; and the condition on $P$, $Q$ and $R$ is very
restrictive. These facts seems to hint that parallel composition
cannot live well in the quantum world where entanglement is present
and cloning is forbidden. Parallel composition is definitely the
most important combinator in any process algebra. Thus,
understanding the behavior of parallel composition of quantum
processes should be one of the key issues in designing a quantum
process algebra. As pointed out in the Introduction, in order to
have a clear understanding of quantum parallel composition, we
decide to make a sharp cleanup in this paper, excluding classical
computation and communication from the previous quantum process
algebras and focusing our attention on pure quantum processes. This
enables us to define a bisimilarity between quantum processes which
is preserved by parallel composition and thus enjoys full
congruence.

\begin{thm}\label{sim-prop}\begin{enumerate}\item If $A\stackrel{def}{=}P$ then $A\sim P$. \item If $P\sim Q$, then we have: \begin{enumerate}\item $\tau.P\sim
\tau.Q$;
\item $\mathcal{E}[X].P\sim \mathcal{E}[X].Q$; \item $c!x.P\sim
c!x.Q $; \item $c?x.P\sim c?x.Q$; \item $P+R\sim Q+R$; \item
$P\|R\sim Q\|R$; \item $P\backslash L\sim Q\backslash L$.
\end{enumerate}\end{enumerate}
\end{thm}

\textit{Proof}. The proofs of (1), (2.a)-(2.c) and (2.e) are routine
applications of Lemma~\ref{sim}, and (2.d) may be proved by using
Lemmas~\ref{sim} and~\ref{sub-sim}. For (2.g), we only need to show
that $$\mathcal{R}=\{(\langle P\backslash L, \rho\rangle, \langle
Q\backslash L, \sigma\rangle): \langle P,\rho\rangle\sim \langle
Q,\sigma\rangle\}$$ is a strong bisimulation, and the routine
details are omitted.

The proof of (2.f) is not a straightforward generalization of the
proof for classical processes, and it requires a new idea in
constructing a strong bisimulation equating $P\|R$ and $Q\|R$. The
major difficulty arises from interference between sequential quantum
computation and communicating quantum systems. The key technique is
inserting a properly chosen quantum operation into an existing
sequence of quantum operations. Furthermore, the proof requires a
very careful analysis of interplay between quantum variables,
sequential applications of super-operators, and communication of
quantum systems as well as subtle treatment of substitution of
quantum variables. We define $\mathcal{R}$ to be a binary relation
between configurations, consisting of the pairs:
\begin{equation*}\begin{split}( \langle P\|R,
\mathcal{F}^{(n)}_{Y_n}\mathcal{E}^{(n)}_{X_n}\mathcal{F}^{(n-1)}_{Y_{n-1}}&\mathcal{E}^{(n-1)}_{X_{n-1}}
...\mathcal{F}^{(1)}_{Y_1}\mathcal{E}^{(1)}_{X_1}\mathcal{F}^{(0)}_{Y_0}(\rho)\rangle,\\
& \langle Q\|R,
\mathcal{F}^{(n)}_{Y_n}\mathcal{E}^{(n)}_{X_n}\mathcal{F}^{(n-1)}_{Y_{n-1}}\mathcal{E}^{(n-1)}_{X_{n-1}}
...\mathcal{F}^{(1)}_{Y_1}\mathcal{E}^{(1)}_{X_1}\mathcal{F}^{(0)}_{Y_0}(\sigma)\rangle),\end{split}
\end{equation*} where $n\geq 0,$ $R\in \mathcal{P}$, $X_i\ (1\leq i\leq n)$ and $Y_i\ (0\leq
i\leq n)$ are finite subsets of $Var$, $\mathcal{E}^{(i)}_{X_i}$ is
a super-operator on $\mathcal{H}_{X_i}$ for each $1\leq i\leq n,$
and $\mathcal{F}^{(i)}_{Y_i}$ is a super-operator on
$\mathcal{H}_{Y_i}$ for each $0\leq i\leq n$, and
$$\langle P,\mathcal{E}^{(n)}_{X_n}\mathcal{E}^{(n-1)}_{X_{n-1}}
...\mathcal{E}^{(1)}_{X_1}(\rho)\rangle\sim \langle Q,
\mathcal{E}^{(n)}_{X_n}\mathcal{E}^{(n-1)}_{X_{n-1}}
...\mathcal{E}^{(1)}_{X_1}(\sigma)\rangle.$$ The idea behind the
definition of $\mathcal{R}$ is that we can insert an arbitrary
quantum operation $\mathcal{F}^{(i)}_{Y_i}$ between two existing
previously quantum operation $\mathcal{E}^{(i)}_{X_i}$ and
$\mathcal{E}^{(i+1)}_{X_{i+1}}$ for any $1\leq i\leq n-1$, and we
can also insert an arbitrary quantum operation
$\mathcal{F}^{(n)}_{Y_n}$ after the last operation
$\mathcal{E}^{(n)}_{X_n}$ and insert $\mathcal{F}^{(0)}_{Y_0}$
before the first operation $\mathcal{E}^{(1)}_{X_1}$. The technique
of insertion is unnecessary in the classical value-passing CCS from
which sequential computation is abstracted by assuming value
expressions (see~\cite{M89}, page 55). However, it is indispensable
in qCCS where one has to consider interference between sequential
quantum computation and communicating quantum systems.

For simplicity, we write
$\mathcal{A}=\mathcal{E}^{(n)}_{X_n}\mathcal{E}^{(n-1)}_{X_{n-1}}
...\mathcal{E}^{(1)}_{X_1}$ and
$$\mathcal{B}=\mathcal{F}^{(n)}_{Y_n}\mathcal{E}^{(n)}_{X_n}\mathcal{F}^{(n-1)}_{Y_{n-1}}\mathcal{E}^{(n-1)}_{X_{n-1}}
...\mathcal{F}^{(1)}_{Y_1}\mathcal{E}^{(1)}_{X_1}\mathcal{F}^{(0)}_{Y_0}.$$
If $P\sim Q$, then for each $\rho$, $\langle P,\rho\rangle\sim
\langle Q,\rho\rangle$, and it implies $\langle
P\|R,\rho\rangle\mathcal{R}\langle Q\|R,\rho\rangle$ by taking $n=0$
and $\mathcal{F}^{(0)}_{Y_0}=\mathcal{I}_{\mathcal{H}_{Y_0}}$ in
$\mathcal{B}$. Therefore, it suffices to show that $\mathcal{R}$ is
a strong bisimulation. To this end, suppose that $\langle
P,\mathcal{A}(\rho)\rangle\sim \langle Q,\mathcal{A}(\sigma)\rangle$
and
\begin{equation}\label{eq-bisi}\langle P\|R,\mathcal{B}(\rho)\rangle\rto{\alpha}{\langle
S,\rho^{\prime}\rangle}.\end{equation}Our aim is to find a
transition of $\langle Q\|R,\mathcal{B}(\rho)\rangle$ which matches
transition Eq.~(\ref{eq-bisi}) according to
Definition~\ref{strongdef}. We consider the following four cases:

Case 1. $\alpha =\tau$. We have $\rho^{\prime}=\mathcal{B}(\rho)$,
and this case is divided into the following four subcases:

Subcase 1.1. The transition~(\ref{eq-bisi}) is derived by
\textbf{Intl2} from $\langle P,\mathcal{B}(\rho)\rangle$
$\rto{\tau}{\langle P^{\prime},\mathcal{B}(\rho)\rangle}$. Then
$S=P^{\prime}\|R$. By Lemma~\ref{action} we obtain $\langle
P,\mathcal{A} (\rho)\rangle\rto{\tau}{\langle
P^{\prime},\mathcal{A}(\rho)\rangle}$. Since $\langle
P,\mathcal{A}(\rho)\rangle\sim \langle
Q,\mathcal{A}(\sigma)\rangle$, it holds that $\langle
Q,\mathcal{A}(\sigma)\rangle\rto{\tau}{\langle
Q^{\prime},\mathcal{A}(\sigma)\rangle}$ for some $Q^{\prime}$ with
$\langle P^{\prime},\mathcal{A}(\rho)\rangle\sim \langle
Q^{\prime},\mathcal{A}(\sigma)\rangle.$ Applying Lemma~\ref{action}
once again we have $\langle
Q,\mathcal{B}(\sigma)\rangle\rto{\tau}{\langle
Q^{\prime},\mathcal{B}(\sigma)\rangle}$, and the \textbf{Intl2} rule
allows us to assert that $\langle Q\|R,\mathcal{B}(\sigma)\rangle$
$\rto{\tau}{\langle Q^{\prime}\|R,\mathcal{B}(\sigma)\rangle}.$ It
is easy to see that $\langle S,\rho^{\prime}\rangle
\mathcal{R}\langle Q^{\prime}\|R,\mathcal{B}(\sigma)\rangle$ from
the definition of $\mathcal{R}$.

Subcase 1.2. The transition~(\ref{eq-bisi}) is derived by
\textbf{Intl2} from $\langle R, \mathcal{B}(\rho)\rangle$
$\rto{\tau}{\langle R^{\prime},\mathcal{B}(\rho)\rangle}$. Then
$S=P\|R^{\prime}$, and from Lemma~\ref{action} and the
\textbf{Intl2} rule it follows that $\langle R,
\mathcal{B}(\sigma)\rangle\rto{\tau}{\langle
R^{\prime},\mathcal{B}(\sigma)\rangle}$ and $\langle Q\|R,
\mathcal{B}(\sigma)\rangle\rto{\tau}{\langle
Q\|R^{\prime},\mathcal{B}(\sigma)\rangle}.$ In addition, we have
$\langle S,\rho^{\prime}\rangle \mathcal{R}\langle Q\|$ $R^{\prime},
\mathcal{B}(\sigma)\rangle$ because $\langle
P,\mathcal{A}(\rho)\rangle\sim \langle
Q,\mathcal{A}(\sigma)\rangle$.

Subcase 1.3. The transition~(\ref{eq-bisi}) is derived by
\textbf{Comm} from $\langle P,
\mathcal{B}(\rho)\rangle\rto{c?x}{\langle
P^{\prime},\mathcal{B}(\rho)\rangle}$ and $\langle
R,\mathcal{B}(\rho)\rangle\rto{c!x}{\langle
R^{\prime},\mathcal{B}(\rho)\rangle}$. First, we have $\langle P,
\mathcal{A}(\rho)\rangle\rto{c?x}{\langle
P^{\prime},\mathcal{A}(\rho)\rangle}$ and $\langle R,
\mathcal{B}(\sigma)\rangle\rto{c!x}{\langle
R^{\prime},\mathcal{B}(\sigma)\rangle}$ by using Lemma~\ref{action}.
With Lemma~\ref{var} we see $x\in fv(R)$. Note that $fv(P)\cap
fv(R)=fv(Q)\cap fv(R)=\emptyset$. Thus, $x\notin fv(P)\cup fv(Q)$.
Since $\langle P,\mathcal{A}(\rho)\rangle\sim \langle
Q,\mathcal{A}(\sigma)\rangle$, it follows that $\langle
Q,\mathcal{A}(\sigma)\rangle \rto{c?x}{\langle
Q^{\prime},\mathcal{A}(\sigma)\rangle}$ for some $Q^{\prime}$ with
$\langle P^{\prime},\mathcal{A}(\rho)\rangle\sim\langle
Q^{\prime},\mathcal{A}(\sigma)\rangle$. By Lemma~\ref{action} and
the \textbf{Comm} rule we obtain $\langle Q,
\mathcal{B}(\sigma)\rangle$ $\rto{c?x}{\langle
Q^{\prime},\mathcal{B}(\sigma)\rangle}$ and $\langle Q\|R,
\mathcal{B}(\sigma)\rangle\rto{\tau}{\langle
Q^{\prime}\|R^{\prime},\mathcal{B}(\sigma)\rangle}.$ Moreover, it
holds that $\langle S,\rho^{\prime}\rangle\mathcal{R}\langle
Q^{\prime}\|R^{\prime},\mathcal{B}(\sigma)\rangle$.

Subcase 1.4. The transition~(\ref{eq-bisi}) is derived by
\textbf{Comm} from $\langle P,
\mathcal{B}(\rho)\rangle\rto{c!x}{\langle
P^{\prime},\mathcal{B}(\rho)\rangle}$ and $\langle R,
\mathcal{B}(\rho)\rangle\rto{c?x}{\langle
R^{\prime},\mathcal{B}(\rho)\rangle}$. Similar to Subcase 1.3.

Case 2. $\alpha =\mathcal{G}[Z]$, where $Z$ is a finite subset of
$Var$, and $\mathcal{G}$ is a super-operator on $\mathcal{H}_Z$. We
have $\rho^{\prime}=\mathcal{G}_Z\mathcal{B}(\rho)$, and this case
is divided into the following two subcases:

Subcase 2.1. The transition~(\ref{eq-bisi}) is derived by
\textbf{Intl2} from $\langle P, \mathcal{B}(\rho)\rangle$
$\rto{\mathcal{G}[Z]}{\langle P^{\prime},
\mathcal{G}_Z\mathcal{B}(\rho)\rangle}$. Then $S=P^{\prime}\|R$. It
follows from Lemma~\ref{action} that $\langle
P,\mathcal{A}(\rho)\rangle\rto{\mathcal{G}[Z]}{\langle P^{\prime},
\mathcal{G}_Z\mathcal{A}(\rho)\rangle}$, and $\langle Q,
\mathcal{A}(\sigma)\rangle\rto{\mathcal{G}[Z]}{\langle Q^{\prime},
\mathcal{G}_Z\mathcal{A}(\sigma)\rangle}$ for some $Q^{\prime}$ with
$\langle P^{\prime}, \mathcal{G}_Z\mathcal{A}(\rho)\rangle\sim
\langle Q^{\prime}, \mathcal{G}_Z\mathcal{A}(\sigma)\rangle$ because
$\langle P,\mathcal{A}(\rho)\rangle$ $\sim \langle
Q,\mathcal{A}(\sigma)\rangle$. Hence, using Lemma~\ref{action} once
again we obtain $\langle
Q,\mathcal{B}(\sigma)\rangle\rto{\mathcal{G}[Z]}{\langle Q^{\prime},
\mathcal{G}_Z\mathcal{B}(\sigma)\rangle}$. Consequently, using the
\textbf{Intl2} rule leads to $$\langle Q\|R,
\mathcal{B}(\sigma)\rangle\rto{\mathcal{G}[Z]}{\langle
Q^{\prime}\|R, \mathcal{G}_Z\mathcal{B}(\sigma)\rangle}.$$ Comparing
carefully $\mathcal{G}_Z\mathcal{A}$ and $\mathcal{G}_Z\mathcal{B}$,
we see that $\mathcal{G}_Z\mathcal{B}$ results from inserting
$$\mathcal{F}^{(n+1)}_{Y_{n+1}}=\mathcal{I}_{\mathcal{H}_{Y_{n+1}}}=\mathcal{I}_\mathcal{H},
\mathcal{F}^{(n)}_{Y_n},
\mathcal{F}^{(n-1)}_{Y_{n-1}},...,\mathcal{F}^{(1)}_{Y_1},\mathcal{F}^{(0)}_{Y_0}$$
at appropriate positions in $\mathcal{G}_Z\mathcal{A}$, where
$Y_{n+1}$ is an arbitrary finite subset of $Var$. This implies
$\langle S,\rho^{\prime}\rangle\mathcal{R}\langle
Q^{\prime}\|R,\mathcal{G}_Z\mathcal{B}(\sigma)\rangle$.

Subcase 2.2. The transition~(\ref{eq-bisi}) is derived by
\textbf{Intl2} from $\langle R,\mathcal{B}(\rho)\rangle$
$\rto{\mathcal{G}[Z]}{\langle
R^{\prime},\mathcal{G}_Z\mathcal{B}(\rho)\rangle}.$ Then
$S=P\|R^{\prime}$, and $\langle
R,\mathcal{B}(\sigma)\rangle\rto{\mathcal{G}[Z]}{\langle
R^{\prime},\mathcal{G}_Z\mathcal{B}(\sigma)\rangle}$ follows
immediately by using Lemma~\ref{action}. Hence, it holds that
$$\langle Q\|R,\mathcal{B}(\sigma)\rangle\rto{\mathcal{G}[Z]}{\langle
Q\|R^{\prime},\mathcal{G}_Z\mathcal{B}(\sigma)\rangle}.$$ Let
$$\mathcal{K}^{(n)}_{Y_n\cup Z}=(\mathcal{F}^{(n)}_{Y_n}\otimes
\mathcal{I}_{\mathcal{H}_{Z-Y_n}})\circ (\mathcal{G}_Z\otimes
\mathcal{I}_{\mathcal{H}_{Y_n-Z}}).$$ Then
$\mathcal{K}^{(n)}_{Y_n\cup Z}$ is a super-operator on
$\mathcal{H}_{Y_n\cup Z}$, and $\mathcal{G}_Z\mathcal{B}$ is
obtained by inserting appropriately $$\mathcal{K}^{(n)}_{Y_n\cup Z},
\mathcal{F}^{(n-1)}_{Y_{n-1}},
...,\mathcal{F}^{(1)}_{Y_1},\mathcal{F}_{Y_0}^{(0)}$$ in
$\mathcal{A}$. Now it follows that $\langle
S,\rho^{\prime}\rangle\mathcal{R}\langle
Q\|R,\mathcal{G}_Z\mathcal{B}(\sigma)\rangle$ from $\langle
P,\mathcal{A}(\rho)\rangle\sim \langle
Q,\mathcal{A}(\sigma)\rangle$.

At the first glance, one may think that the full generality of
$\mathcal{A}$ is not necessary because in the above two subcases we
only add a quantum operation at the beginning of a sequence of
quantum operations. However, this is not the case. It should be
noted that our proof is carried by induction on the depth of
inference~(\ref{eq-bisi}), and quantum operation inserted at the
beginning of a sequence will be moved to the middle of a lager
sequence in the later steps.

Case 3. $\alpha=c!x$. We need to consider the following two
subcases:

Subcase 3.1. The transition~(\ref{eq-bisi}) is derived by
\textbf{Intl2} from $\langle P,\mathcal{B}(\rho)\rangle$
$\rto{c!x}{\langle P^{\prime},\mathcal{B}(\rho)\rangle}$. Similar to
Subcase 1.1.

Subcase 3.2. The transition~(\ref{eq-bisi}) is derived by
\textbf{Intl2} from $\langle R,\mathcal{B}(\rho)\rangle$
$\rto{c!x}{\langle R^{\prime},\mathcal{B}(\rho)\rangle}$. Similar to
Subcase 1.2.

Case 4. $\alpha=c?x$ and $$x\notin fv(P\|R)\cup fv(Q\|R)=fv(P)\cup
fv(Q)\cup fv(R).$$ We have $\rho^{\prime}=\mathcal{B}(\rho)$, and
this case is divided into the following two subcases:

Subcase 4.1. The transition~(\ref{eq-bisi}) is derived by
\textbf{Intl1} from $\langle P,\mathcal{B}(\rho)\rangle$
$\rto{c?x}{\langle P^{\prime},\mathcal{B}(\rho)\rangle}$. Then
$S=P^{\prime}\|R$, and using Lemma~\ref{action} we obtain $\langle
P,\mathcal{A}(\rho)\rangle\rto{c?x}{\langle
P^{\prime},\mathcal{A}(\rho)\rangle}$. From $\langle
P,\mathcal{A}(\rho)\rangle\sim \langle Q,\mathcal{A}(\sigma)\rangle$
and $x\notin fv(P)\cup fv(Q)$, it follows that $\langle
Q,\mathcal{A}(\sigma)\rangle\rto{c?x}{\langle
Q^{\prime},\mathcal{A}(\sigma)\rangle}$ for some $Q^{\prime}$ with
for all $y\notin fv(P^{\prime})\cup fv(Q^{\prime})-\{x\}$, and
$$\langle P^{\prime}\{y/x\},\mathcal{A}(\rho)\rangle\sim \langle
Q^{\prime}\{y/x\},\mathcal{A}(\sigma)\rangle.$$ Furthermore, we have
$\langle Q,\mathcal{B}(\sigma)\rangle \rto{c?x}{\langle
Q^{\prime},\mathcal{B}(\sigma)\rangle}$ by using Lemma~\ref{action}
once again. Note that $x\notin fv(R)$. Thus, applying the
\textbf{Intl1} rule yields $\langle Q\|R,\mathcal{B}(\sigma)\rangle
\rto{c?x}{\langle Q^{\prime}\|R,\mathcal{B}(\sigma)\rangle}.$ What
remains is to verify that $$\langle
(P^{\prime}\|R)\{z/x\},\mathcal{B}(\rho)\rangle\mathcal{R}\langle
(Q^{\prime}\|R)\{z/x\}, \mathcal{B}(\sigma)\rangle$$ for all
$z\notin fv(P^{\prime}\|R)\cup fv(Q^{\prime}\|R)-\{x\}.$ To this
end, we only need to note that
$$(P^{\prime}\|R)\{z/x\}=P^{\prime}\{z/x\}\|R\{z/x\},$$
$$(Q^{\prime}\|R)\{z/x\}=Q^{\prime}\{z/x\}\|R\{z/x\},$$ and $z\notin
fv(P^{\prime}\|R)\cup fv(Q^{\prime}\|R)-\{x\}$ implies $z\notin
fv(P^{\prime})\cup fv(Q^{\prime})-\{x\}.$

Subcase 4.2. The transition~(\ref{eq-bisi}) is derived by
\textbf{Intl1} from $\langle R,\mathcal{B}(\rho)\rangle$
$\rto{c?x}{\langle R^{\prime},\mathcal{B}(\rho)\rangle}$. Then
$S=P\|R^{\prime}$, and $\langle
R,\mathcal{B}(\sigma)\rangle\rto{c?x}{\langle
R^{\prime},\mathcal{B}(\sigma)\rangle}$ follows from
Lemma~\ref{action}. Consequently, we may obtain $\langle
Q\|R,\mathcal{B}(\sigma)\rangle$ $\rto{c?x}{\langle
Q\|R^{\prime},\mathcal{B}(\sigma)\rangle}$ by using the
\textbf{Intl1} rule, because $x\notin fv(Q)$. So, we only need to
show that $$\langle
(P\|R^{\prime})\{y/x\},\mathcal{B}(\rho)\rangle\mathcal{R}\langle (Q
\|R^{\prime})\{y/x\},\mathcal{B}(\sigma)\rangle$$ for all $y\notin
fv(P\|R^{\prime})\cup fv(Q\|R^{\prime})-\{x\}.$ Note that $x\notin
fv(P)\cup fv(Q)$. Thus, it holds that
$(P\|R^{\prime})\{y/x\}=P\|R^{\prime}\{y/x\}$ and
$(Q\|R^{\prime})\{y/x\}=Q\|R^{\prime}\{y/x\}$, and the conclusion
follows immediately from the definition of $\mathcal{R}$. $\Box$

\subsection{Recursion}

We now assume a set of process variable schemes, ranged over by
$\mathbf{X}, \mathbf{Y},...$. For each process variable scheme
$\mathbf{X}$, a nonnegative arity $ar(\mathbf{X})$ is assigned to
it. If $\widetilde{x}=x_1,...,x_{ar(\mathbf{X})}$ is a tuple of
distinct quantum variables, $\mathbf{X}(\widetilde{x})$ is called a
process variable.

Process expressions may be defined by adding the following clause
into Definition~\ref{syn} (and replacing the word \textquotedblleft
process\textquotedblright by the phrase \textquotedblleft process
expression\textquotedblright): \begin{itemize}\item\ \textit{each
process variable $\mathbf{X}(\widetilde{x})$ is a process expression
and $fv(\mathbf{X}(\widetilde{x}))=\{\widetilde{x}\}.$}\end{itemize}

We use meta-variables $\mathbf{E},\mathbf{F},...$ to range over
process expressions.

Suppose that $\mathbf{E}$ is a process expression, and $
\{\mathbf{X}_i(\widetilde{x}_i):i\leq m\}$ is a family of process
variables. If $\{P_i:i\leq m\}$ is a family of processes such that
$fv(P_i)\subseteq \{\widetilde{x}_i\}$ for all $i\leq m$, then we
write $$\mathbf{E}\{\mathbf{X}_i(\widetilde{x}_i):=P_i, i\leq m\}$$
for the process obtained by replacing simultaneously
$\mathbf{X}_i\{\widetilde{y}_i\}$ in $\mathbf{E}$ with
$P_i\{\widetilde{y}/\widetilde{x}\}$ for all $i\leq m$.

\begin{defn}Let $\mathbf{E}$ and $\mathbf{F}$ be process expressions
containing at most process variable schemes $\mathbf{X}_i$ $(i\leq
m)$. If for all families $\{P_i\}$ of processes with
$fv(P_i)\subseteq \{\widetilde{x}_i\}$, $i\leq m$,
$$\mathbf{E}\{\mathbf{X}_i(\widetilde{x}_i):=P_i,i\leq m\}\sim
\mathbf{F}\{\mathbf{X}_i(\widetilde{x}_i):=P_i,i\leq m\},$$ then we
say that $\mathbf{E}$ and $\mathbf{F}$ are strongly bisimilar and
write $\mathbf{E}\sim\mathbf{F}$.
\end{defn}

We now present the main results of this subsection, but their proofs
are put into the Appendix to increase readability of the paper. The
next proposition indicates that recursive definition preserves
strong bisimilarity.

\begin{prop}\label{rec-prop}Let $\{A_i:i\leq m\}$ and $\{B_i:i\leq m\}$ be two families of process constant schemes, and let
$\{\mathbf{E}_i:i\leq m\}$ and $\{\mathbf{F}_i:i\leq m\}$ contain at
most process variable schemes $\mathbf{X}_i$ $(i\leq m)$. If for all
$i\leq m$, we have: $\mathbf{E}_i\sim \mathbf{F}_i,$ and
$$A_i(\widetilde{x}_i)\stackrel{def}{=}\mathbf{E}_i\{\mathbf{X}_j(\widetilde{x}_j):=A_j(\widetilde{x}_j),
j\leq m\},$$
$$B_i(\widetilde{x}_i)\stackrel{def}{=}\mathbf{F}_i\{\mathbf{X}_j(\widetilde{x}_j):=B_j(\widetilde{x}_j),j\leq m\},$$
then $A_i(\widetilde{x}_i)\sim B_i(\widetilde{x}_i)$ for all $i\leq
m$.
\end{prop}

A process variable scheme $\mathbf{X}$ is said to be weakly guarded
in a process expression $\mathbf{E}$ if every occurrence of
$\mathbf{X}$ in $\mathbf{E}$ is within a subexpression of the form
$\alpha.\mathbf{F}$.

The following proposition shows uniqueness of solutions of
equations.

\begin{prop}\label{unique}Suppose that process expressions $\mathbf{E}_i$
$(i\leq m)$ contain at most process variable schemes $\mathbf{X}_i$
$(i\leq m)$, and each $\mathbf{X}_i$ is weakly guarded in each
$\mathbf{E}_j$ $(i,j\leq m)$. If processes $P_i$ and $Q_i$ $(i\leq
m)$ satisfy that, for all $i\leq m$, $fv(P_i),fv(Q_i)\subseteq
\{\widetilde{x}_i\},$ and
$$P_i\sim\mathbf{E}_i\{\mathbf{X_j}(\widetilde{x}_j):=P_j, j\leq m\},$$
$$Q_i\sim\mathbf{E}_i\{\mathbf{X_j}(\widetilde{x}_j):=Q_j, j\leq
m\},$$ then $P_i\sim Q_i$ for all $i\leq m$.
\end{prop}

\section{Strong Reduction-Bisimilarity}

Quantum operations describe sequential computation in quantum
processes. It is obvious that two different sequences of quantum
operations may have the same effect, but they are distinguished from
each other in defining strong bisimulation
(Definition~\ref{strongdef}). To overcome this objection, we need to
introduce the notion of operation reduction. Operation reduction
between strings of actions is defined by the following two rules: if
$X_i$ is a finite subset of $Var,$ $\mathcal{E}^{(i)}$ is a
super-operator on $\mathcal{H}_{X_i}$ for all $1\leq i\leq n,$
$X=\bigcup_{i=1}^{n}X_i,$ and
$$\mathcal{E}=(\mathcal{E}^{(n)}\otimes \mathcal{I}_{X-X_n}) \circ
...\circ (\mathcal{E}^{(2)}\otimes \mathcal{I}_{X-X_2})\circ
(\mathcal{E}^{(1)}\otimes \mathcal{I}_{X-X_1}),$$ then we have

\[
\begin{array}{rl}
\mbox{\textbf{Oper-Red}}: & \frac{}{\displaystyle
\mathcal{E}^{(1)}[X_1]\mathcal{E}^{(2)}[X_2]...\mathcal{E}^{(n)}[X_n]
\rto{}{\mathcal{E}[X]}}\\
\\
\mbox{\textbf{String-Struct}}: & \frac{\displaystyle
t\rto{}{t^{\prime}}}{\displaystyle
t_1tt_2\rto{}{t_1t^{\prime}t_2}}\end{array}\]
$$$$
where $t, t^{\prime}, t_1,t_2\in Act^{\ast}$ are any strings of
actions.

Operation reduction between processes is a natural extension of
reduction between strings of actions, and it is defined by the
following structural rules:

\[
\begin{array}{rl}
\mbox{\textbf{Act-Red}}: & \frac{\displaystyle
\alpha_1...\alpha_m\rto{}{\beta_1...\beta_n}}{\displaystyle
\alpha_1....\alpha_m.P\rto{}{\beta_1....\beta_n.P}}
\\
\\
\mbox{\textbf{Pre-Struct}}: & \frac{\displaystyle
P\rto{}{P^{\prime}}}{\displaystyle
\alpha.P\rto{}{\alpha.P^{\prime}}}
\\
\\
\mbox{\textbf{Sum-Struct}}: & \frac{\displaystyle
P\rto{}{P^{\prime}}}{\displaystyle
P+Q\rto{}{P^{\prime}+Q}}\\
\\
\mbox{\textbf{Par-Struct}}: & \frac{\displaystyle
P\rto{}{P^{\prime}}}{\displaystyle P\|Q\rto{}{P^{\prime}\|Q}}
\\
\\
\mbox{\textbf{Res-Struct}}: & \frac{\displaystyle
P\rto{}{P^{\prime}}}{\displaystyle P\backslash
L\rto{}{P^{\prime}\backslash L}}
\\
\\
\mbox{\textbf{Ref}}: & \frac{}{\displaystyle P\rto{}{P}}
\\
\\
\mbox{\textbf{Trans}}: & \frac{\displaystyle
P\rto{}{Q}\hspace{2em}Q\rto{}{R} }{\displaystyle P\rto{}{R}}
\end{array}\]

\smallskip\

\smallskip\

The symmetric forms of the \textbf{Sum-Struct} and
\textbf{Par-Struct} rules are omitted in the above table.

\begin{lem}\begin{enumerate}\item For any $P\in \mathcal{P}$, there
exists a unique process, written $\lceil P\rceil$, such that
$P\rightarrow \lceil P\rceil$, and $\lceil P\rceil\rightarrow Q$
does not hold for all $Q\in \mathcal{P}$ except $\lceil P\rceil$
itself.
\item If $P\rightarrow P^{\prime}$, then $P^{\prime}\rightarrow
\lceil P\rceil$.
\end{enumerate}
\end{lem}

\textit{Proof}. Induction on the structure of $P$. $\Box$

\smallskip\

By ignoring different decompositions of a quantum operation, we
have:

\begin{defn}\label{r-sim}Strong reduction-bisimilarity $\stackrel{\ast}{\sim}$ is defined to
be the transitive closure of $\simeq$, i.e.,
$$\stackrel{\ast}{\sim}\ =\bigcup_{n=1}^{\infty}\simeq^{n},$$ where for
any $P,Q\in \mathcal{P}$, $P\simeq Q$ if there are $P_1, P_2, Q_1$
and $Q_2$ such that $P\sim P_1\rto{}{P_2}$, $Q\sim Q_1\rto{}{Q_2}$
and $P_2\sim Q_2$. $$$$
$$
\begin{array}{ccccc}
P & \sim & P_1 & \rightarrow & P_2 \\
& & & & \\
\simeq  & & & & \sim \\
& & & & \\
Q & \sim & Q_1 & \rightarrow & Q_2
\end{array}$$ $$$$
\end{defn}

Strong reduction-bisimilarity provides us with a framework in which
we can observe interaction between sequential quantum computation
and communication of quantum systems. Some basic properties of
strong reduction-bisimilarity are presented in the following:

\begin{thm}\label{red-prop}\begin{enumerate}\item If $P\sim Q$ then $P\stackrel{\ast}{\sim} Q$.
\item If $P\rto{}{P^{\prime}}$ then $P\stackrel{\ast}{\sim} P^{\prime}$. In
particular, if $X=\bigcup_{i=1}^{n}X_n$ and
$$\mathcal{E}=(\mathcal{E}^{(n)}\otimes \mathcal{I}_{X-X_n})\circ
...\circ (\mathcal{E}^{(2)}\otimes \mathcal{I}_{X-X_2})\circ
(\mathcal{E}^{(1)}\otimes \mathcal{I}_{X-X_1}),$$ then we have:
\begin{enumerate}\item
$\mathcal{E}^{(1)}[X_1].\mathcal{E}^{(2)}[X_2]....\mathcal{E}^{(n)}[X_n].P\stackrel{\ast}{\sim}
\mathcal{E}[X].P;$ \item $A(\widetilde{x})\stackrel{\ast}{\sim}
\mathcal{E}[X].A(\widetilde{x})$ when process constant scheme $A$ is
defined by
$$A(\widetilde{x})\stackrel{def}{=}\mathcal{E}^{(1)}[X_1].\mathcal{E}^{(2)}[X_2]....\mathcal{E}^{(n)}[X_n].A(\widetilde{x}),$$
where $\{\widetilde{x}\}=\bigcup_{i=1}^{n}X_i$.
\end{enumerate}
\item $\stackrel{\ast}{\sim}$ is an equivalence relation.
\item If $P\stackrel{\ast}{\sim} Q$ then \begin{enumerate}
\item $\alpha.P\stackrel{\ast}{\sim} \alpha.Q$; \item $P+R\stackrel{\ast}{\sim} Q+R$;
\item $P\|R\stackrel{\ast}{\sim} Q\|R$; and \item $P\backslash L\stackrel{\ast}{\sim} Q\backslash L$.
\end{enumerate}
\end{enumerate}
\end{thm}

\textit{Proof}. (1), (2) and (3) are immediately from
Definition~\ref{r-sim}, and (4) may be easily proved by using
Theorem~\ref{sim-prop}. $\Box$

\section{Approximate Strong Bisimulations}

It is required in the definition of strong bisimulation that two
bisimilar processes must perform exactly the same sequences of
quantum operations. This condition is obviously over-discriminative
because two different sequences of quantum operations may have the
same effect. Such an observation motivated us to introduce the
notion of strong reduction-bisimilarity in the last section. In many
cases, however, strong reduction-bisimilarity still may not make
sense because quantum operations form a continuum and their minor
changes can violate strong reduction-bisimilarity between two
quantum processes. Thus, an approximate variant of bisimilarity
should be vital in a quantum process algebra. Let $\lambda$ be a
nonnegative real number, and let $\mathcal{R}$ be a binary relation
between quantum processes. If for any $P\in \mathcal{P}$ and
$\rho,\sigma\in \mathcal{D}(\mathcal{H})$, $D(\rho,\sigma)\leq
\lambda$ implies $\langle P,\rho\rangle\mathcal{R}\langle P,
\sigma\rangle$, where $D(\cdot,\cdot)$ stands for trace distance,
then $\mathcal{R}$ is said to be $\lambda-$closed. Now we are able
to define approximate strong bisimulation.

\begin{defn}A symmetric, $\lambda-$closed relation $\mathcal{R}\subseteq Con\times
Con$ is called a strong $\lambda-$bisimulation if for any $\langle
P,\rho\rangle, \langle Q,\sigma\rangle\in Con$, $\langle
P,\rho\rangle\mathcal{R}\langle Q,\sigma\rangle$ implies,
\begin{enumerate}\item whenever $\alpha$ is $\tau$ or an output and $\langle P,\rho\rangle\rto {\alpha}{\langle
P^{\prime}, \rho\rangle}$, then for some $Q^{\prime},$ $\langle
Q,\sigma\rangle\rto{\alpha}{\langle Q^{\prime},\sigma\rangle}$ and
$\langle P^{\prime},\rho \rangle\mathcal{R}\langle
Q^{\prime},\sigma\rangle;$
\item whenever $\langle P,\rho\rangle\rto {\mathcal{E}[X]}{\langle
P^{\prime}, \rho^{\prime}\rangle}$, then for some $\mathcal{F}$,
$Q^{\prime}$ and $\sigma^{\prime}$, $\langle
Q,\sigma\rangle\rto{\mathcal{F}[X]}{\langle
Q^{\prime},\sigma^{\prime}\rangle}$, $\langle
P^{\prime},\rho^{\prime}\rangle\mathcal{R}\langle
Q^{\prime},\sigma^{\prime}\rangle$, and
$D_\diamond(\mathcal{E},\mathcal{F})\leq \lambda,$ where diamond
distance $D_\diamond(\cdot,\cdot)$ between super-operators is
defined as in Subsection~\ref{dis};
\item whenever $\langle P,\rho\rangle\rto {c?x}{\langle
P^{\prime}, \rho\rangle}$ and $x\notin fv(P)\cup fv(Q)$, then for
some $Q^{\prime}$, $\langle Q,\sigma\rangle\rto{c?x}{\langle
Q^{\prime},\sigma\rangle}$ and for all $y\notin fv(P^{\prime})\cup
fv(Q^{\prime})-\{x\}$, $\langle
P^{\prime}\{y/x\},\rho\rangle\mathcal{R}\langle
Q^{\prime}\{y/x\},\sigma\rangle.$
\end{enumerate}
\end{defn}

\begin{defn}For any $\langle P,\rho\rangle,
\langle Q,\sigma\rangle\in Con$, we say that $\langle P,\rho\rangle$
and $\langle Q,\sigma\rangle$ are strongly $\lambda-$bisimilar,
written $\langle P,\rho\rangle\sim_\lambda \langle Q,\sigma\rangle$,
if $\langle P,\rho\rangle\mathcal{R}\langle Q,\sigma\rangle$ for
some strong $\lambda-$bisimulation $\mathcal{R}$. In other words,
strong $\lambda-$bisimilarity on $Con$ is defined by
$$\sim_\lambda \ =\bigcup \{\mathcal{R}: \mathcal{R}\ is\
a\ strong\ \lambda-bisimulation\}.$$\end{defn}

\begin{defn}Let $P,Q\in \mathcal{P}$. Then: \begin{enumerate}\item We say that $P$
and $Q$ are strongly $\lambda-$bisimilar, written $P\sim_\lambda Q$,
if $\langle P,\rho\rangle \sim_\lambda \langle Q,\rho\rangle$ for
all $\rho\in \mathcal{D}(\mathcal{H})$.
\item The strong bisimulation distance between $P$ and $Q$ is
defined by $$D_{sb}(P,Q)=\inf \{\lambda\geq 0: P\sim_\lambda Q\}.$$
\end{enumerate}
\end{defn}

The following characterization of $\lambda-$bisimilarity between
configurations is useful, and its proof is easy.

\begin{lem}\label{app-def}For any $\langle P,\rho\rangle, \langle Q,\sigma\rangle\in Con$,
$\langle P,\rho\rangle\sim_\lambda \langle Q,\sigma\rangle$ if and
only if,
\begin{enumerate}\item whenever $\alpha$ is $\tau$ or an output and $\langle P,\rho\rangle\rto {\alpha}{\langle
P^{\prime}, \rho\rangle}$, then for some $ Q^{\prime},$ $\langle
Q,\sigma\rangle\rto{\alpha}{\langle Q^{\prime},\sigma\rangle}$ and
$\langle P^{\prime},\rho\rangle\sim_\lambda\langle
Q^{\prime},\sigma\rangle;$
\item whenever $\langle P,\rho\rangle\rto {\mathcal{E}[X]}{\langle
P^{\prime}, \rho^{\prime}\rangle}$, then for some $\mathcal{F}$ and
$Q^{\prime}$ and $\sigma^{\prime}$, $\langle
Q,\sigma\rangle\rto{\mathcal{F}[X]}{\langle
Q^{\prime},\sigma^{\prime}\rangle}$, $\langle
P^{\prime},\rho^{\prime}\rangle\sim_\lambda\langle
Q^{\prime},\sigma^{\prime}\rangle$, and
$D_\diamond(\mathcal{E},\mathcal{F})\leq \lambda;$
\item whenever $\langle P,\rho\rangle\rto {c?x}{\langle
P^{\prime}, \rho\rangle}$ and $x\notin fv(P)\cup fv(Q)$, then for
some $Q^{\prime}$, $\langle Q,\sigma\rangle\rto{c?x}{\langle
Q^{\prime},\sigma\rangle}$ and for all $y\notin fv(P^{\prime})\cup
fv(Q^{\prime})-\{x\}$, $\langle
P^{\prime}\{y/x\},\rho\rangle\sim_\lambda\langle
Q^{\prime}\{y/x\},\sigma\rangle,$
\end{enumerate} and the symmetric forms of 1, 2 and 3.
\end{lem}

We shall need the following simple lemma in the proof of
Theorem~\ref{app-prop}(1) below.

\begin{lem}If $R_i$ is a strong $\lambda_i-$bisimulation $(i=1,2)$,
then $\mathcal{R}_1\circ \mathcal{R}_2$ is a strong
$(\lambda_1+\lambda_2)-$bisimulation.
\end{lem}

\textit{Proof}. We first show that $\mathcal{R}_1\circ
\mathcal{R}_2$ is $(\lambda_1+\lambda_2)-$closed. If
$D(\rho,\sigma)\leq \lambda_1+\lambda_2$, then there must be
$\delta$ such that $D(\rho,\delta)\leq\lambda_1$ and
$D(\delta,\sigma)\leq\lambda_2$. Since $\mathcal{R}_i$ is
$\lambda_i-$closed for $i=1,2$, it holds that $\langle
P,\rho\rangle\mathcal{R}_1\langle P,\delta\rangle$ and $\langle
P,\delta\rangle\mathcal{R}_2\langle P,\sigma\rangle$. This implies
$\langle P,\rho\rangle\mathcal{R}_1\circ \mathcal{R}_2\langle
P,\sigma\rangle$.

Now suppose that $\langle P,\rho\rangle \mathcal{R}_1\circ
\mathcal{R}_2\langle Q,\sigma\rangle$. Then $\langle P,\rho\rangle
\mathcal{R}_1\langle R,\delta\rangle$ $\mathcal{R}_2\langle
Q,\sigma\rangle$ for some $R$ and $\delta$. We only need to consider
the following case: if $\langle
P,\rho\rangle\rto{\mathcal{E}[X]}{\langle
P^{\prime},\rho^{\prime}\rangle}$, then for some $\mathcal{G},
R^{\prime}$ and $\delta^{\prime}$, $\langle
R,\delta\rangle\rto{\mathcal{G}[X]}{\langle
R^{\prime},\delta^{\prime}}\rangle$, $\langle
P^{\prime},\rho^{\prime}\rangle \mathcal{R}_1\langle
R^{\prime},\delta^{\prime}\rangle$ and
$D_\diamond(\mathcal{E},\mathcal{G})\leq \lambda_1$, and
furthermore, for some $\mathcal{F}, Q^{\prime}$ and
$\sigma^{\prime}$, $\langle Q,\sigma\rangle
\rto{\mathcal{F}[X]}{\langle Q^{\prime},\sigma^{\prime}\rangle}$,
$\langle R^{\prime},\delta^{\prime}\rangle \mathcal{R}_2\langle
Q^{\prime},\sigma^{\prime}\rangle$ and
$D_\diamond(\mathcal{G},\mathcal{F})\leq \lambda_2$. Then $\langle
P^{\prime},\rho^{\prime}\rangle \mathcal{R}_1\circ \mathcal{R}_2
\langle Q^{\prime},\sigma^{\prime}\rangle$ and
$$D_\diamond(\mathcal{E},\mathcal{F})\leq
D_\diamond(\mathcal{E},\mathcal{G})+D_\diamond(\mathcal{G},\mathcal{F})\leq
\lambda_1+\lambda_2.\ \Box$$

\smallskip\

The next proposition shows that the process constructors introduced
in qCCS are all non-expansive according to pseudo-metric $D_{sb}$.

\begin{thm}\label{app-prop}\begin{enumerate}\item
Strong bisimulation distance $D_{sb}$ is a pseudo-metric on
$\mathcal{P}$.
\item
For any quantum processes $P, Q$, we have:
\begin{enumerate}\item $D_{sb}(\alpha .P, \alpha .Q)\leq
D_{sb}(P,Q)$ if $\alpha$ is $\tau$, an output or an input; \item
$D_{sb}(\mathcal{E}[X].P, \mathcal{F}[Y].Q)\leq \max \{\eta_{X,Y},
D_\diamond(\mathcal{E},\mathcal{F})+
D_{sb}(P,$ $Q)\}$, where $$\eta_{X,Y}=\begin{cases}0, & \mbox{if }X=Y,\\
\infty, & \mbox{otherwise};\end{cases}$$ \item $D_{sb}(P+R,Q+R)\leq
D_{sb}(P,Q)$; \item $D_{sb}(P\|R,Q\|R)\leq D_{sb}(P,Q)$ if all
super-operators occurring in $P,Q$ and $R$ are trace-preserving;
\item $D_{sb}(P\backslash L,Q\backslash L)\leq D_{sb}(P,Q)$.
\end{enumerate}\end{enumerate}
\end{thm}

\textit{Proof.} To prove (1), we only need to check the triangle
inequality
$$D_{sb}(P,R)\leq D_{sb}(P,Q)+D_{sb}(Q,R)$$ for any quantum processes $P,Q$ and $R$. It suffices to show
that for any $\lambda_1,\lambda_2>0$, if $D_{sb}(P,Q)<\lambda_1$ and
$D_{sb}(Q,R)<\lambda_2$, then $D_{sb}(P,R)<\lambda_1+\lambda_2$. In
fact, it follows from $D_{sb}(P,Q)<\lambda_1$ and
$D_{sb}(Q,R)<\lambda_2$ that for some $\mu_1<\lambda_1$ and
$\mu_2<\lambda_2$, we have $P\sim_{\mu_1}Q\sim_{\mu_2} R$. Thus, for
all $\rho$, $\langle P,\rho\rangle\sim_{\mu_1}\langle
Q,\rho\rangle\sim_{\mu_2} \langle R, \rho\rangle$, and there are
strong $\mu_1-$bisimulation $\mathcal{R}_1$ and strong
$\mu_2-$bisimulation $\mathcal{R}_2$ such that $\langle
P,\rho\rangle\mathcal{R}_1\langle Q,\rho\rangle \mathcal{R}_2\langle
R,\rho\rangle$. This leads to $\langle P,\rho\rangle
\mathcal{R}_1\circ \mathcal{R}_2 \langle R,\rho\rangle$. The above
lemma asserts that $\mathcal{R}_1\circ \mathcal{R}_2$ is a strong
$(\mu_1+\mu_2)-$bisimulation, and thus $\langle
P,\rho\rangle\sim_{\mu_1+\mu_2}\langle R,\rho\rangle$. Hence,
$P\sim_{\mu_1+\mu_2}R$, and $D_{sb}(P,R)\leq
\mu_1+\mu_2<\lambda_1+\lambda_2$.

(2.a) is immediate from Lemma~\ref{app-def}. The proofs of (2.c) and
(2.e) are easy.

(2.b) It is obvious for the case of $X\neq Y$. Now assume $X=Y$. If
$D_\diamond(\mathcal{E},\mathcal{F})<\lambda$ and $D_{sb}(P,Q)<\mu$,
then there is $\mu^{\prime}<\mu$ such that $P\sim_{\mu^{\prime}}Q$;
that is, $\langle P,\sigma\rangle\sim_{\mu^{\prime}}\langle
Q,\sigma\rangle$ for all $\sigma$. For each $\rho$, we have $\langle
\mathcal{E}[X].P,\rho\rangle\rto{\mathcal{E}[X]}{\langle P,
\mathcal{E}_X(\rho)\rangle}$ and $\langle
\mathcal{F}[Y].Q,\rho\rangle\rto{\mathcal{F}[X]}{\langle Q,
\mathcal{F}_X(\rho)\rangle}.$ Note that
$$D(\mathcal{E}_X(\rho),\mathcal{F}_X(\rho))=D(\mathcal{E}(\rho),\mathcal{F}(\rho))\leq
D_\diamond(\mathcal{E},\mathcal{F})<\lambda$$ and $\sim_\lambda$ is
$\lambda-$closed. Then $$\langle
P,\mathcal{E}_X(\rho)\rangle\sim_{\mu^{\prime}}\langle
Q,\mathcal{E}_X(\rho)\rangle\sim_{\lambda}\langle
Q,\mathcal{F}_X(\rho)\rangle,$$ and $\langle
P,\mathcal{E}_X(\rho)\rangle\sim_{\lambda+\mu^{\prime}}\langle Q,
\mathcal{F}_X(\rho)\rangle$. From Lemma~\ref{app-def} we see that
$\langle\mathcal{E}[X].P,\rho\rangle\sim_{\lambda
+\mu^{\prime}}\langle \mathcal{F}[Y].Q,\rho\rangle$. Hence
$$D_{sb}(\mathcal{E}[X].P,\mathcal{F}[X].Q)\leq \lambda +\mu^{\prime}
<\lambda +\mu.$$ This completes the proof by noting that $\lambda$
and $\mu$ are arbitrary.

(2.d) For arbitrary $\lambda>0$, if $D_{sb}(P,Q)<\lambda$, then
there is $\mu<\lambda$ such that $P\sim_\mu Q$; that is, $\langle
P,\rho\rangle\sim_\mu \langle Q,\rho\rangle$ for all $\rho$. Our
purpose is to show that $D_{sb}(P\|R,Q\|R)\leq \lambda$. To do this,
we only need to find a strong $\mu-$bisimulation $\mathcal{R}_\mu$
containing $(\langle P\|R,\rho\rangle,\langle Q\|R,\rho\rangle)$ for
all $\rho$. This can be carried out by a modification of the
technique used in the proof of Theorem~\ref{sim-prop}(2.f). We put
the technical details into the Appendix. $\Box$

\smallskip\

An approximate version of strong reduction-bisimilarity can be
defined in a natural way:

\begin{defn}Let $P,Q\in \mathcal{P}$. Then: \begin{enumerate}
\item We say that $P$ and $Q$ are strongly $\lambda-$reduction-bisimilar, written $P\stackrel{\ast}{\sim}_\lambda Q$, if
there are $n\geq 0$, $\lambda_1,...,\lambda_n\geq 0$ and $R_1,
R_1^{\prime},...,R_n,R_n^{\prime}\in \mathcal{P}$ such that
$\sum_{i=1}^{n}\lambda_i\leq \lambda$ and
$$P\stackrel{\ast}{\sim}R_1\sim_{\lambda_1}R_1^{\prime}\stackrel{\ast}{\sim}
...\stackrel{\ast}{\sim}R_n\sim_{\lambda_n}R^{\prime}_n\stackrel{\ast}{\sim}Q.$$
\item The strong reduction-bisimulation distance between $P$ and $Q$
is defined by $$D_{srb}(P,Q)=\inf \{\lambda\geq 0:
P\stackrel{\ast}{\sim}_\lambda Q\}$$
\end{enumerate}
\end{defn}

Similar to Theorem~\ref{app-prop}, we have:

\begin{thm}\label{red-dis}\begin{enumerate}\item Strong reduction-bisimulation distance $D_{srb}$
is a pseudo-metric on $\mathcal{P}$.
\item For any quantum processes $P, Q$, we have:
\begin{enumerate}\item $D_{srb}(\alpha .P, \alpha .Q)\leq
D_{srb}(P,Q)$ if $\alpha$ is $\tau$, an output or an input; \item
$D_{srb}(\mathcal{E}[X].P, \mathcal{F}[Y].Q)\leq \max \{\eta_{X,Y},
D_\diamond(\mathcal{E},\mathcal{F})+ D_{srb}(P,$ $Q)\}$, where
$\eta_{X,Y}$ is as in Proposition~\ref{app-prop}(2.b); \item
$D_{srb}(P+R,Q+R)\leq D_{srb}(P,Q)$;
\item $D_{srb}(P\|R,Q\|R)\leq D_{srb}(P,Q)$ if all super-operators
occurring in $P,Q$ and $R$ are trace-preserving;
\item $D_{srb}(P\backslash L,Q\backslash L)\leq D_{srb}(P,Q)$.
\end{enumerate}\end{enumerate}
\end{thm}

\textit{Proof}. (1) To show the triangle inequality:
$$D_{srb}(P,Q)+D_{srb}(Q,R)\geq D_{srb}(P,R),$$ it suffices to note
that for any $\lambda,\mu\geq 0$,
$P\stackrel{\ast}{\sim}_{\lambda}Q$ and
$Q\stackrel{\ast}{\sim}_{\mu} R$ implies
$P\stackrel{\ast}{\sim}_{\lambda+\mu} R$. This is immediate from the
definition of strong $\lambda-$reduction-bisimilarity.

(2) We choose to prove (2.b), and the proofs of the other items are
similar. Assume that $X=Y$. For any $\lambda\geq 0$, if
$P\stackrel{\ast}{\sim}_{\lambda} Q$, then we have
$$P\stackrel{\ast}{\sim}R_1\sim_{\lambda_1}R_1^{\prime}\stackrel{\ast}{\sim}
...\stackrel{\ast}{\sim}R_n\sim_{\lambda_n}R^{\prime}_n\stackrel{\ast}{\sim}Q$$
for some $R_1,R_1^{\prime},...,R_n,R_n^{\prime}$ and
$\lambda_1,...,\lambda_n$ with $\sum_{i=1}^{n}\lambda_i\leq
\lambda$. Then it follows from Theorems~\ref{red-prop}(4)
and~\ref{app-prop}(2) that
\begin{equation*}\begin{split}\mathcal{E}[X].P \stackrel{\ast}{\sim}  \mathcal{E}[X].R_1&\sim_{D(\mathcal{E},\mathcal{F})+\lambda_1}
\mathcal{F}[X].R_1^{\prime}\stackrel{\ast}{\sim} ...\\
&
\stackrel{\ast}{\sim}\mathcal{F}[X].R_n\sim_{\lambda_n}\mathcal{F}[X].
R^{\prime}_n\stackrel{\ast}{\sim}\mathcal{F}[X].Q\end{split}\end{equation*}
On the other hand, we have
$$(D_\diamond(\mathcal{E},\mathcal{F})+\lambda_1)+
\lambda_2+...+\lambda_n\leq
D_\diamond(\mathcal{E},\mathcal{F})+\lambda.$$ Thus,
$\mathcal{E}[X].P
\stackrel{\ast}{\sim}_{D(\mathcal{E},\mathcal{F})+\lambda}\mathcal{F}[X].Q$.
Therefore,
\begin{equation*}\begin{split}D_{srb}(\mathcal{E}[X].P,\mathcal{F}[X].Q) & \leq \inf
\{D_\diamond(\mathcal{E},\mathcal{F})+\lambda:
P\stackrel{\ast}{\sim}_{\lambda}Q\}\\ & =
D_\diamond(\mathcal{E},\mathcal{F}) +D_{srb}(P,Q).\ \Box \end{split}
\end{equation*}

A quantum process $P\in \mathcal{P}$ is said to be finite if it
contains no process constants. We write $\mathcal{P}_{fin}$ for the
set of finite quantum processes. For any set $\Omega$ of quantum
gates, we write $\mathcal{P}_{fin}[\Omega]$ for the set of finite
quantum processes in which only gates from $\Omega$ and measurements
in computational bases are used as quantum operations (see Clause 4
in Definition~\ref{syn} and Example~\ref{ex-op}). By combining
Theorems~\ref{red-prop}(2.a) and~\ref{red-dis}(2) we obtain:

\begin{cor}If $\Omega$ is an approximately universal set of quantum gates (e.g., the Hadamard gate,
phase gate, CNOT, and $\pi / 8$ gate (or the Toffoli gate))
(\cite{NC00}, Chapter 4), then $\mathcal{P}_{fin}[\Omega]$ is dense
in $\mathcal{P}_{fin}$ according to pseudo-metric $D_{srb}$.
\end{cor}

\section{Conclusion}

This paper defines an algebra qCCS of quantum processes and presents
its transitional semantics. The strong bisimulation semantics of
qCCS is established, and its modification by reduction of quantum
operations is given. Furthermore, approximate versions of strong
bisimulation and reduction bisimulation are introduced.

We conclude this paper by mentioning some topics for further
studies. Only the strong bisimulation semantics of qCCS has been
established in the present paper, and a weak bisimulation semantics
is still to be exploited for qCCS. However, it is more interesting
to consider some problems about quantum processes that are
irrelevant in classical quantum process algebras. Several authors
started to examine the role of entanglement in quantum sequential
computation (see for example~\cite{Jo03},~\cite{D05}). It seems that
entanglement is much more essential in quantum concurrent
computation. So, an interesting topic is to understand the role of
entanglement in computation within the framework of qCCS. The most
spectacular result in fault-tolerant quantum computation is the
threshold theorem that it is possible to efficiently perform an
arbitrarily large quantum computation provided the noise in
individual quantum gates is below a certain constant (cf.
\cite{NC00}, Section 10.6). This theorem considers only the case of
quantum sequential computation. Its generalization in quantum
concurrent computation would be a great challenge. The bisimulation
distances $D_{sb}$ and $D_{srb}$ introduced in this paper can be
used to express certain fault-tolerance criteria.

It is should be pointed out that qCCS is a purely quantum process
algebra in the sense that no classical information is explicitly
involved in it. The motivation for striping out classical
computation and communication is that the combination of classical
and quantum information gives rise to major difficulties when
attempting to define a bisimilarity which is a congruence with
respect to parallel composition. The main purpose and relevance of
designing a quantum process algebra is to provide a formal model for
distributed quantum computations and quantum communication
protocols, typical instances of which are teleportation, super-dense
coding and BB84. These paradigmatic protocols rely on both quantum
and classical computation and communication. So, one of the most
important topics for further studies would be to find a suitable
extension of qCCS in which both quantum and classical information
can be accommodated well.

\appendixhead{URLend}

\section*{Acknowledgement}

The authors are very grateful to the anonymous referees for their
invaluable comments and suggestions which helped to improve
considerably the presentation of this paper.

\elecappendix

\setcounter{section}{1}

\medskip\

\subsection{Proof of Lemma~\ref{var}}

This is carried out by induction on the depth of inference $\langle
P,\rho\rangle\rto{\alpha}{\langle P^{\prime},\rho^{\prime}\rangle}$.
We only consider the following cases:

Case 1. The last rule is \textbf{Intl2}. Let $P=P_1\|Q$, $\langle
P_1,\rho\rangle\rto{\alpha}{\langle
P_1^{\prime},\rho^{\prime}\rangle}$ and
$P^{\prime}=P_1^{\prime}\|Q$. Then the induction hypothesis
indicates that $fv(\alpha)\subseteq fv(P_1)-fv(P_1^{\prime})$ and
$fv(P_1^{\prime})\subseteq fv(P_1)\cup \{bv(\alpha)\}$. It follows
immediately that
\begin{equation*}\begin{split}fv(P^{\prime}) & =fv(P_1^{\prime})\cup fv(Q)\subseteq
fv(P_1)\cup \{bv(\alpha)\}\cup fv(Q)\\
& =fv(P)\cup\{bv(\alpha)\}.\end{split}\end{equation*} On the other
hand, we have
\begin{equation*}\begin{split}fv(P_1)-fv(P_1^{\prime}) & \subseteq fv(P_1)\cup
fv(Q)-fv(P_1^{\prime})\cup fv(Q)\\
& =fv(P)-fv(P^{\prime})\end{split}\end{equation*} because
$fv(P_1)\cap fv(Q)=\emptyset$. This implies $fv(\alpha)\subseteq
fv(P)-fv(P^{\prime})$.

Case 2. The last rule is \textbf{Comm}. Suppose that $P=P_1\|Q$,
$\langle P_1,\rho\rangle\rto{c?x}{\langle
P_1^{\prime},\rho\rangle}$, $\langle Q,\rho\rangle\rto{c!x}{\langle
Q^{\prime},\rho\rangle}$ and $P^{\prime}=P_1^{\prime}\|Q^{\prime}$.
Then $\alpha =\tau$ and $fv(\alpha)=\emptyset\subseteq
fv(P)-fv(P^{\prime})$. In addition, the induction hypothesis leads
to $$fv(P^{\prime})=fv(P_1^{\prime})\cup fv(Q^{\prime})\subseteq
fv(P_1)\cup \{x\}\cup fv(Q).$$ We also have $x\in
fv(Q)-fv(Q^{\prime})\subseteq fv(Q)$. Thus,
$$fv(P^{\prime})\subseteq fv(P_1)\cup fv(Q)=fv(P)=fv(P)\cup
\{bv(\alpha)\}.\ \Box$$

\subsection{Proof of Lemma~\ref{sub1}}

We prove the conclusion by induction on the depth of inference
$\langle P,\rho\rangle\rto{\alpha}{\langle
P^{\prime},\rho^{\prime}\rangle}$. The cases that the last rule is
\textbf{Tau, Output, Choice, Intl2} or \textbf{Res} are easy, and
the cases that the last rule is \textbf{Oper} or \textbf{Comm} are
similar to those in the proof of Lemma~\ref{sub2} below. So, we only
consider the following three cases:

Case 1. The last rule is \textbf{Input}. Let $p=c?x.Q$. Then
$Pf=c?y.Q\{y/x\}f_y$, where $y\notin fv(c?x.Q)\cup fv(Qf)$,
$f_y(y)=y$ and $f_y(u)=u$ for all $u\neq x$. Suppose that $$\langle
P,\rho\rangle \rto{\alpha=c?z}{\langle
P^{\prime}=Q\{z/x\},\rho^{\prime}=\rho\rangle},$$ where $z\notin
fv(c?x.Q)$. We now need the following:

\textit{Claim.} $z\notin fv(c?x.Q)$ implies $z\notin
fv(c?y.Q\{y/x\}f_y)$.

Indeed, if $z\in fv(c?y.Q\{y/x\}f_y)$, then $z\in fv(Q\{y/x\}f_y)$
and $z\neq y$. It follows that $$fv(Q\{y/x\}f_y)\subseteq
f(fv(c?x.Q))\cup \{y\}$$ because $f_y(y)=y$ and $f_y(u)=u$ for all
$u\neq x$. Thus, we have $z\in f(fv(c?x.Q))$ since $z\neq y$, and
there exists $v\in fv(c?x.Q)$ such that $z=f(v)$. Note that
$z=bv(\alpha)$ and $f(bv(\alpha))=bv(\alpha)$. This leads to
$f(z)=z=f(v)$. Since $f$ is one-to-one, it holds that $z=v\in
fv(c?x.Q).$

By the above claim and the \textbf{Input} rule we obtain $$\langle
Pf,f(\rho)\rangle \rto{c?z=\alpha}{\langle
Q\{y/x\}f_y\{z/y\},f(\rho)\rangle}.$$ Finally, we have to show that
$$Q\{y/x\}f_y\{z/y\}\equiv_\alpha Q\{z/x\}f=P^{\prime}f.$$ In fact,
$x$ is substituted by $y$ in $Q\{y/x\}$, and $f_y(y)=y$. Then $x$ is
substituted by $z$ in $Q\{y/x\}f_y\{z/y\}$. This is also true in
$Q\{z/x\}f$ because $f(z)=z$. If $u\in fv(Q)$ and $u\neq x$, then
$u$ becomes $f_y(u)=f(u)$ in $Q\{y/x\}f_y$. Note that $f(u)\neq y$.
Otherwise, $f_y(u)=y=f_y(y)$ and $u=y$ because $f_y$ is one-to-one.
This contradicts to $y\notin fv(c?x.Q)$. Therefore, $u$ is
substituted by $f(u)$ in $Q\{y/x\}f_y\{z/y\}$. The same happens in
$Q\{z/x\}f$.

Case 2. The last rule is \textbf{Intl1}. Suppose that $P=P_1\|P_2$
and
\[\begin{array}{rl} \frac{\displaystyle \langle P_1,\rho\rangle
\rto{c?x}{\langle
 P_1^{\prime},\rho^{\prime}\rangle}} {\displaystyle
\langle P,\rho\rangle \rto{\alpha=c?x} {\langle
P^{\prime}=P_1^{\prime}\|P_2,\rho^{\prime}\rangle}} \hspace{1em}
x\notin fv(Q)\end{array}
\]
$$$$Since $f(bv(\alpha))=bv(\alpha)$, it follows from the induction
hypothesis that $\langle P_1f,f(\rho)\rangle \rto{c?x}{\langle
Q_1,f(\rho)\rangle}$ with $Q_1\equiv_\alpha P_1^{\prime}f$. We
assert that $x\notin fv(P_2f)$. If not so, then there exists $u\in
fv(P_2)$ such that $x=f(u)$. Note that $x=bv(\alpha)$ and $f(x)=x$.
It holds that $f(x)=f(u)$ and $x=u\in fv(P_2)$ because $f$ is
one-to-one. This is a contradiction. Thus, we can use the
\textbf{Intl1} rule to derive $$\langle
Pf=P_1f\|P_2f,f(\rho)\rangle\rto{\alpha=c?x}{\langle
Q_1\|P_2f,f(\rho)\rangle},$$ and $Q_1\|P_2f\equiv_\alpha
(P_1^{\prime}\|P_2)f=P^{\prime}f$.

Case 3. The last rule is \textbf{Comm}. Let $P=P_1\|P_2$ and
\[
\begin{array}{rl}
\frac{\displaystyle \langle P_1,\rho\rangle\rto{c?x}{\langle
P^{\prime}_1,\rho\rangle}\hspace{2em}\langle
P_2,\rho\rangle\rto{c!x}{\langle
P^{\prime}_2,\rho\rangle}}{\displaystyle \langle P,\rho\rangle
\rto{\tau} {\langle P^{\prime}_1\|P^{\prime}_2,\rho\rangle}}
\end{array}\]
$$$$ Then by the induction hypothesis we have $\langle
P_2f,f(\rho)\rangle\rto{c!f(x)}{\langle
P_2^{\prime}f,f(\rho)\rangle}$. This together with Lemma~\ref{var}
implies $f(x)\in fv(P_2f)$. On the other hand, we can find $y\notin
fv(P_1)$ with $f(y)=y$ because $f$ is almost everywhere the identity
in the sense that $f(u)=u$ for all except a finite number of
variables $u$. Then using Lemma~\ref{input} we obtain $\langle
P_1,\rho\rangle\rto{c?y}{\langle Q_1,\rho\rangle}$ with
$Q_1\equiv_\alpha P_1^{\prime}\{y/x\}.$ Now it follows from the
induction hypothesis that $\langle
P_1f,f(\rho)\rangle\rto{c?y}{\langle Q_1^{\prime},f(\rho)\rangle}$
for some $Q_1^{\prime}\equiv_\alpha Q_1f$. Since $f(x)\in fv(P_2f)$
and $fv(P_1f)\cap fv(P_2f)=\emptyset$, it holds that $f(x)\notin
fv(P_1f)$, and with Lemma~\ref{input} we are able to assert that
$\langle P_1f,f(\rho)\rangle\rto{c?f(x)}{\langle
Q_1^{\prime\prime},f(\rho)\rangle}$ with
$Q_1^{\prime\prime}\equiv_\alpha Q_1^{\prime}\{f(x)/y\}$. Then by
applying the \textbf{Comm} rule we have $$\langle
Pf=P_1f\|P_2f,f(\rho)\rangle\rto{\tau}{\langle
Q_1^{\prime\prime}\|P_2^{\prime},f(\rho)\rangle}.$$

Now it holds that
\begin{equation*}\begin{split}Q_1^{\prime\prime} & \equiv_\alpha
Q_1^{\prime}\{f(x)/y\}\equiv_\alpha Q_1f\{f(x)/y\}\\ & \equiv_\alpha
P_1^{\prime}\{y/x\}f\{f(x)/y\}\equiv_\alpha P_1^{\prime}f.
\end{split}
\end{equation*}
The last $\alpha-$conversion is verified as follows: $x$ becomes $y$
in $P_1^{\prime}\{y/x\}$, and it is still $y$ in
$P_1^{\prime}\{y/x\}f$ because $f(y)=y$. Then $x$ is substituted by
$f(x)$ in $P_1^{\prime}\{y/x\}f\{f(x)/y\}$. For any $u\in
fv(P_1^{\prime})-\{x\}$, $u$ is not changed in
$P_1^{\prime}\{y/x\}$, and it becomes $f(u)$ in
$P_1^{\prime}\{y/x\}f$. If $f(u)\neq y$, then $u$ is substituted by
$f(u)$ in $P_1^{\prime}\{y/x\}f\{f(x)/y\}$. So, it suffices to show
that $f(u)\neq y$. If not so, then $f(u)=y=f(y)$ and $u=y$ because
$f$ is one-to-one. Using Lemma~\ref{var} we assert that
$fv(P_1^{\prime})\subseteq fv(P_1)\cup \{x\}$ since $\langle
P_1,\rho\rangle\rto{c?x}{\langle P_1^{\prime},\rho\rangle}$. This
leads to $u\in fv(P_1^{\prime})-\{x\}\subseteq fv(P_1).$ However,
$y\notin fv(P_1)$. This is a contradiction. $\Box$

\subsection{Proof of Lemma~\ref{sub2}}

We proceed by induction on the depth of inference $\langle
Pf,f(\rho)\rangle\rto{\alpha}{\langle Q,\sigma\rangle}$. We only
consider the following three cases:

Case 1. The last rule is \textbf{Oper}. Then $P=\mathcal{E}[X].R$,
$Pf=\mathcal{E}f[f(X)].Rf$ and
$$\langle Pf,f(\rho)\rangle\rto{\mathcal{E}f[f(X)]}{\langle
Rf,(\mathcal{E}f)_{f(X)}(f(\rho))}.$$ On the other hand, we have
$\langle P,\rho\rangle\rto{E[X]}{\langle
R,\mathcal{E}_X(\rho)\rangle}$. It suffices to show that
$f(\mathcal{E}_X(\rho))=(\mathcal{E}f)_{f(X)}(f(\rho))$. In fact,
\begin{equation*}\begin{split}(\mathcal{E}f)_{f(X)} & =\mathcal{E}f\otimes
\mathcal{I}_{Var-f(X)}\\ & = (f|_X\circ \mathcal{E}\circ
(f|_X)^{-1})\otimes \mathcal{I}_{Var-f(X)}\\ & = f\circ
(\mathcal{E}\otimes \mathcal{I}_{Var-X})\circ f^{-1}\\ & =f\circ
\mathcal{E}_X\circ f^{-1}.
\end{split}
\end{equation*} Thus, we obtain
$$(\mathcal{E}f)_{f(X)}(f(\rho))=(f\circ\mathcal{E}_X\circ
f^{-1})(f(\rho))=f(\mathcal{E}_X(\rho)).$$

Case 2. The last rule is \textbf{Input}. Then $P=c?x.R$,
$Pf=c?y.R\{y/x\}f_y$, where $y\notin fv(c?x.R)\cup fv(Rf)$,
$f_y(y)=y$ and $f_y(u)=u$ for all $u\neq x$, and $$\langle Pf,
f(\rho)\rangle\rto{\alpha=c?z}{\langle
R\{y/x\}f_y\{z/y\},f(\rho)\rangle}$$ where $z\notin
fv(c?y.R\{y/x\}f_y)$.

We first prove the following:

\textit{Claim.} $z\notin fv(c?y.R\{y/x\}f_y)$ implies $z\notin
fv(c?x.R)$

In fact, if $z\in fv(c?x.R)$, then $z\in fv(R)$ and $z\neq x$. This
leads to $z\in fv(R\{y/x\})$. Since $z=bv(\alpha)$, we have
$z=f(z)=f_y(z)\in fv(R\{y/x\}f_y)$. Note that $y\notin fv(c?x.R)$.
Then $z\neq y$, and $z\in fv(c?y.R\{y/x\}f_y)$.

Now using the \textbf{Input} rule we have $\langle
P,\rho\rangle\rto{c?z}{\langle R\{z/x\},\rho\rangle}$, and it
suffices to note that $R\{y/x\}f_y\{z/y\}=R\{z/x\}f.$

Case 3. The last rule is \textbf{Comm}. Suppose that $P=P_1\|P_2$
and we have
\[
\begin{array}{rl}
\frac{\displaystyle \langle P_1f,f(\rho)\rangle\rto{c?x}{\langle
Q_1,f(\rho)\rangle}\hspace{2em}\langle
P_2f,f(\rho)\rangle\rto{c!x}{\langle
Q_2,f(\rho)\rangle}}{\displaystyle \langle
Pf=P_1f\|P_2f,f(\rho)\rangle \rto{\tau} {\langle Q_1\|Q_2,
f(\rho)\rangle}}
\end{array}\] Then by the induction hypothesis we obtain $\langle
P_2,\rho\rangle\rto{c!y}{\langle P_2^{\prime},\rho\rangle}$,
$f(y)=x$ and $Q_2\equiv_\alpha P_2^{\prime}f$ for some $y$ and
$P_2^{\prime}$.

We can find variable $z\notin fv(P_1f)$ such that $f(z)=z$ because
$f$ is almost everywhere the identity. Thus by Lemma~\ref{input} we
assert that $\langle P_1f,f(\rho)\rangle\rto{c!z}{\langle
Q_1^{\prime},f(\rho)\rangle}$ for some $Q_1^{\prime}\equiv_\alpha
Q_1\{z/x\}.$ Now using the induction hypothesis we have $\langle
P_1,\rho\rangle\rto{c!z}{\langle P_1^{\prime},\rho\rangle}$ for some
$P_1^{\prime}$ with $Q_1^{\prime}\equiv_\alpha P_1^{\prime}f$. From
Lemma~\ref{var} we see that $y\in fv(P_2)$, which implies $y\notin
fv(P_1)$. Then using Lemma~\ref{input} once again we obtain $\langle
P_1,\rho\rangle\rto{c?y}{\langle P_1^{\prime\prime},\rho\rangle}$
for some $P_1^{\prime\prime}\equiv_\alpha P_1^{\prime}\{y/z\}$.
Therefore, it is derived by the \textbf{Comm} rule that $\langle
P,\rho\rangle\rto{\tau}{\langle
P_1^{\prime\prime}\|P_2^{\prime},\rho\rangle}$. What remains is to
show that $$Q_1\|Q_2\equiv_\alpha
(P_1^{\prime\prime}\|P_2^{\prime})f=P_1^{\prime\prime}f\|P_2^{\prime}f.$$
Note that we already have $Q_2\equiv_\alpha P_2^{\prime}f$. On the
other hand, since $P_1^{\prime\prime}\equiv_\alpha
P_1^{\prime}\{y/z\}$, $Q_1^{\prime}\equiv_\alpha P_1^{\prime}f$ and
$Q_1^{\prime}\equiv_\alpha Q_1\{z/x\}$, it follows that
\begin{equation*}\begin{split}P_1^{\prime\prime}f & \equiv_\alpha
P_1^{\prime}\{y/z\}f \equiv_\alpha P_1^{\prime}f\{x/z\}\\ &
\equiv_\alpha Q_1^{\prime}\{x/z\} \equiv_\alpha
Q_1\{z/x\}\{x/z\}\equiv_\alpha Q_1\end{split}
\end{equation*}because $x=f(y).\ \Box$

\subsection{Proof of Lemma~\ref{sub-sim}}

We first show that $P\sim Q$ implies $Pf\sim Qf$. Put
\begin{equation*}\begin{split}\mathcal{R}=\{(\langle P^{\prime}, \rho\rangle, \langle
Q^{\prime},\sigma\rangle): P^{\prime}\equiv_\alpha Pf,
Q^{\prime}\equiv_\alpha Qf {\rm and}\ \langle
P,f^{-1}(\rho)\rangle\sim \langle
Q,f^{-1}(\sigma)\rangle\}.\end{split}\end{equation*} It suffices to
show that $\mathcal{R}$ is a strong bisimulation. Suppose that
$P^{\prime}\equiv_\alpha Pf$, $Q^{\prime}\equiv_\alpha Qf$ and
$\langle P,f^{-1}(\rho)\rangle\sim \langle Q,f^{-1}(\sigma)\rangle$.

If $\langle P^{\prime},\rho\rangle\rto{c?x}{\langle R,\rho\rangle}$
and $x\notin fv(P^{\prime})\cup fv(Q^{\prime})$, we can choose
$y\notin fv(P^{\prime})\cup fv(Q^{\prime})\cup fv(R)$ such that
$f(y)=y$ because $f$ is almost everywhere the identity, and
$fv(P^{\prime})$, $fv(Q^{\prime})$ and $fv(R)$ are all finite. Since
$y\notin fv(P^{\prime})$, and $P^{\prime}\equiv_\alpha Pf$ implies
$fv(Pf)=fv(P^{\prime})$, we have $x\notin fv(Pf)$. Then it follows
from Lemma~\ref{alpha} that $\langle Pf,
\rho\rangle\rto{c?y}{\langle R_1,\rho\rangle}$ for some
$R_1\equiv_\alpha R\{y/x\}$. Now we can use Lemma~\ref{sub2} to
derive that $$\langle P,f^{-1}(\rho)\rangle\rto{c?y}{\langle
R_2,f^{-1}(\rho)\rangle}$$ for some $R_2$ with $R_1\equiv_\alpha
R_2f$ because $f(y)=y$. Note that $y\notin fv(P)\cup fv(Q)$.
Otherwise, we have $$y=f(y)\in fv(Pf)\cup fv(Qf)=fv(P^{\prime})\cup
fv(Q^{\prime}),$$ which contradicts to the assumption about $y$.
Thus, $\langle P,f^{-1}(\rho)\rangle\sim \langle Q,
f^{-1}(\sigma)\rangle$, together with Lemma~\ref{sim}, leads to
$$\langle Q,f^{-1}(\sigma)\rangle\rto{c?y}{\langle
S_2,f^{-1}(\sigma)\rangle}$$ for some $S_2$ such that $$\langle
R_2\{z/y\},f^{-1}(\rho)\rangle \sim\langle
S_2\{z/y\},f^{-1}(\sigma)\rangle$$ for all $z\notin fv(R_2)\cup
fv(S_2)-\{y\}$. Then, using Lemma~\ref{sub1} we obtain $\langle
Qf,\sigma\rangle\rto{c?y}{\langle S_1,\sigma\rangle}$ for some
$S_1\equiv_\alpha S_2f$, and an application of Lemma~\ref{alpha}
yields $\langle Q^{\prime},\sigma\rangle\rto{c?x}{\langle
S,\sigma\rangle}$ for some $S\equiv_\alpha S_1\{x/y\}$ because
$f(y)=y$ and $x\notin fv(Q^{\prime})$. So, what we still need to
prove is that $$\langle R\{u/x\},\rho\rangle \mathcal{R}\langle
S\{u/x\},\sigma\rangle$$ for each $u\notin fv(R)\cup fv(S)-\{x\}$.
This comes immediately from the following three items:

(i) Since $R_1\equiv_\alpha R\{y/x\}$, it holds that $x\notin
fv(R_1)$. Note that $y\notin fv(R)$. This implies $$R\equiv_\alpha
R\{y/x\}\{x/y\}\equiv_\alpha R_1\{x/y\}\equiv_\alpha R_2f\{x/y\}$$
because $R_1\equiv_\alpha R_2f$. Then $$R\{u/x\}\equiv_\alpha
R_2f\{x/y\}\{u/x\}\equiv_\alpha R_2f\{u/y\}$$ since $x\notin
fv(R_1)=fv(R_2f)$. Furthermore, we obtain $$R\{u/x\}\equiv_\alpha
R_2f\{u/y\}\equiv_\alpha R_2\{f^{-1}(u)/y\}f$$ because $f(y)=y$, $f$
is one-to-one, and $f(v)\neq y$ when $v\neq y$.

(ii) Similarly, we have $S\{u/x\}\equiv_\alpha S_2\{f^{-1}(u)/y\}f.$

(iii) $f^{-1}(u)\notin fv(R_2)\cup fv(S_2)-\{y\}$. Otherwise, we
have $u\in fv(R_2f)\cup fv(S_2f)$ and $y\neq u$ because $f(y)=y$.
Since $R_2f\equiv_\alpha R\{y/x\}$ and $S_2f\equiv_\alpha S\{y/x\}$,
it holds that $u\in fv(R\{y/x\})\cup fv(S\{y/x\})$. This implies
that $u\neq x$ and $u\in fv(R)\cup fv(S)$, or $x\in fv(R)\cup fv(S)$
and $u=y$. However, we already know that $y\neq u$. Then it must be
the case that $u\neq x$ and $u\in fv(R)\cup fv(S)$, which
contradicts to the assumption about $u$.

For the case that $\langle P,\rho\rangle\rto{\alpha}{\langle
R,\rho^{\prime}\rangle}$ and $\alpha$ is $\tau$ or of the form
$c!x$, the argument is similar and much easier. Thus, we complete
proof of the conclusion that $P\sim Q$ implies $Pf\sim Qf$.

Conversely, we show that $Pf\sim Qf$ implies $P\sim Q$. Note that
$f^{-1}$ is also a substitution. Then it holds that $(Pf)f^{-1}\sim
(Qf)f^{-1}$. Since $P\equiv_\alpha (Pf)f^{-1}$ and $Q\equiv_\alpha
(Qf)f^{-1}$, we obtain $P\sim (Pf)f^{-1}$ and $Q\sim (Qf)f^{-1}$ by
using Proposition~\ref{alpha-sim}, and it follows that $P\sim Q$.
$\Box$

\subsection{Proof Technique of \textquoteleft Strong Bisimulation up to\textquoteright}

The \textquoteleft up to\textquoteright\ technique widely used in
process algebras will be needed in proving
Propositions~\ref{rec-prop} and~\ref{unique}. As a preparation of
the next two subsections, this section briefly develops such a
technique. For any $\mathcal{R}\subseteq Con\times Con$, we set
\begin{equation*}\begin{split}sub(\mathcal{R})=\{(\langle Pf,  f(\rho)\rangle, \langle Qf,
& f(\sigma)\rangle):\\ & \langle P,\rho\rangle\mathcal{R}\langle
Q,\sigma\rangle\ {\rm and}
 f\ {\rm is\ a\ substitution}\}.\end{split}
\end{equation*}

\begin{defn}A symmetric relation $\mathcal{R}\subseteq Con\times
Con$ is called a strong bisimulation up to substitution if for any
$\langle P,\rho\rangle, \langle Q,\sigma\rangle\in Con$, $\langle
P,\rho\rangle\mathcal{R}\langle Q,\sigma\rangle$ implies,
\begin{enumerate}\item whenever $\langle P,\rho\rangle\rto {\alpha}{\langle
P^{\prime}, \rho^{\prime}\rangle}$ and $\alpha$ is not an input,
then for some $\langle Q^{\prime},\sigma^{\prime}\rangle,$ $\langle
Q,\sigma\rangle\rto{\alpha}{\langle
Q^{\prime},\sigma^{\prime}\rangle}$ and $\langle
P^{\prime},\rho^{\prime}\rangle sub(\mathcal{R})  \langle
Q^{\prime},\sigma^{\prime}\rangle;$ and
\item whenever $\langle P,\rho\rangle\rto {c?x}{\langle
P^{\prime}, \rho\rangle}$ and $x\notin fv(P)\cup fv(Q)$, then for
some $Q^{\prime}$, $\langle Q,\sigma\rangle\rto{c?x}{\langle
Q^{\prime},\sigma\rangle}$ and for all $y\notin fv(P^{\prime})\cup
fv(Q^{\prime})-\{x\}$, $\langle P^{\prime}\{y/x\},\rho\rangle
sub(\mathcal{R}) \langle Q^{\prime}\{y/x\},\sigma\rangle.$
\end{enumerate}
\end{defn}

\begin{lem}\label{up-sub} If $\mathcal{R}$ is a strong bisimulation up to
substitution then $\mathcal{R}\subseteq\ \sim$.
\end{lem}

\textit{Proof.} Similar to the proof of Lemma 6 in~\cite{MPW92}.
$\Box$

\begin{defn}\label{up-sim-df}A symmetric relation $\mathcal{R}\subseteq Con\times
Con$ is called a strong bisimulation up to $\sim$ if for any
$\langle P,\rho\rangle, \langle Q,\sigma\rangle\in Con$, $\langle
P,\rho\rangle\mathcal{R}\langle Q,\sigma\rangle$ implies,
\begin{enumerate}\item whenever $\langle P,\rho\rangle\rto {\alpha}{\langle
P^{\prime}, \rho^{\prime}\rangle}$ and $\alpha$ is not an input,
then for some $\langle Q^{\prime},\sigma^{\prime}\rangle,$ $\langle
Q,\sigma\rangle\rto{\alpha}{\langle
Q^{\prime},\sigma^{\prime}\rangle}$ and $\langle
P^{\prime},\rho^{\prime}\rangle\sim \mathcal{R}\sim \langle
Q^{\prime},\sigma^{\prime}\rangle;$ and
\item whenever $\langle P,\rho\rangle\rto {c?x}{\langle
P^{\prime}, \rho\rangle}$ and $x\notin fv(P)\cup fv(Q)$, then for
some $Q^{\prime}$, $\langle Q,\sigma\rangle\rto{c?x}{\langle
Q^{\prime},\sigma\rangle}$ and for all $y\notin fv(P^{\prime})\cup
fv(Q^{\prime})-\{x\}$, $\langle P^{\prime}\{y/x\},\rho\rangle\sim
\mathcal{R}\sim \langle Q^{\prime}\{y/x\},\sigma\rangle.$
\end{enumerate}
\end{defn}

\begin{lem}\label{up-sim} If $\mathcal{R}$ is a strong bisimulation up to $\sim$
then $\mathcal{R}\subseteq\ \sim$.
\end{lem}

\textit{Proof}. Similar to the proof of Lemma 9 in~\cite{MPW92}
(note that Lemma~\ref{up-sub} is needed here). $\Box$

\subsection{Proof of Proposition~\ref{rec-prop}}

For simplicity, we write $\mathbf{E}(A)$ for
$\mathbf{E}\{\mathbf{X}(\widetilde{x}):=A(\widetilde{x})\}$ for any
process expression $\mathbf{E}$, process variable scheme
$\mathbf{X}$ and process constant scheme $A$.

We only present the proof for the simplest case where
$A(\widetilde{x})\stackrel{def}{=}\mathbf{E}(A)$,
$B(\widetilde{x})\stackrel{def}{=}\mathbf{F}(B)$ and
$\mathbf{E}\sim\mathbf{F}$, and it can be generalized to the general
case without any essential difficulty.

We set \begin{equation*}\begin{split}\mathcal{R}= \{(\langle
\mathbf{G}(A), \rho\rangle, &\langle \mathbf{G}(B) , \rho\rangle):\
\mathbf{G}\ {\rm contains\ at\ most}\\ & {\rm the\ process\
variable\ scheme}\ \mathbf{X}\ {\rm and}\
\rho\in\mathcal{D}(\mathcal{H})\}.\end{split}\end{equation*} With
Lemma~\ref{up-sim}, it suffices to show that $\mathcal{R}$ is a
strong bisimulation up to $\sim$. Suppose that
\begin{equation}\label{rec-df}\langle
\mathbf{G}(A),\rho\rangle\rto{\alpha}{\langle
P,\rho^{\prime}\rangle}.\end{equation} We are going to prove the
following two claims:

\textit{Claim 1}. If $\alpha$ is not an input, then for some $Q$,
$\langle\mathbf{G}(B),\rho\rangle\rto{\alpha}{\langle
Q,\rho^{\prime}\rangle}$, and $\langle
P,\rho^{\prime}\rangle\sim\langle
P_1,\rho^{\prime}\rangle\mathcal{R}\langle
Q_1,\rho^{\prime}\rangle\sim \langle Q,\rho^{\prime}\rangle$ for
some $P_1,Q_1$;

\textit{Claim 2}. If $\alpha=c?x$ and $x\notin fv(\mathbf{G}(A))\cup
fv(\mathbf{G}(B))$, then for some $Q$,
$\langle\mathbf{G}(B),\rho\rangle\rto{c?x}{\langle Q,\rho\rangle}$,
and for all $y\notin fv(P)\cup fv(Q)-\{x\}$, $$\langle
P\{y/x\},\rho^{\prime}=\rho\rangle\sim \langle
P_1,\rho\rangle\mathcal{R}\langle Q_1,\rho\rangle\sim\langle
Q\{y/x\},\rho\rangle$$ for some $P_1,Q_1$.

Note that the above claims are a little bit stronger than the two
conditions in Definition~\ref{up-sim-df}, where the environments of
the configurations involved in $\sim\mathcal{R}\sim$ are not
required to be the same.

We proceed by induction on the depth of inference~(\ref{rec-df}).
For simplicity, we only consider the following five cases, and the
others are similar or easy and thus omitted.

Case 1. $\mathbf{G}=\mathbf{X}(\widetilde{y})$, $\alpha =c?u$ and
$u\notin fv(\mathbf{G}(A))\cup fv(\mathbf{G}(B))$. Then
$\mathbf{G}(A)=A(\widetilde{y})$, $\mathbf{G}(B)=B(\widetilde{y})$,
$u\notin \{\widetilde{y}\}$, and $\rho^{\prime}=\rho$.

We want to find some $Q$ such that $\langle
\mathbf{G}(B)=B(\widetilde{y}),\rho\rangle\rto{c?u}{\langle
Q,\rho\rangle,}$ and for all $z\notin fv(P)\cup fv(Q)-\{u\}$,
$\langle P\{z/u\},\rho\rangle\sim\mathcal{R}\sim \langle
Q\{z/u\},\rho\rangle$.

First, we choose some $v_0\notin \{\widetilde{y}\}$. Then for each
$z\notin fv(P)\cup fv(Q)-\{u\}$, from~(\ref{rec-df}) and
Lemma~\ref{action}.2 we obtain
\begin{equation}\label{rec-df1}\langle\mathbf{G}(A),\rho\{v_0/z\}\rangle
\rto{c?u}{\langle P, \rho\{v_0/z\}\rangle}.\end{equation} Since
$A(\widetilde{x})\stackrel{def}{=}\mathbf{E}(A)$,
transition~(\ref{rec-df1}) must be derived by the \textbf{Def} rule
from
\begin{equation}\label{rec-df2}\langle
\mathbf{E}(A)\{\widetilde{y}/\widetilde{x}\},\rho\{v_0/z\}\rangle\rto{c?u}{\langle
P,\rho\{v_0/z\}\rangle}.
\end{equation} On the other hand, we have
$fv(\mathbf{E}(A))\subseteq \{\widetilde{x}\}$. Thus,
$fv(\mathbf{E}(A)\{\widetilde{y}/\widetilde{x}\})$ $\subseteq
\{\widetilde{y}\}$ and $u\notin
fv(\mathbf{E}(A)\{\widetilde{y}/\widetilde{x}\})$. Note that
$\mathbf{E}(A)\{\widetilde{y}/\widetilde{x}\}=\mathbf{E}\{\widetilde{y}/\widetilde{x}\}(A)$,
and the depth of inference~(\ref{rec-df2}) is smaller than that of
inference~(\ref{rec-df1}), which is equal to the depth of
inference~(\ref{rec-df}). So, the induction hypothesis leads to, for
some $R$, \begin{equation}\label{rec-df3}\langle
\mathbf{E}(B)\{\widetilde{y}/\widetilde{x}\}=\mathbf{E}\{\widetilde{y}/\widetilde{x}\}(B),\rho\{v_0/z\}
\rangle\rto{c?u}{\langle R,\rho\{v_0/z\}\rangle}\end{equation} and
for all $v\notin fv(P)\cup fv(R)-\{u\}$,
\begin{equation}\label{rec-df4}\langle P\{v/u\},\rho\{v_0/z\}\rangle\sim\mathcal{R}\sim
\langle R\{v/u\},\rho\{v_0/z\}\rangle.\end{equation}

It follows from $\mathbf{E}\sim\mathbf{F}$ that
$\mathbf{E}(B)\sim\mathbf{F}(B)$. Furthermore, we obtain
$\mathbf{E}(B)\{\widetilde{y}/\widetilde{x}\}\sim\mathbf{F}(B)\{\widetilde{y}/\widetilde{x}\}$
by using Lemma~\ref{sub-sim}. Since
$B(\widetilde{x})\stackrel{def}{=}\mathbf{F}(B)$, it holds that
$u\notin fv(\mathbf{F}(B)\{\widetilde{y}/\widetilde{x}\})\subseteq
\{\widetilde{y}\}$. Consequently, for some $Q$,
\begin{equation}\label{rec-df5}\langle
\mathbf{F}(B)\{\widetilde{y}/\widetilde{x}\},\rho\{v_0/z\}\rangle\rto{c?u}{\langle
Q, \rho\{v_0/z\}\rangle}\end{equation} and for all $v\notin
fv(R)\cup fv(Q)-\{u\}$, \begin{equation}\label{rec-df6}\langle
Q\{v/u\},\rho\{v_0/z\}\rangle\sim\langle
R\{v/u\},\rho\{v_0/z\}\rangle.\end{equation} Using the \textbf{Def}
rule, we obtain $\langle
B(\widetilde{y}),\rho\{v_0/z\}\rangle\rto{c?u}{\langle
Q,\rho\{v_0/z\}\rangle}$ from~(\ref{rec-df5}), and
Lemma~\ref{action}.2 yields $\langle B(\widetilde{y}),\rho
\rangle\rto{c?u}{\langle Q,\rho \rangle}.$

Now we have to show that $$\langle
P\{z/u\},\rho\rangle\sim\mathcal{R}\sim\langle
Q\{z/u\},\rho\rangle$$ for all $z\notin fv(P)\cup fv(Q)-\{u\}$. In
fact, from~(\ref{rec-df}), (\ref{rec-df3}), (\ref{rec-df5}) and
Lemma~\ref{var}.2 we see that $fv(P)\subseteq fv(\mathbf{G}(A))\cup
\{u\}$, $fv(R)\subseteq fv(\mathbf{E}(B))\cup \{u\}$, and
$fv(Q)\subseteq fv(\mathbf{F}(B))\cup \{u\}$. Then $fv(P), fv(R),
fv(Q)\subseteq \{\widetilde{y}\}\cup \{u\}$, and $v_0\notin
\{\widetilde{y}\}$ implies $v_0\notin fv(P)\cup fv(R)\cup
fv(Q)-\{u\}$. Furthermore, it follows from~(\ref{rec-df4}) and
(\ref{rec-df6}) that $$\langle
P\{v_0/u\},\rho\{v_0/z\}\rangle\sim\mathcal{R}\sim\langle
R\{v_0/u\},\rho\{v_0/z\}\rangle\sim \langle
Q\{v_0/u\},\rho\{v_0/z\}\rangle.$$ With the observation
$\mathbf{G}(A)f=\mathbf{G}f(A)$ for all substitutions $f$, we see
that $sub(\mathcal{R})=\mathcal{R}$. Therefore, we obtain
\begin{equation*}\begin{split} \langle P\{z/u\},\rho\rangle &=\langle
P\{v_0/u\}\{z/v_0\},\rho\{v_0/z\}\{z/v_0\}\rangle\\
& \sim\mathcal{R}\sim\langle
Q\{v_0/u\}\{z/v_0\},\rho\{v_0/z\}\{z/v_0\}\rangle=\langle
Q\{z/u\},\rho\rangle\end{split}\end{equation*} by using
Lemma~\ref{sub-sim} once again.

Case 2. $\mathbf{G}=\mathcal{E}[X].\mathbf{G}_1$. Then
$\mathbf{G}(A)=\mathcal{E}[X].\mathbf{G}_1(A)$,
$\mathbf{G}(B)=\mathcal{E}[X].\mathbf{G}_1(B)$,
$\alpha=\mathcal{E}[X]$, $P=\mathbf{G}_1(A)$ and
$\rho^{\prime}=\mathcal{E}_X(\rho)$. We have $$\langle
\mathbf{G}(B),\rho\rangle\rto{\alpha =\mathcal{E}[X]}{\langle
\mathbf{G}_1(B),\mathcal{E}_X(\rho)\rangle}$$ and $\langle P,
\rho^{\prime}\rangle\mathcal{R}\langle
\mathbf{G}_1(B),\mathcal{E}_X(\rho)\rangle.$

Case 3. $\mathbf{G}=c?x.\mathbf{G}_1$. Then
transition~(\ref{rec-df}) must be as follows: $$\langle
\mathbf{G}(A)=c?x.\mathbf{G}_1(A),\rho\rangle\rto{\alpha
=c?y}{\langle P=\mathbf{G}_1(A)\{y/x\},\rho^{\prime}=\rho\rangle}$$
where $y\notin fv(\mathbf{G}_1(A))-\{x\}$. In this case, we have the
assumption that $y\notin fv(\mathbf{G}(A))\cup fv(\mathbf{G}(B))$.
Since $\mathbf{G}(B)=c?x.\mathbf{G}_1(B)$, we obtain $\langle
\mathbf{G}(B),\rho\rangle\rto{c?y}{\langle
\mathbf{G}_1(B)\{y/x\},\rho\rangle}$ by the \textbf{Input} rule.
Moreover, for any $z\notin fv(P)\cup
fv(\mathbf{G}_1(B)\{y/x\})-\{y\}$, we have
$$P\{z/y\}=\mathbf{G}_1(A)\{y/x\}\{z/y\}=\mathbf{G}_1(A)\{z/x\}=\mathbf{G}_1\{z/x\}(A),$$
$$\mathbf{G}_1(B)\{y/x\}\{z/y\}=\mathbf{G}_1(B)\{z/x\}=\mathbf{G}_1\{z/x\}(B).$$
So, it follows that $\langle
P\{z/y\},\rho^{\prime}\rangle\mathcal{R}\langle
\mathbf{G}_1(B)\{y/x\}\{z/y\},\rho\rangle$.

Case 4. $\mathbf{G}=\mathbf{G}_1\|\mathbf{G}_2$, $\alpha=c?x$,
$x\notin fv(\mathbf{G}_1(A))\cup fv(\mathbf{G}_2(A))$ and
transition~(\ref{rec-df}) is derived by the \textbf{Intl1} from
$\langle \mathbf{G}_1(A),$ $\rho\rangle\rto{c?x}{\langle
P_1,\rho\rangle}$. Then
$\mathbf{G}(A)=\mathbf{G}_1(A)\|\mathbf{G}_2(A)$,
$P=P_1\|\mathbf{G}_2(A)$ and $\rho^{\prime}=\rho$. By the induction
hypothesis we have, for some $Q_1$, $\langle
\mathbf{G}_1(B),\rho\rangle\rto{c?x}{\langle Q_1,\rho\rangle}$, and
for all $y\notin fv(P_1)\cup fv(Q_1)-\{x\}$, $$\langle
P_1\{y/x\},\rho\rangle\sim \langle
P_1^{\prime},\rho\rangle\mathcal{R}\langle
Q_1^{\prime},\rho\rangle\sim \langle Q_1\{y/x\},\rho\rangle$$ for
some $P_1^{\prime},Q_1^{\prime}$. It is clear that $x\notin
fv(\mathbf{G}_2(B))$. Thus, we obtain $$\langle
\mathbf{G}(B)=\mathbf{G}_1(B)\|\mathbf{G}_2(B),\rho\rangle\rto{c?x}{\langle
Q_1\|\mathbf{G}_2(B),\rho\rangle}$$ by the \textbf{Intl1} rule. For
any $z\notin fv(P)\cup fv(Q_1\|\mathbf{G}_2(B))-\{x\}$, we have
$z\notin fv(P_1)\cup fv(Q_1)-\{x\}$, and $$\langle
P_1\{z/x\},\rho\rangle\sim\langle
P_1^{\prime},\rho\rangle\mathcal{R}\langle
Q_1^{\prime},\rho\rangle\sim\langle
Q_1^{\prime}\{z/x\},\rho\rangle.$$ This, together with
Proposition~\ref{sim-prop}.2.f, leads to
\begin{equation*}\begin{split} \langle  P\{z/x\},\rho\rangle =&\langle
P_1\{z/x\}\|\mathbf{G}_2(A),\rho\rangle \sim\langle
P_1^{\prime}\|\mathbf{G}_2(A),\rho\rangle\mathcal{R}\\ & \langle
Q_1^{\prime}\|\mathbf{G}_2(B),\rho\rangle\sim\langle
Q_1\{z/x\}\|\mathbf{G}_2(B)
=(Q_1\|\mathbf{G}_2(B))\{z/x\},\rho\rangle.\end{split}\end{equation*}

Case 5. $\mathbf{G}=\mathbf{G}_1\|\mathbf{G}_2$, $\alpha=\tau$ and
transition~(\ref{rec-df}) is derived by the \textbf{Comm} rule from
$\langle \mathbf{G}_1(A),\rho\rangle\rto{c?x}{\langle
P_1,\rho\rangle}$ and $\langle
\mathbf{G}_2(A),\rho\rangle\rto{c!x}{\langle P_2,\rho\rangle}$. Then
$\rho^{\prime}=\rho$ and $P=P_1\|P_2$. With the induction hypothesis
we have, for some $Q_1, Q_2$, $\langle
\mathbf{G}_1(B),\rho\rangle\rto{c?x}{\langle Q_1,\rho\rangle}$ and
$\langle \mathbf{G}_2(B),\rho\rangle\rto{c!x}{\langle
Q_2,\rho\rangle}$, $$\langle P_1,\rho\rangle\sim \langle
P_1^{\prime},\rho\rangle\mathcal{R}\langle
Q_1^{\prime},\rho\rangle\sim\langle Q_1,\rho\rangle$$ for some
$P_1^{\prime},Q_1^{\prime}$, and $$\langle P_2,\rho\rangle\sim
\langle P_2^{\prime},\rho\rangle\mathcal{R}\langle
Q_2^{\prime},\rho\rangle\sim\langle Q_2,\rho\rangle$$ for some
$P_2^{\prime},Q_2^{\prime}$. Then $$\langle
\mathbf{G}(B)=\mathbf{G}_1(B)\|\mathbf{G}_2(B),\rho\rangle\rto{\tau}{\langle
Q_1\|Q_2,\rho\rangle},$$ and by Proposition~\ref{sim-prop}.2.f it
follows that $$\langle P,\rho\rangle\sim \langle
P_1^{\prime}\|P_2^{\prime},\rho\rangle\mathcal{R}\langle
Q_1^{\prime}\|Q_2^{\prime},\rho\rangle\sim \langle
Q_1\|Q_2,\rho\rangle.\ \Box$$

\subsection{Proof of Proposition~\ref{unique}}

We first have the following familiar lemma for the actions of weakly
guarded process expressions:

\begin{lem}\label{gu-act}If $\mathbf{X}_i$ $(i\leq m)$ are weakly guarded in
$\mathbf{E}$, and $$\langle
\mathbf{E}\{\mathbf{X}_i(\widetilde{x}_i):=P_i,i\leq m\},
\rho\rangle\rto{}{\langle P^{\prime},\rho^{\prime}\rangle},$$ then
for some $\mathbf{E}^{\prime}$, we have:
\begin{enumerate}\item $P^{\prime}=\mathbf{E}^{\prime}\{\mathbf{X}_i(\widetilde{x}_i):=P_i,i\leq
m\}$; and \item $\langle \mathbf{E}\{\mathbf{X}_i(\widetilde{x}_i):
=Q_i, i\leq m\}, \rho\rangle \rto{}{\langle \mathbf{E}^{\prime}\{
\mathbf{X}_i(\widetilde{x}_i):=Q_i, i\leq
m\},\rho^{\prime}\rangle}.$
\end{enumerate}
\end{lem}

\textit{Proof}. Induction on the structure of $\mathbf{E}$. $\Box$

\smallskip\

Now we begin to prove Proposition~\ref{unique}. For simplicity, we
write $\mathbf{G}(\widetilde{P})$ for
$\mathbf{G}\{\mathbf{X}_i(\widetilde{x}_i):=P_i, 1\leq i\leq m\}$.
Let \begin{equation*}\begin{split}\mathcal{R}= \{(\langle
\mathbf{G}(\widetilde{P}),  \rho\rangle,\langle
\mathbf{G}(&\widetilde{Q}),\rho\rangle): \mathbf{G}\ {\rm contains\
at\ most}\ \mathbf{X}_i\\ & (1\leq i\leq m)\ {\rm and}\
\rho\in\mathcal{D}(\mathcal{H})\}\cup
Id_{Con},\end{split}\end{equation*} where $Id_{Con}$ is the identity
relation on configurations. Our purpose is to show that
$\mathcal{R}$ is a strong bisimulation up to $\sim$. Assume that
\begin{equation}\label{uniq}\langle \mathbf{G}(\widetilde{P}),\rho\rangle
\rto{\alpha}{\langle P,\rho^{\prime}\rangle.}\end{equation} By
induction on the depth of inference~(\ref{uniq}) we are going to
prove the following:

\textit{Claim 1}. If $\alpha$ is not an input, then for some $Q$,
$\langle\mathbf{G}(\widetilde{Q}),\rho\rangle\rto{\alpha}{\langle
Q,\rho^{\prime}\rangle}$, and $\langle
P,\rho^{\prime}\rangle\sim\langle
P_1,\rho^{\prime}\rangle\mathcal{R}\langle
Q_1,\rho^{\prime}\rangle\sim \langle Q,\rho^{\prime}\rangle$ for
some $P_1,Q_1$;

\textit{Claim 2}. If $\alpha=c?x$ and $x\notin
fv(\mathbf{G}(\widetilde{P}))\cup fv(\mathbf{G}(\widetilde{Q}))$,
then for some $Q$,
$\langle\mathbf{G}(\widetilde{Q}),\rho\rangle\rto{c?x}{\langle
Q,\rho\rangle}$, and for all $y\notin fv(P)\cup fv(Q)-\{x\}$,
$$\langle P\{y/x\},\rho^{\prime}=\rho\rangle\sim \langle
P_1,\rho\rangle\mathcal{R}\langle Q_1,\rho\rangle\sim\langle
Q\{y/x\},\rho\rangle$$ for some $P_1,Q_1$

We only consider the following case as a sample:

Case 1. $\mathbf{G}=\mathbf{Y}(\widetilde{y})$, $\alpha=c?x$ and
$x\notin fv(\mathbf{G}(\widetilde{P}))\cup
fv(\mathbf{G}(\widetilde{Q}))$. Then $\mathbf{Y}=\mathbf{X}_i$ for
some $i\leq m$,
$\mathbf{G}(\widetilde{P})=P_i\{\widetilde{y}/\widetilde{x}_i\}$,
$\mathbf{G}(\widetilde{Q})=Q_i\{\widetilde{y}/\widetilde{x}_i\}$,
and $\rho=\rho^{\prime}$. We choose some $$u\notin
\bigcup_{i=1}^{m}(fv(P_i)\cup fv(Q_i)\cup fv(\mathbf{E}_i))\cup
\{\widetilde{y}\}.$$ Then $u\notin fv(\mathbf{G}(\widetilde{P}))$,
and $\langle P_i\{\widetilde{y}/\widetilde{x}_i\},\rho\rangle$
$\rto{c?u}{\langle P^{\prime},\rho\rangle}$ for some
$P^{\prime}\equiv_{\alpha}P\{u/x\}$. Since
$P_i\sim\mathbf{E}_i(\widetilde{P})$, we obtain
$$P_i\{\widetilde{y}/\widetilde{x}_i\}\sim
\mathbf{E}_i(\widetilde{P})\{\widetilde{y}/\widetilde{x}_i\}=\mathbf{E}_i\{\widetilde{y}/\widetilde{x}_i\}
(\widetilde{P})$$ by Lemma~\ref{sub-sim}. It holds that $$u\notin
fv(P_i\{\widetilde{y}/\widetilde{x}_i\})\cup
fv(\mathbf{E}_i\{\widetilde{y}/\widetilde{x}_i\}(\widetilde{P})).$$
So, we have $$\langle
\mathbf{E}_i\{\widetilde{y}/\widetilde{x}_i\}(\widetilde{P}),\rho\rangle\rto{c?u}{\langle
P^{\prime\prime},\rho\rangle}$$ for some $P^{\prime\prime}$, and
\begin{equation}\label{uniq1}\langle
P^{\prime}\{z/u\},\rho\rangle\sim\langle
P^{\prime\prime}\{z/u\},\rho\rangle\end{equation} for all $z\notin
fv(P^{\prime})\cup fv(P^{\prime\prime})-\{u\}$. By
Lemma~\ref{gu-act} we obtain for some $\mathbf{E}^{\prime}$,
$P^{\prime\prime}=\mathbf{E}^{\prime}(\widetilde{P})$ and
$$\langle
\mathbf{E}_i(\widetilde{Q})\{\widetilde{y}/\widetilde{x}_i\}=\mathbf{E}_i
\{\widetilde{y}/\widetilde{x}_i\}(\widetilde{Q}),\rho\rangle\rto{c?u}{\langle\mathbf{E}^{\prime}(\widetilde{Q}),\rho\rangle.}$$
Note that
$\mathbf{G}(\widetilde{Q})\sim\mathbf{E}_i(\widetilde{Q})\{\widetilde{y}/\widetilde{x}_i\}$,
and $$u\notin fv(\mathbf{G}(\widetilde{Q}))\cup
fv(\mathbf{E}_i(\widetilde{Q})\{\widetilde{y}/\widetilde{x}_i\}).$$
Then for some $Q^{\prime}$,
$\langle\mathbf{G}(\widetilde{Q}),\rho\rangle\rto{c?u}{\langle
Q^{\prime},\rho\rangle}$, and
\begin{equation}\label{uniq2}\langle
\mathbf{E}^{\prime}(\widetilde{Q})\{z/u\},\rho\rangle\sim\langle
Q^{\prime}\{z/u\},\rho\rangle\end{equation} for all $z\notin
fv(\mathbf{E}^{\prime}(\widetilde{Q}))\cup fv(Q^{\prime})-\{u\}$.
Since $x\notin fv(\mathbf{G}(\widetilde{Q}))$, we have $\langle
\mathbf{G}(\widetilde{Q}),\rho\rangle\rto{c?x}{\langle
Q,\rho\rangle}$, where $Q\equiv_\alpha Q^{\prime}\{x/u\}$.

It follows from Lemma~\ref{var} that $fv(P)\subseteq
fv(P_i\{\widetilde{y}/\widetilde{x}_i\})\cup \{x\}$. Then $u\notin
fv(P)-\{x\}$. Since $P^{\prime}\equiv_\alpha P\{u/x\}$, it holds
that $P\equiv_\alpha P^{\prime}\{x/u\}$. We now choose $$v_0\notin
\bigcup_{i=1}^{m}(fv(P_i)\cup fv(Q_i)\cup fv(\mathbf{E}_i))\cup
\{\widetilde{y}\}\cup fv(P)\cup fv(Q).$$ It is obvious that
$fv(P^{\prime})\subseteq fv(P)\cup \{u\}$. On the other hand, we see
that $fv(P^{\prime\prime})\subseteq
fv(\mathbf{E}_i\{\widetilde{y}/\widetilde{x}_i\}(\widetilde{P}))$ by
Lemma~\ref{var}. Thus, $v_0\notin fv(P^{\prime})\cup
fv(P^{\prime\prime})-\{u\}$, and from~(\ref{uniq1}) we obtain
$\langle P^{\prime}\{v_0/u\},\rho\rangle\sim \langle
P^{\prime\prime}\{v_0/u\},\rho\rangle$. Furthermore, it follows from
Lemma~\ref{sub-sim} that $\langle
P^{\prime},\rho\{u/v_0\}\rangle\sim \langle
P^{\prime\prime},\rho\{u/v_0\}\rangle$ and $$\langle
P^{\prime}\{x/u\},\rho\{u/v_0\}\{x/u\}=\rho\{x/v_0\}\rangle\sim
\langle P^{\prime\prime}\{x/u\},\rho\{x/v_0\}\rangle.$$ Then using
Proposition~\ref{alpha-sim} we obtain
\begin{equation*}\begin{split}\langle P,\rho\{x/v_0\}\rangle & \sim\langle
P^{\prime\prime}\{x/u\}=\mathbf{E}^{\prime}(\widetilde{P})\{x/u\}\\
& =
\mathbf{E}^{\prime}\{x/u\}(\widetilde{P}),\rho\{x/v_0\}\rangle\\
& \mathcal{R}\langle
\mathbf{E}^{\prime}\{x/u\}(\widetilde{Q})=\mathbf{E}^{\prime}(\widetilde{Q})\{x/u\},
\rho\{x/v_0\}.\end{split}\end{equation*} Again, we see that
$$fv(\mathbf{E}^{\prime}(\widetilde{Q}))\subseteq
fv(\mathbf{E}_i(\widetilde{Q})\{\widetilde{y}/\widetilde{x}_i\})\cup
\{u\}$$ and $fv(Q^{\prime})\subseteq
fv(Q_i\{\widetilde{y}/\widetilde{x}_i\})$ by Lemma~\ref{var}. Hence,
$v_0\notin fv(\mathbf{E}^{\prime}(\widetilde{Q}))\cup
fv(Q^{\prime})-\{u\}$. This, together with~(\ref{uniq2}), implies
$\langle \mathbf{E}^{\prime}(\widetilde{Q})\{v_0/u\}, \rho \rangle
\sim \langle Q^{\prime}\{v_0/u\},\rho\rangle$. Using
Lemma~\ref{sub-sim} again we obtain
\begin{equation*}\begin{split}\langle
\mathbf{E}^{\prime}\{x/u\} &
=\mathbf{E}^{\prime}\{v_0/u\}\{x/v_0\},\rho\{x/v_0\}\rangle\\ & \sim
\langle
Q^{\prime}\{x/u\}=Q^{\prime}\{v_0/u\}\{x/v_0\},\rho\{x/v_0\}\rangle\\
& \sim\langle Q,\rho\{x/v_0\}\rangle\end{split}\end{equation*}
because $Q\equiv_\alpha Q^{\prime}\{x/u\}$.

Now it suffices to show that for all $y\notin fv(P)\cup
fv(Q)-\{x\}$, we have $$\langle P\{y/x\},\rho\rangle\sim\langle
P^{\prime},\rho\rangle\mathcal{R}\langle
Q^{\prime},\rho\rangle\sim\langle Q\{y/x\},\rho\rangle$$ for some
$P^{\prime},Q^{\prime}$. This can be carried out in a way similar to
that at the end of Case 1 in the proof of
Proposition~\ref{rec-prop}. $\Box$

\subsection{Proof of Theorem~\ref{app-prop}(2.d)}

We can construct $\mathcal{R}_\mu$ by modifying $\mathcal{R}$ in the
proof of Theorem~\ref{sim-prop}(2.f) (see Appendix~\ref{thmpf}). Let
$\mathcal{R}_\mu =\mathcal{B}_\mu \cup \mathcal{R}_\mu^{\prime}$,
where
$$\mathcal{B}_\mu =\{(\langle P,\rho\rangle, \langle
P,\sigma\rangle): D(\rho,\sigma)\leq \mu\},$$ and
$\mathcal{R}_\mu^{\prime}$ consists of the pairs:
\begin{equation*}\begin{split}( \langle P\|R,
\mathcal{F}^{(n)}_{Y_n}\mathcal{E}^{(n)}_{X_n}\mathcal{F}^{(n-1)}_{Y_{n-1}}&\mathcal{E}^{(n-1)}_{X_{n-1}}
...\mathcal{F}^{(1)}_{Y_1}\mathcal{E}^{(1)}_{X_1}\mathcal{F}^{(0)}_{Y_0}(\rho)\rangle,\\
& \langle Q\|R, \mathcal{F}^{(n)}_{Y_n}\mathcal{E}^{\prime
(n)}_{X_n}\mathcal{F}^{(n-1)}_{Y_{n-1}}\mathcal{E}^{\prime
(n-1)}_{X_{n-1}} ...\mathcal{F}^{(1)}_{Y_1}\mathcal{E}^{\prime
(1)}_{X_1}\mathcal{F}^{(0)}_{Y_0}(\sigma)\rangle),\end{split}
\end{equation*} in which it holds that $$\langle P,\mathcal{E}^{(n)}_{X_n}\mathcal{E}^{(n-1)}_{X_{n-1}}
...\mathcal{E}^{(1)}_{X_1}(\rho)\rangle\sim_\mu \langle Q,
\mathcal{E}^{\prime (n)}_{X_n}\mathcal{E}^{\prime (n-1)}_{X_{n-1}}
...\mathcal{E}^{\prime (1)}_{X_1}(\sigma)\rangle.$$ The difference
between $\mathcal{R}$ in the proof of Theorem~\ref{sim-prop}(2.f)
and $\mathcal{R}_\mu^{\prime}$ is that, in
$\mathcal{R}_\mu^{\prime}$, $\mathcal{E}_{X_i}^{(i)}$ and
$\mathcal{E}_{X_i}^{\prime (i)}$ are allowed to be different $(i\leq
n)$. Now it suffices to show that $\mathcal{R}_\mu$ is a strong
$\mu-$bisimulation. It is obvious that $\mathcal{R}_\mu$ is
$\mu-$closed.

Suppose that $\langle P,\rho\rangle\mathcal{B}_\mu \langle
P,\sigma\rangle$ and $\langle P,\rho\rangle\rto{\alpha}{\langle
P^{\prime},\rho^{\prime}\rangle}$. If $\alpha\notin Act_{op}$, then
$\rho^{\prime}=\rho$, and with Lemma~\ref{action} we have $\langle
P,\sigma\rangle\rto{\alpha}{\langle P^{\prime},\sigma\rangle}$ and
$\langle P^{\prime},\rho^{\prime}\rangle\mathcal{B}_\mu\langle
P^{\prime},\sigma\rangle$ (for the case that $\alpha =c?x$ and
$x\notin fv(P),$ it holds that $\langle
P^{\prime}\{y/x\},\rho^{\prime}\rangle\mathcal{B}_\mu\langle
P^{\prime}\{y/x\},\sigma\rangle$ for all $y\notin
fv(P^{\prime})-\{x\}$). If $\alpha =\mathcal{E}[X]$, then
$\rho^{\prime}=\mathcal{E}_X(\rho)$, and by Lemma~\ref{action} we
obtain $\langle P,\sigma\rangle\rto{\alpha}{\langle
P^{\prime},\mathcal{E}_X(\sigma)\rangle}$. It follows from
Lemma~\ref{op-dist} that $D(\rho^{\prime},\mathcal{E}_X(\sigma))\leq
D(\rho,\sigma)\leq \mu$ and $\langle
P^{\prime},\rho^{\prime}\rangle\mathcal{B}_\mu \langle
P^{\prime},\mathcal{E}_X(\sigma)\rangle$.

Finally, we use the symbols $\mathcal{A}$ and $\mathcal{B}$ in the
same way as in the proof of Theorem~\ref{sim-prop}(2.f), and let
$\mathcal{A}^{\prime}$ and $\mathcal{B}^{\prime}$ be obtained by
replacing $\mathcal{E}_{X_i}^{(i)}$ with $\mathcal{E}_{X_i}^{\prime
(i)}$ $(i\leq n)$ in $\mathcal{A}$ and $\mathcal{B}$, respectively.
Suppose that $\langle P,\mathcal{A}(\rho)\rangle\sim_\mu \langle
Q,\mathcal{A}^{\prime}(\sigma)\rangle$ and $\langle P\|R,
\mathcal{B}(\rho)\rangle\rto{\alpha}{\langle
S,\rho^{\prime}\rangle}$. We only consider the case that $\alpha
=\mathcal{G}[Z]$ and the transition is derived by \textbf{Intl2}
from $\langle P,\mathcal{B}(\rho)\rangle\rto{\mathcal{G}[Z]}{\langle
P^{\prime},\mathcal{G}_Z\mathcal{B}(\rho)\rangle}$ (and the other
cases are the same as in the proof of Theorem~\ref{sim-prop}(2.f)).
It holds that $S=P^{\prime}\|R$ and
$\rho^{\prime}=\mathcal{G}_Z\mathcal{B}(\rho)$. An application of
Lemma~\ref{action} leads to $$\langle
P,\mathcal{A}(\rho)\rangle\rto{\mathcal{G}[Z]}{\langle
P^{\prime},\mathcal{G}_Z\mathcal{A}(\rho)\rangle}.$$ Since $\langle
P,\mathcal{A}(\rho)\rangle\sim_\mu \langle
Q,\mathcal{A}^{\prime}(\rho)\rangle$, there are
$\mathcal{G}^{\prime}$ and $Q^{\prime}$ such that $$\langle
Q,\mathcal{A}^{\prime}(\sigma)\rangle\rto{\mathcal{G}^{\prime}[Z]}{\langle
Q^{\prime},\mathcal{G}^{\prime}_Z\mathcal{A}^{\prime}(\sigma)\rangle}\sim_\mu
\langle P^{\prime},\mathcal{G}_Z\mathcal{A}(\rho)\rangle$$ and
$D_\diamond(\mathcal{G},\mathcal{G}^{\prime})\leq \mu$. Then, using
Lemma~\ref{action} once again, we obtain $$\langle
Q,\mathcal{B}^{\prime}(\sigma)\rangle\rto{\mathcal{G}^{\prime}[Z]}{\langle
Q^{\prime},
\mathcal{G}^{\prime}_Z\mathcal{B}^{\prime}(\sigma)\rangle},$$ and it
follows that $$\langle
Q\|R,\mathcal{B}^{\prime}(\sigma)\rangle\rto{\mathcal{G}^{\prime}[Z]}{\langle
Q^{\prime}\|R,
\mathcal{G}^{\prime}_Z\mathcal{B}^{\prime}(\sigma)\rangle}.$$ It is
easy to see that $\langle
S,\rho^{\prime}\rangle\mathcal{R}_\mu^{\prime}\langle Q^{\prime}\|R,
\mathcal{G}^{\prime}_Z\mathcal{B}^{\prime}(\sigma)\rangle$ from the
definition of $\mathcal{R}_\mu^{\prime}$. $\Box$

\begin{received}
Received October 2007; revised April 2008; accepted May 2008
\end{received}

\end{document}